# Generalized formulation for ideal light-powered systems through energy and entropy flow analysis

# Part 2: Beyond the first-order evaluation under realistic conditions

Tetsuo Yabuki

School of Economics, Hokusei Gakuen University, 2-3-1 Oyachi-nishi,

Atsubetsu-ku, Sapporo,004-8631, Japan

E-mail address: t-yabuki@hokusei.ac.jp

Rakuno Gakuen University,582 Bunkyo-dairi, Ebetsu, Hokkaido, Japan

**Abstract** (250words)

This study formulates the ideal efficiency of *light-powered systems* in the most general form, based on the first principle of energy-entropy flow analysis under the condition of zero entropy generation within the system. A unified formula for the ideal efficiency of *light-powered systems* is presented in this study. The formula incorporates the absorption ratio $|\varepsilon|$ as an indicator beyond the first-order evaluation based on photon number, for light with a dilution indicator $d$, and it is extended to cases where entropy is simultaneously discarded from the system via radiation and heat. Selecting the appropriate $Y$-factors and $p$-parameters from this study for given conditions allows us to accurately and systematically derive the ideal efficiencies of *light-powered systems* and correctly classify the multiple ideal efficiencies that were previously confused, such as the Jeter, Spanner, and Landsberg-Petela efficiencies which form the basis of practical efficiency. This study also classified existing *light-powered systems* into two models: the piston-cylinder radiation model and the flowing radiation model, and demonstrated that the latter model is suitable for micro *light-powered systems*. Finally, this study clarified two issues with the ideal efficiency proposed by Landsberg and Tonge (often referred to as the Landsberg limit) based on the classical flowing radiation model, and derived a new ideal efficiency using a simple mathematical model based on Einstein's theory of radiation and absorption in a two-level system, which assumes quantum transitions, to resolve those problems. The newly obtained ideal efficiency was found to behave very similarly to the Carnot efficiency.

## 1. Introduction

Solar energy and the natural and artificial photovoltaic systems have received significant attention [1-7] concerning agricultural and global warming issues. Their ideal efficiency of these is being discussed since the mid-20th century. In part 1 of this study, the relevant literature was reviewed, starting from the first formulation by Duysens [8] on monochromatic light, and the most general formula for it was constructed. There have also been a lot of previous studies on non-monochromatic light [9-22], and they are consistently referenced in research on solar cells, next-generation solar power generation devices, and artificial photosynthesis [1-7] as the theoretical ideal efficiencies underlying practical techniques [23-28]. However, our understanding of the theory of ideal efficiency for *light-powered systems* remains inadequate, and the derivation conditions for ideal efficiencies were not correctly understood. This prevented correct classification and unification of multiple ideal efficiencies, despite several papers aiming to unify them having been published.

The ideal efficiency of photosynthesis at the surface of the Earth was first formulated by Duysens [8] in the 1950s as a theoretical maximum energy efficiency for monochromatic light. He considered the dilution effect of sunlight emitted by the Sun as blackbody radiation at about 5800 K, reaching the surface of the Earth from a distance of about 150 million km. His formulation involved two steps: 1) Substituting the photon number flux (light intensity) reduced by the dilution effect into the blackbody radiation equation (Planck radiation equation) and solving for temperature; 2) Substituting the resulting temperature (called the effective temperature) into Carnot efficiency formula, the ideal efficiency of a heat engine. However, owing to dilution effects, the sunlight reaching the surface of the Earth does not remain blackbody radiation, making this effective temperature wavelength dependent, represented as $T_\gamma(\lambda)$ in this paper. Although this effective temperature has been widely used in the physicochemical analysis of the theoretical efficiency of photosynthesis, certain questions and critical discussions remain unresolved (for detailed references, see Part I of this study [29], [30]). These questions stem mainly from the fact that this temperature $T_\gamma(\lambda)$, which is not in equilibrium (i.e., not a true temperature), is automatically applied to Carnot efficiency and Boltzmann-type factor formulae, as $\eta_{max} = 1 - T_{out}/T_\gamma(\lambda)$ and $\exp\left(-\Delta E/k_B T_\gamma(\lambda)\right)$, respectively, without proper justification. This approach is not only logically flawed but also fails to accurately calculate ideal efficiencies that consider various variables, such as solid angle, polarization, absorption ratio (the number ratio of photons absorbed by the system to photons irradiated), and entropy changes due to photochemical reactions within the system. Furthermore, this method implicitly assumes a quasi-equilibrium state between the radiation and the system, and so it is not possible to analyze the non-equilibrium contribution between the two, which is analyzed in Part 2 of this study.

Part 1 of this study, instead of using the traditional approach, has constructed and presented the correct formula based on the energy-entropy analysis, and has derived the ideal efficiency and Boltzmann-type factor for arbitrary solid angles and polarization degrees. The method used in Part1 of this study first derived a general formula for the theoretical maximum energy efficiency, $\eta_{max}$, through energy–entropy flow analysis based on the second law of thermodynamics. Second, it extracted the temperature dimensional quantity from this formula. This quantity coincided with the effective temperature $T_\gamma(\lambda)$ within the first-order evaluation (for a quasi-reversible condition) based on the photon number change rate using $\varepsilon = \Delta N^\gamma(\lambda)/N^\gamma(\lambda)$ (where $N^\gamma(\lambda) < 0$ is the number of photons contained in the irradiation, and $\Delta N^{\gamma}(\lambda)$ is its change, which is decreasing), provided that the solid angle $\Omega$ and the degree of polarization $P$ of the photons are 4π and 0, respectively (for details, see Part 1 of this study [29],[30]).

This Part2 formulates the ideal efficiency of *light-powered systems* in the most general form, beyond the first-order evaluation. Following Part1, this formulation is based on the first principle of energy-entropy flow analysis, under the condition of zero entropy generation within the system. The established formula is represented by $Y$-factors, which are functions of the dilution indicator $d$ defined by the relative concentration and the absorption ratio $|\varepsilon|$, the latter of which serves as an indicator beyond the first-order evaluation based on the number of photons $\Delta N^{\gamma}$ flowing into the system. This part 2 solves the problems and issues in the previous studies on the ideal efficiency of light-powered system not only monochromatic light but also with the entire spectrum by employing the new formula. This formula includes cases in which entropy is discarded from the system via both heat and radiation, and the weights of the entropy discarded via radiation and heat are represented by the parameters, $p_{\gamma}$ and $p_Q$, respectively.

This study quantitatively analyzes the non-equilibrium contributions beyond the first-order evaluation range by the following steps. These contributions increase with the light absorption ratio , $|\varepsilon| = |\Delta N^{\gamma}(\lambda)|/N^{\gamma}(\lambda)$, i.e., the ratio of the number of photons absorbed by the *light-powered system* $|\Delta N^{\gamma}(\lambda)|$to the number of irradiated photons $N^{\gamma}(\lambda)$, and reduce the ideal efficiency $\eta_{upper}$ of *a light-powered system*. First, the ideal efficiency $\eta_{upper}(\lambda, |\varepsilon|)$ of a *light-powered system* absorbing monochromatic light with a wavelength $\lambda$ and a photon number flux $n(\lambda)$ at a light absorption ratio $|\varepsilon|$ is formulated. Second, a unified formula is derived in a very simple form for the ideal efficiency $\eta_{upper}(T, T_{out}, d, |\varepsilon|)$, where $T_{out}$ is the ambient temperature, of a *light-powered system* with the absorption ratio |ε|, for non-monochromatic light diluted with the dilution indicator $d$ after being emitted by a blackbody radiation source at a temperature $T$.

The effects of the absorption ratio $|\varepsilon|$ and the dilution indicator $d$ are reflected in the equation: $S_{in}^{\gamma}(T, d, |\varepsilon|) = Y_{in}(d, |\varepsilon|)\frac{E(d,|\varepsilon|)}{T}$, which can be regarded as the extended Clausius equation in a non-equilibrium condition. In the absence of dilution ($d = 1$) and when the light absorption ratio is infinitely small ($|\varepsilon| \to 0$), $Y(d = 1, |\varepsilon| \to 0) = 1$ on the right-hand side of this equation, yielding the Clausius equation of ordinary (equilibrium) thermodynamics. In this study, the ideal efficiency of a *light-powered system* has been derived as the most general formula as follows.

$$\eta_{upper} = 1 - \left(\frac{p_Q}{Y_{out}^{Q}} + \frac{p_{\gamma}}{Y_{out}^{\gamma}}\right) Y_{in}^{\gamma} \frac{T_{out}}{T_{in}}, \tag{1}$$

where $Y_{out}^{Q}$and $Y_{out}^{\gamma}$ are the $Y$-factors for the entropy discarded via heat (**thermal discarded entropy)** and radiation (**radiative discarded entropy)** from the system, respectively, and $p_Q$ and $p_{\gamma}$ are their respective weights (ratios); $Y_{in}^{\gamma}$ corresponds to the entropy flowing via the blackbody radiation into the system. As explained in Section 6, this formulation includes the Jeter efficiency ($\eta_{Jeter} = 1 - \frac{T_{out}}{T}$ [19]) and the Spanner efficiency ($\eta_{Spanner} = 1 - (4/3)\frac{T_{out}}{T}$) [12,13] for the case of using undiluted solar radiation and only thermal discarded entropy with no radiative discarded entropy. In addition, this formulation is also extended to the case where entropy is discarded from the system both by blackbody radiation and waste heat, such as the Landsberg-Petela efficiency $\eta_{Landsberg-Petela} = 1 - (4/3)T_{out}/T + (1/3)(T_{out}/T)^4$ [9,15]. This enabled a correct unified classification of several ideal efficiencies, proposed in previous studies, by flow conditions. These ideal efficiencies (the Jeter, Spanner and Landsberg-Petela efficiencies) are also cited in the recent review article [23,26-28] on renewable and sustainable energy as the basis for the practical efficiency of solar cells.
In addition, this paper categorizes prior frameworks on *light-powered systems* into box-type (**piston-cylinder model, closed photon gas model**) and flow-type **(flowing radiation model**, **open photon gas model**), demonstrating quantitatively that the latter is more suitable for microscopic systems. However, the ideal efficiency derived from the prior flow-type framework (Landsberg and Tonge, 1980 [9]) has two issues: 1) The condition for deriving the radiative discarded entropy is that it is emitted by blackbody radiation within the system. 2) As soon as the energy source's temperature becomes lower than the ambient temperature, the Petela–Landsberg efficiency becomes positive, which probably contradicts the first law of thermodynamics (conservation of energy). Regarding (1), this study assumed that the sink condition was necessary to achieve Landsberg efficiency, which is outer space with an absolute temperature close to zero. Regarding (2), this study analyzed negative waste heat as positive environmental heat energy when the temperature of the energy source radiation was lower than the ambient temperature. Consequently, a modified Landsberg efficiency formula has been presented in Eq. (6.30).These issues were solved more rationally in this study through quantitative analysis using a simplified mathematical model, based on Einstein's absorption and radiation theory in a two-level system that assumes quantum transitions. The new ideal efficiency obtained is shown in Eq. (8.36) and in Figures 11b and 12. This is explained in detail in Chapter 8.

From Section 4 onwards, quantities related to radiation are denoted by a superscript $\gamma$, e.g., $\eta^{\gamma}_{max}(\lambda)$, $E^{\gamma}_{in}(\lambda)$, $S^{\gamma}_{in}(\lambda)$, $N^{\gamma}_{in}(\lambda)$, etc. In addition, the following parts of this paper relate to quantum theory: (1) analyzing radiation as a population of photons, (2) using the indistinguishability of identical quantum particles in the formulation of photon entropy, and (3) making the phase volume dimensionless using the cube of Planck's constant $h$ as the unit. However, this study does not consider quantum coherence and other factors. In other words, this study analyzes radiation under the condition that the correlation between the phases of each photon is close to zero, meaning the phase uncertainty is sufficiently large. Therefore, it treats radiation using the particle picture of the photon, such as blackbody radiation.

Furthermore, in this study, a l*ight-powered system* is defined not only as a system that outputs electrical energy, such as solar power generation, photovoltaics and solar cells, etc., but also as a system from which any kind of physical work is extracted using light energy, including natural and artificial photosynthesis. In this paper, the "theoretical maximum energy efficiency" represented by Carnot efficiency of a heat engine is written as $\eta_{max}$, and the "ideal efficiency" under conditions beyond the first-order evaluable conditions in this study is written as $\eta^{\gamma}_{upper}$. Further details on this point can be found in Section 7 of this paper.

## 2. Flow analysis

The general powered system, including the heat-powered (heat engine) and radiation-powered systems, consists of three parts: an energy source, a system that outputs work, and a heat sink (environment outside the system). The energy flow consists of three flows: 1) flow from the energy source to system ($E_{in}$), 2) flow extracted as work ($W_{out}$), and 3) flow emitted from the system required for entropy discard ($E_{out}$). Here the term 'flow' refers to the flow of energy and entropy into and out of a common time. The flow per unit area and per unit time is called 'flux'.

The entropy flow associated with the energy flow satisfies (Fig. 1)

$$S_{out} = S_{in} + S_g - S_W, \tag{2.1}$$

where the four types of entropy flow are represented as

$S_{in}$: Entropy imported into a powered system by the source energy

$S_g$: Entropy generated within a powered system

$S_{out}$: Entropy discarded from the power system

$S_W$: Entropy contained in the power output from a powered system (2.2)

**Fig. 1. Energy and entropy flows into and out of a powered system, and entropy generated in the**

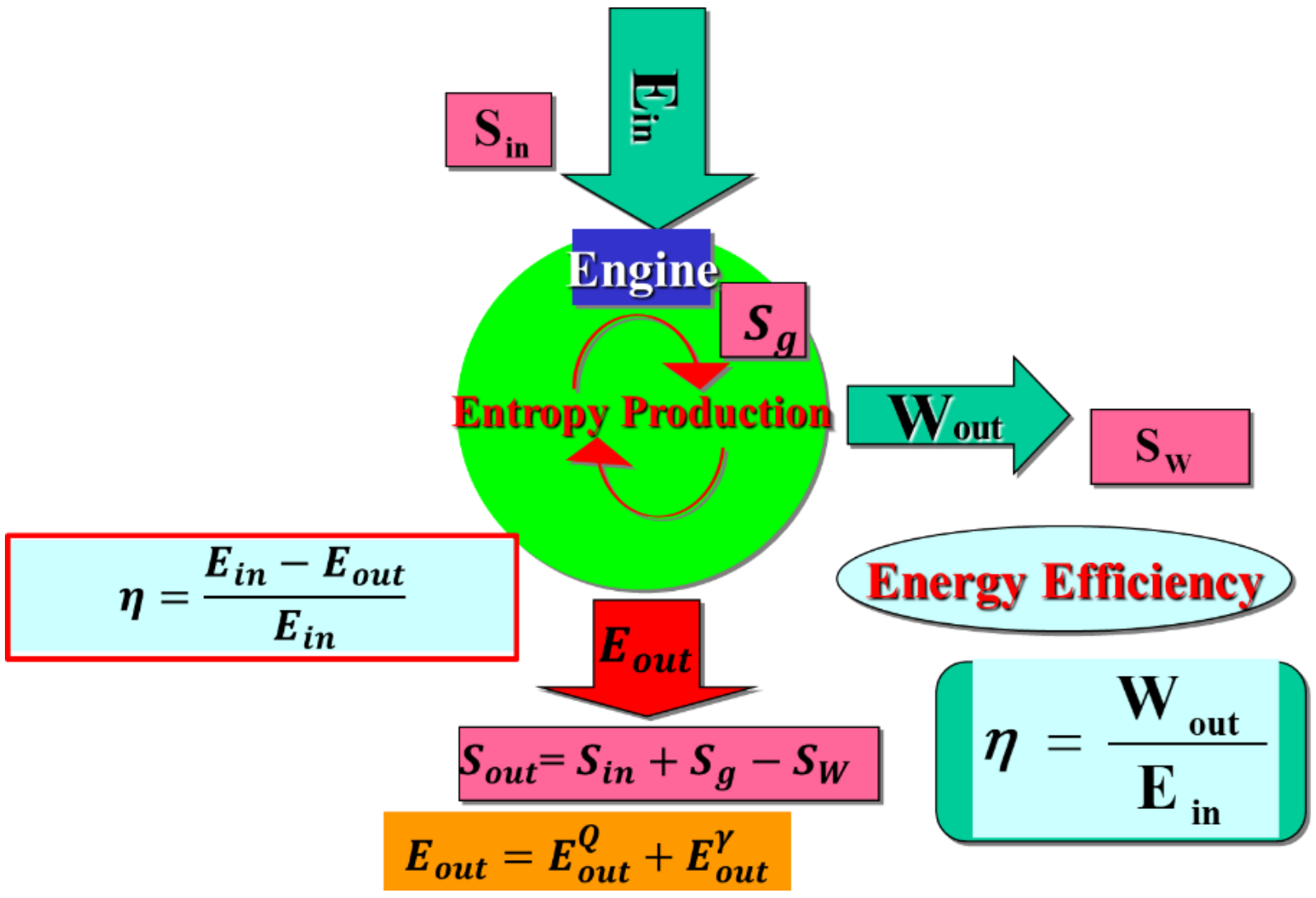

**system.** This figure shows the flow diagram for a powered system, which forms the basis of the analysis in this study. the basis of the analysis in this study. The energy source $E_{in}$ and its entropy $S_{in}$ are analyzed as radiation energy and radiation entropy $E^{\gamma}_{in}$, $S^{\gamma}_{in}$, respectively, and $S_{out}$ has radiation entropy $S^{\gamma}_{out}$ as well as thermal entropy $S^{Q}_{in}$.

According to the second law of thermodynamics, the total entropy can never decrease, i.e., $S_g \geq 0$ is satisfied. Furthermore, according to the first law of thermodynamics, which implies that total energy must be conserved, $W_{out} = E_{in} - E_{out}$ is satisfied. By the definition of energy efficiency $\eta = W_{out}/E_{in}$, we have

$$\eta = \frac{E_{in} - E_{out}}{E_{in}} = 1 - \frac{E_{out}}{E_{in}}. \quad (2.3)$$

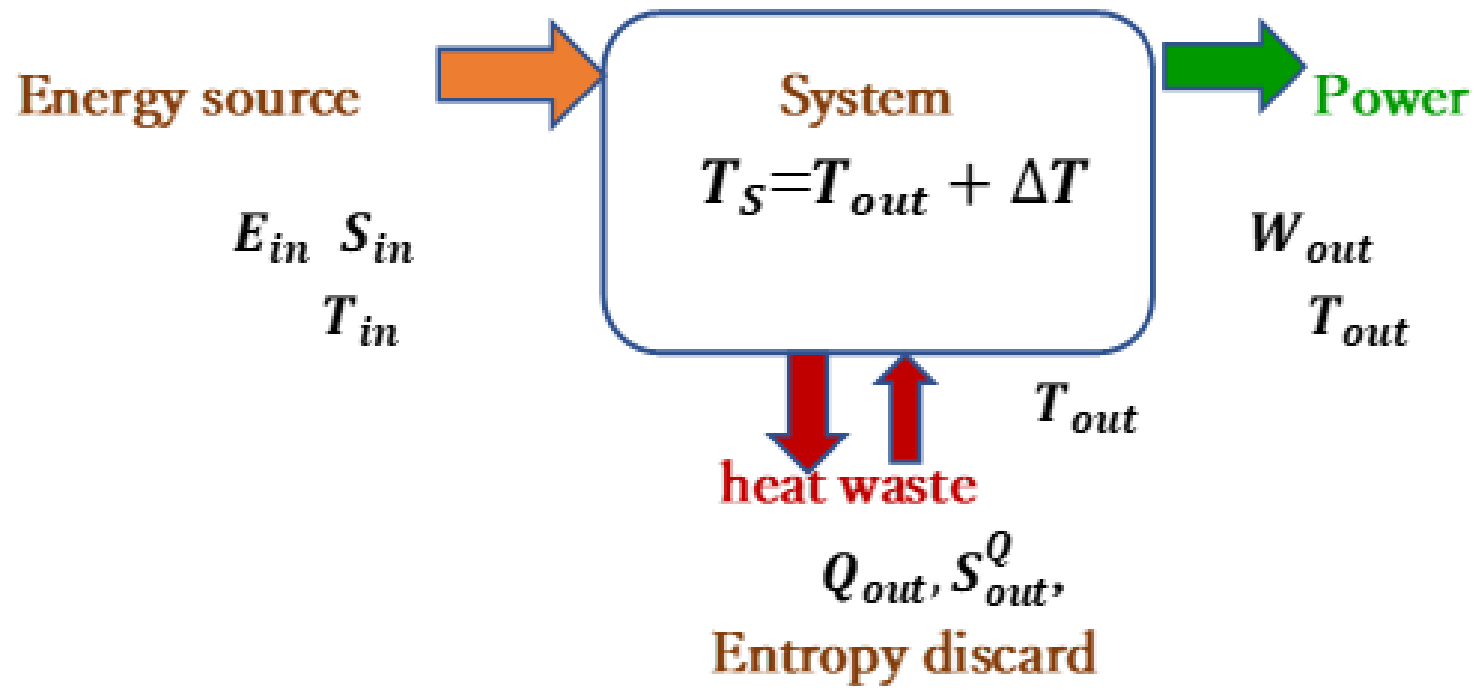

**Fig. 2. Schematic representation of the energy and entropy fluxes into and out of the system:** This diagram corresponds to the first analysis in this paper for a *light-powered system* (such as photosynthesis), where the energy source is radiation (e.g., solar radiation) and entropy is discarded via waste heat. (In this figure, the temperatures of the inflow and outflow radiation of the system are represented by $T_{in}$ and $T_{out}$, $T_{out} + \Delta T_{out}$, respectively).

In this study it is first assumed that: 1) the entropy is discarded from the system only via heat, and 2) the temperature difference $\Delta T$ between the system $\langle S \rangle$ and its external environment $\langle E \rangle$ during the process of entropy discard is sufficiently small ($\Delta T \ll 1$); Therefore, the net waste heat flow for entropy discarding, expressed as the subtraction of the outward flow from the system to the environment and its reverse inward flow, can be formulated by a first-order evaluation of $\Delta T/T$ (Fig. 2). As a result, the net discarding of entropy, $S_{out}^{Q}$, from the system is reduced to the Clausius equation as follows.

$$\begin{aligned} S_{out}^{Q} &= C_V \ln(T_{out} + \Delta T) - C_V \ln T_{out}, \\ &= C_V \ln(1 + \Delta T/T_{out}) \cong C_V \, \Delta T/T_{out} = Q_{out}/T_{out}, \end{aligned} \quad (2.4)$$

where $C_V$ is the heat capacity at constant volume. From Eq. (2.4), $E_{out} = Q_{out}(E_{out}^{Q}) = T_{out} S_{out}^{Q}$ is satisfied, and from Eqs. (2.3) and (2.1) the following formula is obtained:

$$\eta = 1 - \frac{T_{out} S_{out}^{Q}}{E_{in}} = 1 - \frac{T_{out}(S_{in} + S_g - S_W)}{E_{in}} \leq 1 - \frac{T_{out}(S_{in} - S_W)}{E_{in}} = 1 - \frac{T_{out} S_{in}}{E_{in}}. \quad (2.5)$$

The inequality in Eq. (2.5) is derived from $S_g \geq 0$ according to the second law of thermodynamics, and from $S_W = 0$, which is usually assumed for motive power. Thus, the ideal efficiency $\eta_{upper}$ can be expressed as

$$\eta_{upper}^{\gamma} = 1 - \frac{T_{out}S_{in}}{E_{in}}. \tag{2.6}$$

The entropy imported into a powered system, $S_{in}^{\gamma}$, by the radiation as source energy is quantitatively analyzed under conditions where the light absorption ratio $|\varepsilon| = |\Delta N^{\gamma}(\lambda)|/N^{\gamma}(\lambda)$ varies from 0 to 1.

## 3. First-order evaluation

### 3.1. $\eta_{max}^{\gamma}(\lambda)$ under monochromatic light

First, $S_{in}^{\gamma}$ is considered quantitatively for monochromatic light.

The entropy of radiation is obtained by applying the mathematical formula $S = k_B \ln W$ (where $k_B$ is the Boltzmann constant, ln is the natural logarithm, and $W$ is the total number of accessible microscopic states for a particle constituting the macroscopic state of an ensemble), originally formulated by Boltzmann in statistical mechanics, to radiation as a population of photons based on Planck's radiation theory:

As a result, the following Eq. (3.1) is obtained as the formula for the quantum statistical entropy of radiation as a photon ensemble, which is valid in any thermodynamic state, including non-equilibrium:

$$S(\lambda) = k_B G(\lambda)\{(1 + f(\lambda))\ln(1 + f(\lambda)) - f(\lambda)\ln f(\lambda)\}, \tag{3.1}$$

which can be derived from $W = {}_{N+G-1}C_N = \frac{(N+G-1)!}{(G-1)!N!} \sim \frac{(N+G)!}{G!N!}$ based on the Stirling approximation, where $N = N(\lambda, \Omega)$ is the number of photons contained in the radiation or the average number of photons in the case of thermal equilibrium (blackbody radiation) and $G = G(\lambda, \Omega)$ is the number of quantum states accessible to a photon with wavelength $\lambda$, within the solid angle $\Omega$, and $f(\lambda, \Omega)$ is a distribution function expressed as $f(\lambda, \Omega) = \frac{N(\lambda,\Omega)}{G(\lambda,\Omega)}$. In general, the number of quantum states $G$ contained in a phase space volume $\Delta q^3 \Delta p^3$ can be counted by the unit $h^3$ (where $h$ is the Planck constant), based on the uncertainty relation between $q$ and $p$, $\Delta q \Delta p \sim h$, as $G = \Delta q^3 \Delta p^3 / h^3$.

After some calculations on this basis, we obtain:

$$f(\lambda) = \frac{N(\lambda)\,\lambda^4}{8\pi\Delta\lambda V}, \tag{3.2}$$

where $V$ is the volume accessible to a photon. In the case of thermal equilibrium at temperature $T$, i.e., blackbody radiation, $f(\lambda, T)$ is obtained as the following Bose-Einstein distribution function, which maximizes the entropy expressed in Eq. (3.1) under the constraint that the total energy is constant:

$$f_{BB}(\lambda, T) = \frac{1}{\exp\left(\frac{hc}{\lambda k_B T}\right) - 1}, \tag{3.3}$$

where $k_B$ is the Boltzmann constant.

The change in radiation entropy given by$\Delta\boldsymbol{S}^{\gamma}(N^{\gamma}(\lambda),\lambda) = \boldsymbol{S}^{\gamma}(N^{\gamma}(\lambda) + \Delta N^{\gamma}((\lambda),\lambda) - \boldsymbol{S}^{\gamma}(N^{\gamma}(\lambda),\lambda)$ due to the change in photon number can be represented as the first-order evaluation based on $\Delta N^{\gamma}/N^{\gamma}$ using Eq. (3.1) as

$$\Delta S^{\gamma(1)}(N^{\gamma}(\lambda),\lambda) = \frac{\partial \boldsymbol{S}^{\gamma}}{\partial N^{\gamma}}\Delta N^{\gamma} = \frac{\partial \boldsymbol{S}^{\gamma}}{\partial f}\frac{\partial f}{\partial N^{\gamma}}\Delta N^{\gamma} = k_B \ln\left(1+\frac{1}{f}\right)\Delta N^{\gamma}(\lambda), \tag{3.4}$$

where $\Delta N^{\gamma}(< 0)$ is the reduction in the number of photons of the irradiating radiation due to the light absorption of a system. In this paper, quantities that can be obtained as a first-order evaluation based on the photon numbers are denoted by a superscript (1) in the upper right-hand corner (e.g., $S_{in}^{\gamma(1)}$, $\Delta S^{\gamma(1)}$ etc). When the number of photons imported into a *light-powered system* is expressed by $N_{in}^{\gamma}(\lambda)$,then $N_{in}^{\gamma}(\lambda) = -\Delta N^{\gamma}(\lambda)$, and $S_{in}^{(1)}(\lambda,\Omega) = -\Delta S^{(1)}(\lambda,\Omega)$. Therefore, Eq. (3.4) gives:

$$S_{in}^{(1)}(\lambda) = k_B \ln\left(1+\frac{1}{f(\lambda)}\right) N_{in}^{\gamma}(\lambda). \tag{3.5}$$

Substitution of Eq. (3.5) into Eq. (2.6) and $E_{in}(\lambda) = (hc/\lambda)N_{in}^{\gamma}(\lambda)$ gives

$$\eta_{upper}{}^{(1)}(\lambda) = 1 - \frac{T_{out}S_{in}^{(1)}(\lambda)}{E_{in}(\lambda)} = 1 - \frac{T_{out}k_B \ln\left(1+\frac{1}{f(\lambda)}\right)}{hc/\lambda}. \tag{3.6}$$

If we define the temperature dimensional quantity in Eq. (3.6) as

$$T_{\gamma}(\lambda) = \frac{E_{in}^{\gamma}(\lambda)}{S_{in}^{\gamma(1)}(\lambda)} = \frac{hc/\lambda}{k_B \ln\left(1++\frac{1}{f(\lambda)}\right)}, \tag{3.7}$$

then, we get

$$\eta_{upper}^{\gamma}{}^{(1)}(\lambda) = 1 - T_{out}/T_{\gamma}(\lambda) = \eta_{max}^{\gamma}(\lambda), \tag{3.8}$$

where $\eta_{max}^{\gamma}(\lambda)$ is denoted as the normal theoretical maximum efficiency, obtained implicitly using the first-order approximation condition, and is distinguished in this study from the ideal efficiency $\eta_{upper}^{\gamma}(\lambda)$ for the following reasons.

In the case of a heat engine, if the first-order evaluation condition is not satisfied (if $\Delta T/T \ll 1$ is not satisfied), even if the entropy generation in the system is 0 ($S_g = 0$), the entropy will increase during the process of heat transfer from the high temperature energy source to the engine. In the case of a light-powered system, a similar increase in entropy is considered in the Closed Photon Gas model defined in Section7 in this paper, and there is also the possibility of this occurring in the case of the Flowing Radiation model defined in Section7 in this paper. For this reason, in this study which goes beyond the first-order evaluation condition, i.e., breaks the equilibrium condition between the energy source and system, the ideal efficiency $\eta_{upper}^{\gamma}$ is distinguished from the maximum efficiency $\eta_{max}$ under the first-order evaluation conditions of a conventional heat engine. Therefore

$$\eta_{upper}^{\gamma} \leq \eta_{max} \tag{3.9}$$

generally holds.

$T_\gamma(\lambda)$, defined by Eq. (3.7), coincides with the source temperature $T$ in the case of the distribution function $f_{BB}(\lambda, T)$ for blackbody radiation given by Eq. (3.3) and with the conventional radiation temperature $T_\gamma(\lambda)$ for other distribution functions $f(\lambda)$. For solar radiation as blackbody radiation, $T_\gamma(\lambda) = T_{sun}$, i.e., the solar temperature. For terrestrial solar radiation diluted with the dilution indicator $d = (R/D)^2$ (where $R$ is the radius of the Sun and $D$ is the Sun–Earth distance), substituting $f(\lambda) = (R/D)^2 f_{BB}(\lambda, T)$ into Eq. (3.7) gives $T_\gamma(\lambda) = T_D(\lambda)$, i.e., the terrestrial effective temperature [8,31-33] (for detailed references, see Part I of this study [29], [30] ), as one of the conventional radiation temperatures. In any case, the radiation temperature defined by Eq. (3.7) assumes a quasi-equilibrium condition (quasi-reversibility condition) between the radiation and the system, as can be seen from the derivation process in this study.

### 3.2. Boltzmann-type factor for the concentration ratio in the pigment system under monochromatic light

In this section, we consider a two-level pigment system consisting of the ground state pigment $P$ and the excited state pigment $P^*$, where light absorption promotes $P$ to $P^*$. The specific considerations and equation development are fully described in the previous work of the present author [29,30]. Here, only the essential points are described, focusing on the first-order evaluation of the entropy changes of the pigment molecules that make up the light-harvesting aggregates of the photosynthetic system, which are Fermi molecules. The entropy of the pigment composition of such a system, constituting a two-level system of ground and excited states, is obtained in the final form as follows [29,30]:

$$S_P(n_P) = -k_B N_P\{(1 - f_P)\ln(1 - f_P) + f_P \ln f_P\}, \tag{3.10}$$

where $f_P$ is the ratio of the number of ground state pigments, $n_P$, to the total number of pigments, $N_P$, i.e., $f_P = n_P/N_P$. Consequently, $(1 - f_P)$=$(N_P - n_P)/\ N_P = n_{P^*}/N_P$, where $n_{P^*}$ is the number of excited state pigments. When such a pigment system is exposed to a certain intensity of radiation (a certain photon number flux per wavelength $\hat{n}_\gamma(\lambda)$), the entropy of radiation is transferred to the pigment system. Using Eq. (3.10), the first-order evaluation of the entropy change $\Delta S_p(n_P)$ due to the reduction $\Delta n_P(< 0)$ in the ground state pigment molecule caused by photon absorption in the system can be obtained as

$$\begin{aligned}\Delta S_P^{(1)}(n_P) &= \frac{\partial S_p(n_P)}{\partial n_P}\Delta n_P \\ &= \frac{\partial S_p(n_P)}{\partial f_P}\frac{1}{N_P}\Delta n_P \\ &= k_B \ln\left(-1 + \frac{1}{f_P}\right)\Delta n_P.\end{aligned} \tag{3.11}$$

Eq. (3.11) can be represented in terms of $n_{P^*}$ and $n_P$ as

$$\Delta S_p^{(1)}(n_P) = k_B \ln\left(\frac{n_{P^*}}{n_P}\right)\Delta n_P = k_B \ln\left(\frac{[P^*]}{[P]}\right)\Delta n_P \tag{3.12}$$

Eq. (3.7) is used to obtain the following formula for the entropy of a photon population (radiation) absorbed by the pigments:

$$S_{in}^{\gamma(1)}(\lambda) = \frac{hc/\lambda}{T_\gamma(\lambda,\hat{n}(\lambda))} N_{in}^{\gamma}(\lambda). \quad (3.13)$$

Thus, assuming that when radiation energy is absorbed by a pigment system, its entropy is also absorbed without producing additional entropy in the process, i.e., the reversibility condition, the absolute decrease in radiation entropy $\Delta S_\gamma$ is equal to the increase in the entropy of the system $\Delta S_p$, i.e., $\Delta S_\gamma + \Delta S_p = 0$. Therefore, $\Delta S_p(n_P) = S_{in}(\lambda)$ is satisfied because $S_{in}(\lambda,\Omega)$ corresponds to $(-\Delta S_\gamma)$. Furthermore, the following equation is satisfied by considering Eqs. (3.12) and (3.13):

$$\Delta S_p^{(1)}(n_P) = k_B \ln\left(\frac{[P^*]}{[P]}\right) \Delta n_P = S_{in}^{\gamma(1)}(\lambda) = \frac{hc/\lambda}{T_\gamma(\lambda,\hat{n}(\lambda))} N_{in}^{\gamma}(\lambda). \quad (3.14)$$

The number of photons absorbed by a pigment system is equal to the number of pigments excited by photon absorption. This is because the process of absorbing photons into the pigment occurs through a one-to-one interaction between a photon and an electron in the pigment. Thus, $N_{in}^{\gamma}(\lambda) = \Delta n_{P^*} = -\Delta n_P$ is satisfied, and Eq. (3.14) gives

$$-k_B \ln\left(\frac{[P^*]}{[P]}\right) = \frac{\frac{hc}{\lambda}}{T_\gamma(\lambda,\hat{n}(\lambda))}, \quad (3.15)$$

and Eq. (3.7) gave

$$\frac{[P^*]}{[P]} = \exp\left(-\frac{hc/\lambda}{k_B T_\gamma(\lambda,\hat{n}(\lambda))}\right), \quad (3.16)$$

which is the Boltzmann-type factor [29,30,34,35] modified by the radiation temperature obtained from the first-order evaluation.

## 4. Deviation from the first-order evaluation for $\eta_{upper}^{\gamma}(\lambda)$ and for the Boltzmann-type factor, due to a finite light absorption ratio $|\varepsilon| = -\Delta N_\gamma / \Delta N_\gamma$

As mentioned in the previous section, the ideal efficiency and Boltzmann-type factor have been analyzed under the assumption of sufficiently small ratio of absorbed photons in the *light-powered system*, $-\Delta N^\gamma / N^\gamma$, enabling a first-order evaluation based on $\varepsilon = \Delta N^\gamma / N^\gamma$. However, the real *light-powered systems* can be reasonably assumed to operate with a finite photon absorption ratio $|\varepsilon|$. Next, the ideal efficiency and Boltzmann-type factor are analyzed in turn under realistic conditions that deviate from the first-order evaluation.

### 4.1. $\eta_{upper}^{\gamma}(\lambda, |\varepsilon|)$ under monochromatic light and the deviation index $\gamma(f(\lambda), \varepsilon(\lambda))$ from the first-order evaluation condition

Here we define $\gamma(f(\lambda), \varepsilon(\lambda))$ as a relative proportion of the deviation from the first-order evaluation conditiont that ensures the quasi-equilibrium condition, for the radiation entropy decrease $\Delta S^\gamma(\lambda)$, as follows.

$$\gamma\big(f(\lambda),\varepsilon(\lambda)\big) \equiv \frac{\Delta S^{\gamma}(\lambda)-\Delta S^{\gamma(1)}(\lambda)}{\Delta S^{\gamma(1)}(,\lambda,)} = \frac{S^{\gamma}\big((1+\varepsilon(\lambda))f(\lambda)\big)-S^{\gamma}(f(\lambda))}{\Delta S^{\gamma(1)}(,\lambda,)} - 1, \tag{4.1}$$

where $\varepsilon(\lambda) = \Delta N^{\gamma}(\lambda)/N^{\gamma}(\lambda)$ is used, and the identity $S^{\gamma}\{(N^{\gamma}(\lambda)+\Delta N^{\gamma}(\lambda))/G\} = S^{\gamma}\Big((1+\varepsilon(\lambda))f(\lambda)\Big)$ derived by the formula $f(\lambda) = N^{\gamma}(\lambda)/G(\lambda)$ is used. Later in this section, the argument wavelength $\lambda$ is noted where necessary, but is often omitted for space reasons. Based on the second law of thermodynamics, the following equation holds

$$S_{in}^{\gamma}(\lambda) \geq -\Delta S^{\gamma}(\lambda) = -\Big\{S^{\gamma}\Big(\big(1+\varepsilon(\lambda)\big)f\Big) - S^{\gamma}(f)\Big\}. \tag{4.2}$$

The inequality sign in Eq. (4.2) is owing to the entropy production that might occur with the photons transferring from the radiation into the system during the light absorption process, similar to the entropy production occurring with the heat transfer. However, here we assume that the entropy generation rate is 0, as in previous studies, and express the inequality in (4.2) as an equality as follows.
From Eq. (4.2), where the inequality sign is replaced by an equality sign (4.2), and Eq. (4.1), Eq. (3.5), the following formula is derived.

$$\begin{aligned} S_{in}^{\gamma}(\lambda) &= S_{in}^{\gamma(1)}(\lambda)\{1+\gamma(f(\lambda),\varepsilon(\lambda))\} \\ &= k_B \ln\left(1+\frac{1}{f(\lambda)}\right)\Delta N^{\gamma}\{1+\gamma(f(\lambda),\varepsilon(\lambda))\}. \end{aligned} \tag{4.3}$$

Using $\varepsilon(\lambda) = \Delta N^{\gamma}(\lambda)/N^{\gamma}(\lambda)$, the following formula can be derived from Eq. (4.1):

$$\gamma(f(\lambda),\varepsilon) = \frac{1}{\left\{f\ln\left(1+\frac{1}{f}\right)\right\}|\varepsilon|}\left[f(1-|\varepsilon|)\ln(1-|\varepsilon|) - \{1+f(1-|\varepsilon|)\}\ln\left(\frac{1+f(1-|\varepsilon|)}{1+f}\right)\right], \tag{4.4}$$

The decrease in irradiated photons $\Delta N^{\gamma}$ resulting from absorption in the *light-powered system* satisfies $-N^{\gamma} \leq \Delta N^{\gamma} \leq 0$, i.e.,$-1 \leq \varepsilon \leq 0$. $|\varepsilon| = -\Delta N^{\gamma}/N^{\gamma} = N_{in}^{\gamma}(\lambda)/N^{\gamma}(\lambda)$ is the ratio of the absorbed photon number flux to the irradiated photon number flux, i.e., the light absorption ratio. From $\eta_{max} = 1 - T_{out}S_{in}/E_{in}$ (Eq. (2.6)) and Eqs. (3.7) and (4.3), we obtain

$$\eta_{upper}^{\gamma}(\lambda,\varepsilon) \leq 1 - \frac{T_{out}}{T_{\gamma}(\lambda)}\{1+\gamma(f(\lambda),\varepsilon)\} \leq 1 - \frac{T_{out}}{T_{\gamma}(\lambda)} = \eta_{max}^{\gamma}(\lambda), \tag{4.5}$$

where the inequality sign is due to Eq. (4.3). Thus, the right side of Eq. (4.5) is called the upper bound efficiency in this paper and is denoted by $\eta_{upper}^{\gamma}(\lambda,\varepsilon)$. Based on analytical calculations, $\gamma(f,\varepsilon)$ is a monotonically decreasing function of the variable $\varepsilon$; thus, the larger the photon absorption ratio $|\varepsilon|$, the larger the $\gamma(f(\lambda),\varepsilon)$ and the smaller the $\eta_{upper}(\lambda,\varepsilon)$. In particular, the condition $\varepsilon \to 0$ gives $\gamma(f(\lambda),\varepsilon \to 0) \to 0$, and hence, $\eta_{upper}^{\gamma}(\lambda,\varepsilon \to 0) = 1 - T_{out}/\,T_{\gamma}(\lambda) = \eta_{max}^{\gamma}(\lambda)$, which reproduces the maximum efficiency obtained by a first-order evaluation. When $\varepsilon = -1$ (the absorption ratio $|\varepsilon| = 1$) is satisfied, i.e., all photons of monochromatic light are absorbed by the *light-powered system*, $\gamma$ becomes maximum, and it can be easily calculated from Eq. (4.4) as

$$\gamma(f, \varepsilon = -1) = \frac{\ln(1+f)}{f\ln\left(1+\frac{1}{f}\right)}. \tag{4.6}$$

For the solar radiation as the blackbody radiation with $\lambda = 670$ nm at $T_{sun}$ = 5800 K, $f_{BB} = 1/(e^{\frac{hc}{\lambda k_B T_{sun}}} - 1 = 2.5 \times 10^{-2}$ is obtained, and $\gamma(f = 2.5 \times 10^{-2}, \varepsilon = -1) = 2.7 \times 10^{-1}$ is obtained from Eq. (4.6). The dependence of $\gamma(= 2.5 \times 10^{-2}, \varepsilon)$ on ε under these conditions is shown in Fig. 3. The result shows that $\eta^{\gamma}_{upper}(\lambda = 670\ \text{nm}, \varepsilon)$ decreases from 0.95 to 0.93 with ε changing from 0 to $-1$. For the terrestrial solar radiation diluted with the dilution indicator $d = (R/D)^2$ with $\lambda = 670$ nm at $T_{sun}$ = 5800 K, $f = (R/D)^2 f_{BB}$=4.3× $10^{-7}$ is obtained, and $\gamma(f = 4.3 \times 10^{-7}, \varepsilon = -1) = 6.8 \times 10^{-2}$ is obtained from Eq. (4.6). Consequently, $\eta^{\gamma D}_{upper}(\lambda = 670\ \text{nm}, \varepsilon)$ decreases from 0.80 to 0.78 with ε decreasing from 0 to $-1$.

In both cases the decrease in the upper bound $\eta^{\gamma}_{upper}(\lambda = 670\ \text{nm}, \varepsilon)$ is negligible, indicating that the first-order evaluation given by Eq. (3.6) is valid as an approximation, at least for the ideal efficiency. This is because $\eta^{\gamma}_{upper}$ is a linear function of the radiation temperature modified by $1/(1 + \gamma(\lambda, \varepsilon))$. However, quantities that depend nonlinearly on the radiation temperature, such as the Boltzmann-type factor $\exp\left(-\Delta E/k_B T_{\gamma}(\lambda)\right)$, can be significantly reduced by the light absorption ratio .

**Fig. 3. Dependence of the index $\gamma(f(\lambda), \varepsilon(\lambda))$ of the deviation from the first-order evaluation on $\varepsilon = \Delta N^{\gamma}/N^{\gamma}$:** This figure shows the dependence of $\gamma(f = 2.48 \times 10^{-2}, \varepsilon)$, corresponding to $T_{sun}$ = 5800 K and $\lambda = 670$ nm on the photon number decrease rate $\varepsilon = \Delta N^{\gamma}/N^{\gamma}$ of the irradiating radiation.

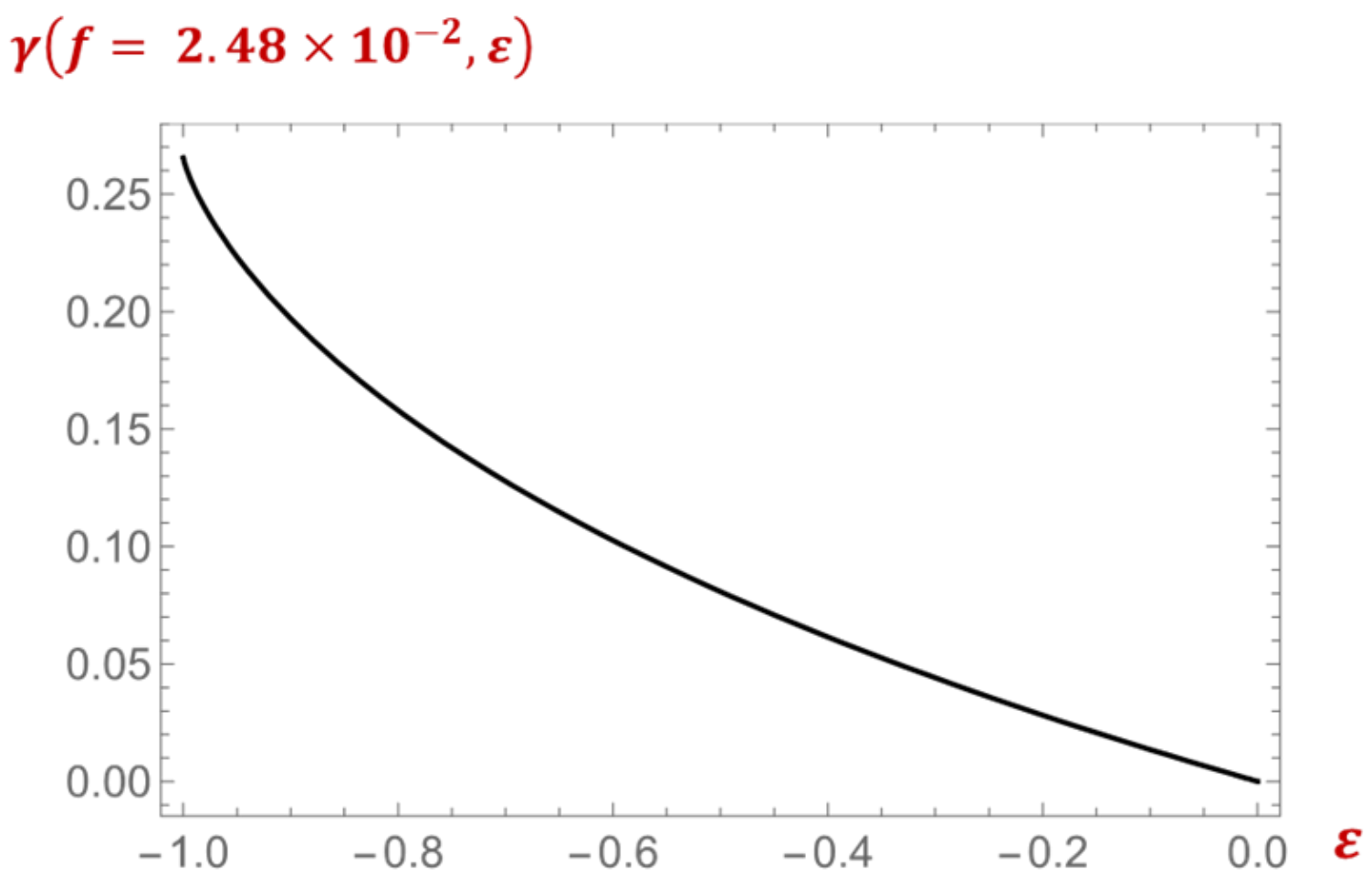

### 4.2. Boltzmann-type factor for the concentration ratio in the pigment system under monochromatic light

In the case of radiative entropy, the relative proportion of deviation from the first-order evaluation of the compositional entropy change of the pigment molecule can be defined as $\gamma(f_P, \varepsilon_P)$ as

$$\gamma(f_P, \varepsilon_P) \equiv \frac{\Delta S_P - \Delta S_P^{(1)}}{\Delta S_P^{(1)}} = \frac{S_P((1+\varepsilon_P)f_P) - S_P(f_P)}{\Delta S_P^{(1)}} - 1, \tag{4.7}$$

where $S_P(n_P + \Delta n_P) = S_P\big((1+\varepsilon_P)f_P\big)$ and $\varepsilon_P = \Delta n_P / n_P (< 0)$.

Eq. (4.7) gives

$$\Delta S_p(n_P) = S_P\big((1+\varepsilon_P)f_P\big) - S_P(f_P) = \Delta S_P^{(1)}\{1 + \gamma(f_P, \varepsilon_P)\}. \tag{4.8}$$

The following formula can be derived from Eq. (4.7) after some calculations:

$$\gamma(f_P, \varepsilon_P) =$$

$$= \frac{1}{\left\{f_P \ln\left(-1 + \frac{1}{f_P}\right)\right\}|\varepsilon_P|}\left[f_P(1-|\varepsilon_P|)\ln(1-|\varepsilon_P|) + \{1 - f_P(1-|\varepsilon_P|)\}\ln\left(\frac{1-f_P(1-|\varepsilon_P|)}{1-f_P}\right)\right] \tag{4.9}$$

Since $\Delta S_p(n_P) \geq S_{in}^{\gamma}(\lambda)$ holds, we obtain from Eqs. (4.3) and (4.8) the following inequality, consisting of two correction terms $\gamma(f(\lambda), \varepsilon)$ (Eq. (4.4)) and $\gamma(f_P, \varepsilon_P)$ (Eq. (4.9)) due to the contributions beyond the first-order evaluation.

$$\Delta S_P^{(1)}\{1 + \gamma(f_P, \varepsilon_P)\} \geq S_{in}^{\gamma(1)}(\lambda)\{1 + \gamma\big(f(\lambda), \varepsilon(\lambda)\big)\}. \tag{4.10}$$

Eqs. (3.12) and (3.13) give

$$k_B \ln\left(\frac{[P^*]}{[P]}\right)\Delta n_P\{1 + \gamma(f_P, \varepsilon_P)\} \geq \frac{\frac{hc}{\lambda}}{T_\gamma(\lambda, \hat{n}(\lambda))} N_{in}^{\gamma}(\lambda)\{1 + \gamma\big(f(\lambda), \varepsilon(\lambda)\big)\}. \tag{4.11}$$

As $N_{in}^{\gamma}(\lambda) = \Delta n_{P^*} = -\Delta n_P$ is satisfied, we obtain Eq. (4.11), including two correction terms $\gamma(f(\lambda), \varepsilon(\lambda))$ and $\gamma(f_P, \varepsilon_P)$ due to the contributions beyond the first-order evaluation in Eq. (3.16).

$$\frac{[P^*]}{[P]} \leq \exp\left\{-\frac{hc/\lambda}{k_B T_\gamma(\lambda, \hat{n}(\lambda))} \cdot \frac{1 + \gamma\big(f(\lambda), \varepsilon(\lambda)\big)}{1 + \gamma(f_P, \varepsilon_P)}\right\}. \tag{4.12}$$

Calculations show that the two correction terms $\gamma(f(\lambda), \varepsilon(\lambda))$ and $\gamma(f_P, \varepsilon_P)$ are monotonically increasing functions of the variables $|\varepsilon(\lambda)|$ and $|\varepsilon_P|$, respectively. Eq. (4.12) modifies the first-order evaluation, Eq. (3.16), by $|\varepsilon(\lambda)|$ and $|\varepsilon_P|$.

## 5. General unified formula of the ideal efficiency $\eta_{upper}^{\gamma}$ for non-monochromatic light with arbitrary dilution indicator $d$ and absorption ratio $|\varepsilon|$

Based on the above-mentioned results, a unified formula can be derived for the ideal efficiency $\eta_{upper}^{\gamma}$ of a *light-powered system* with absorption ratio $|\varepsilon|$ for non-monochromatic light diluted as the dilution indicator $d$ after being emitted by blackbody radiation at temperature $T$. The dilution indicator $d$ that must be considered at the surface of the Earth is attributed to the dilution effect caused by the reduction of the photon number density from the Sun to Earth, as analyzed in the previous section. Strictly speaking, this dilution effect occurs through the scattering of sunlight in the atmosphere surrounding the Earth. In other words, this effect of entropy increase by dilution does not occur before the sunlight is scattered in the atmosphere. Concrete proof of this, generally guaranteed by Liouville's theorem, is given in Appendix C in Part1 of this study [29,30].

Under the simplifying assumption that both the dilution indicator $d$ and the absorption ratio $|\varepsilon| =$

$-\varepsilon = -\Delta\bar{N}^{\gamma}/\bar{N}^{\gamma}$ are uniform over all wavelengths (or frequencies), Eq. (2.6) is applied and the following analysis is conducted: From Eq. (2.6), the ideal efficiency can be formulated as:

$$\eta_{upper}^{\gamma}(T, T_{out}, d, |\varepsilon|) = 1 - \frac{T_{out} S_{in}^{\gamma}(T,d,|\varepsilon|)}{E_{in}^{\gamma}(T,d,|\varepsilon|)}. \quad (5.1)$$

In this analysis, the wavelength $\lambda$ is converted to frequency $\nu$ and formulated in integral form over all frequencies $\nu$. Furthermore, between the frequencies $\nu$ and $\nu + \mathrm{d}\nu$, $G(\nu)$ is given by $G(\nu) = (8\pi V\nu^2/c^3)d\nu \equiv G_\nu d\nu$, and the average number of photons of the blackbody radiation at temperature $T$ is given by $\bar{n}_\nu(T)d\nu = G_\nu d\nu f(\nu, T)$, where $f(\nu, T)$ is the distribution function of blackbody radiation given by $f(\nu, T) = 1/\left(e^{h\nu/k_B T} - 1\right)$. The effects of the dilution indicator $d$ and absorption ratio $|\varepsilon|$ are incorporated into the formulation by the rewriting operation $\bar{n}_\nu(T) \to |\varepsilon| d\bar{n}_\nu(T)$ and $f(\nu, T) \to df(\nu, T)$. Consequently, $E_{in}^{\gamma}(T, d, |\varepsilon|)$ and $S_{in}^{\gamma}(T, d, |\varepsilon|)$ are obtained as

$$E_{in}^{\gamma}(T, d, |\varepsilon|) = |\varepsilon| d \int_0^\infty d\nu \bar{n}_\nu(T) h\nu, \quad (5.2)$$

$$S_{in}^{\gamma}(T, d, |\varepsilon|) = |\varepsilon| d k_B \int_0^\infty d\nu \bar{n}_\nu(T) \ln(1 + \frac{1}{df(\nu,T)}) \{1 + \gamma(df(\nu, T), |\varepsilon|), \quad (5.3)$$

where $\gamma(df(\nu, T), |\varepsilon|)$ is obtained from Eq (4.4) as

$$\gamma(df(\nu), |\varepsilon|) = \frac{1}{-|\varepsilon| \ln(1+\frac{1}{df(\nu,T)})} \left[ -(1 - |\varepsilon|) \ln(1 - |\varepsilon|) + (\frac{1}{df(\nu,T)} + 1 - |\varepsilon|) \ln(\frac{1+df(\nu,T)(1-|\varepsilon|)}{1+df(\nu,T)}) \right] \quad (5.4)$$

Substituting Eqs. (5.2) and (5.3) into Eq. (5.1) gives the ideal efficiency $\eta_{max}^{\gamma}(T, T_{out}, d, |\varepsilon|)$. If $h/k_B T = a$, $f(a\nu) = 1/(e^{a\nu} - 1)$, and the integral variable $\nu$ is replaced by the dimensionl ess variable $\mu = a\nu$, we obtain

$$\begin{aligned}
\eta_{upper}{}^{\gamma}(T, T_{out}, d, |\varepsilon|) &= 1 - \frac{k_B T_{out} \int_0^\infty d\nu \bar{n}_\nu \ln\left(1 + \frac{e^{a\nu}-1}{d}\right) \{1 + \gamma(df(a\nu), |\varepsilon|)\}}{\int_0^\infty d\nu \bar{n}_\nu h\nu} \\
&= 1 - \frac{k_B T_{out} \int_0^\infty d\nu \frac{\nu^2}{e^{a\nu}-1} \ln\left(1 + \frac{e^{a\nu}-1}{d}\right) \{1 + \gamma(df(a\nu), |\varepsilon|)\}}{k_B T a \int_0^\infty d\nu \frac{\nu^3}{e^{a\nu}-1}} \\
&= 1 - \frac{T_{out} \frac{1}{a^3} \int_0^\infty d\mu \frac{\mu^2}{e^{\mu}-1} \ln\left(1 + \frac{e^{\mu}-1}{d}\right) \{1 + \gamma(df(\mu), |\varepsilon|)\}}{k_B T \frac{a}{a^4} \int_0^\infty d\mu \frac{\mu^3}{e^{\mu}-1}} \quad \text{(Replaced by } \mu = a\nu) \\
&= 1 - \frac{T_{out} \int_0^\infty d\mu \frac{\mu^2}{e^{\mu}-1} \ln\left(1 + \frac{e^{\mu}-1}{d}\right) \{1 + \gamma(df(\mu), |\varepsilon|)\}}{T \Gamma(4) \zeta(4)}, \quad (5.5)
\end{aligned}$$

where $\Gamma(4)$ and $\zeta(4)$ are the gamma function and the Riemann zeta function, respectively. $\eta_{upper}{}^{\gamma}(T, T_{out}, d, |\varepsilon|)$ describes the ideal efficiency of a *light-powered system*, using light from a blackbody radiation source at temperature $T$, as a function of the light dilution indicator $d$ and light absorption ratio $|\varepsilon|$ of the system. In Eq. (5.5), the temperature of the blackbody radiation $T$ can be completely factorized in the form of $T_{out}/T$ ($T_{out}$ is the ambient temperature), and the remaining factor contains only dimensionless variables. Consequently, Eq. (5.5) can be expressed as

$$\eta^{\gamma}_{upper}(T, T_{out}, d, |\varepsilon|) = 1 - \frac{T_{out}}{T} Y(d, |\varepsilon|), \tag{5.6}$$

where the dimensionless factor $Y(d, |\varepsilon|)$ is given by

$$Y(d, |\varepsilon|) = \frac{\int_0^{\infty} d\mu \frac{\mu^2}{e^{\mu}-1} \ln\left(1+\frac{1}{df(\mu)}\right)\{1+\gamma(df(\mu), |\varepsilon|)\}}{\Gamma(4)\zeta(4)}, \tag{5.7}$$

as

$$\gamma(df(\mu), |\varepsilon|) = \frac{1}{-|\varepsilon| \ln(1+\frac{1}{df(\mu)})}\left[-(1-|\varepsilon|)\ln(1-|\varepsilon|) + (\frac{1}{df(\mu)} + 1 - |\varepsilon|)\ln(\frac{1+df(\mu)(1-|\varepsilon|)}{1+df(\mu)})\right]. \tag{5.8}$$

where $f(\mu) = 1/(e^{\mu} - 1)$. Notably, Planck's constant $h$ and Boltzmann's constant $k_B$, which are dimensional constants included in the ideal efficiency of monochromatic light (Eq. (3.6)), are not included in Eqs. (5.4) and (5.8). When $|\varepsilon| = 1$ in Eqs. (5.6) and (5.7), the resultant formula $Y(d, |\varepsilon| = 1)$ is consistent with the formula first presented by Landsberg and Tonge [36] and followed in several subsequent papers [37-39]. Therefore, Eq. (5.6) can be regarded as extrapolating $|\varepsilon|$ from 1 to an arbitrary value in the range $0 \le |\varepsilon| \le 1$.

Without dilution, i.e., $d = 1$, Eq. (5.7) can be expressed as

$$Y(d = 1, |\varepsilon|) = 1 + \frac{\frac{1}{|\varepsilon|}\int_0^{\infty} d\mu \frac{\mu^2}{e^{\mu}-1}\left[(1-|\varepsilon|)\ln(1-|\varepsilon|) - (\frac{1}{f(\mu)} + 1 - |\varepsilon|)\ln(\frac{1+f(\mu)(1-|\varepsilon|)}{1+f(\mu)})\right],}{\Gamma(4)\zeta(4)}. \tag{5.9}$$

The evaluation of $Y(d, |\varepsilon|)$ beyond the first-order evaluation, which is the non-equilibrium contribution between the radiation and the system, is expressed as the second term in Eq. (5.9). A simple calculation shows that this non-equilibrium contribution term in $Y(d, |\varepsilon|)$ is always ≥ 0 and increases monotonically with $|\varepsilon|$ from 0 for $|\varepsilon| = 0$ to 1/3 for $||\varepsilon| = 1$.

Eqs. (5.1) and (5.6) give following relational formula:

$$\frac{Y(d, |\varepsilon|)}{T} = \frac{S^{\gamma}_{in}(T, d, |\varepsilon|)}{E^{\gamma}_{in}(T, d, |\varepsilon|)}. \tag{5.10}$$

Applying Eqs. (5.8) and (5.7) to two special cases of $d = 1$ (no dilution), (1) the first-order evaluation applicable case ($|\varepsilon| \to 0$) and (2) the perfect light absorption case ($|\varepsilon| = 1$)) gives $\gamma(d = 1, |\varepsilon| \to 0) = 0$ (the second term in Eq. (5.9) is $0$) and $Y(d = 1, |\varepsilon| \to 0) = 1$. After some calculations, $Y(d = 1, |\varepsilon| = 1) = 4/3$ is derived (the second term in Eq. (5.9) is $1/3$). Consequently, Eq. (5.6) gives $\eta^{\gamma}_{upper}(T, T_{out}, d = 1, |\varepsilon| \to 0) = 1 - T_{out}/T$ and $\eta^{\gamma}_{upper}(T, T_{out}, d = 1, |\varepsilon| = 1) = 1 - (4/3)T_{out}/T$, the former of which is exactly the Jeter efficiency [19], i.e., the quasi-Carnot efficiency in the *light-powered system*, and the latter of which is exactly the Spanner efficiency [12,13]. Therefore, the formula for the ideal efficiency of *light-powered systems*, compactly expressed by Eq. (5.6) containing Eqs.

(5.7) and (5.8), is the most general formulation that includes the Jeter (Carnot) and Spanner efficiencies as special cases. This enables the evaluation of the ideal efficiency of *light-powered systems* with any light absorption ratio $|\varepsilon|$ under blackbody radiation with any dilution indicator $d$ in a unified manner.

The values of $Y(d,|\varepsilon|)$ and $\eta^{\gamma}_{upper}(T_{sun},T_{out},d,|\varepsilon|)$, corresponding to $T_{sun}=5800$ K (the solar temperature) and $T_{out}=300$ K (the ambient temperature at the ground surface), obtained from the numerical analysis are shown in Fig. 4 as three-dimensional plots with the dilution indicator $d$ and the light absorption ratio $|\varepsilon|$ as $x$ and $y$ coordinates, respectively. The behaviors of $Y(d=1,|\varepsilon|)$ and $\eta^{\gamma}_{upper}(T=5800\text{K},T_{out}=300\text{K},d=1,|\varepsilon|)$ with respect to $|\varepsilon|$, for the undiluted-radiation case are shown in Figs. 5(a) and (c), respectively. The behaviors of $Y(d=(R/D)^2,|\varepsilon|)$ and $\eta^{\gamma}_{upper}(T=5800\text{ K},T_{out}=300\text{ K},d=(R/D)^2,|\varepsilon|)$ with respect to $|\varepsilon|$, for diluted radiation from the Sun to Earth, are shown in Fig. 5 (b) and (d), respectively. The ideal efficiency $\eta^{\gamma}_{max}(T=5800\text{K},T_{out}=300\text{K},d=1,|\varepsilon|)$ without dilution effect (Fig. 5(b)) decreases monotonically from 0.95 ($|\varepsilon|\to 0$) to 0.93 ($|\varepsilon|=1$) with $|\varepsilon|$. Conversely, the ideal efficiency $\eta^{\gamma}_{upper}(T=5800\text{ K},T_{out}=300\text{ K},d=(R/D)^2,|\varepsilon|)$ with the dilution effect of $d=(R/D)^2$ from the Sun to Earth (Fig. 5(d)) decreases monotonically from 0.75 ($|\varepsilon|\to 0$) to 0.73 ($|\varepsilon|=1$) with $|\varepsilon|$. In both cases, the reduction in the ideal efficiency due to a finite light absorption ratio $|\varepsilon|$ is small. However, when using artificial light with a temperature lower than the solar temperature as a radiation source, the non-negligible increase in $Y$ with $|\varepsilon|$ can arise and lead to a significant decrease in the ideal efficiency

The light absorption ratio $|\varepsilon|$ is assumed to be uniform and independent of wavelength (frequency), whereas the actual $|\varepsilon|$ of terrestrial photosynthetic organisms is wavelength-dependent. Based on the first-order evaluation, which does not take into consideration the "decrease in ideal efficiency due to light absorption ratio " revealed in this study, the present author has calculated and reported the ideal efficiencies using the actual absorption spectra of several photosynthetic organisms [33]. The values, before considering the decrease in entropy due to the photochemical reaction of glucose production averaged 0.79, which is higher than the above results. More details can be found in the literature [33].

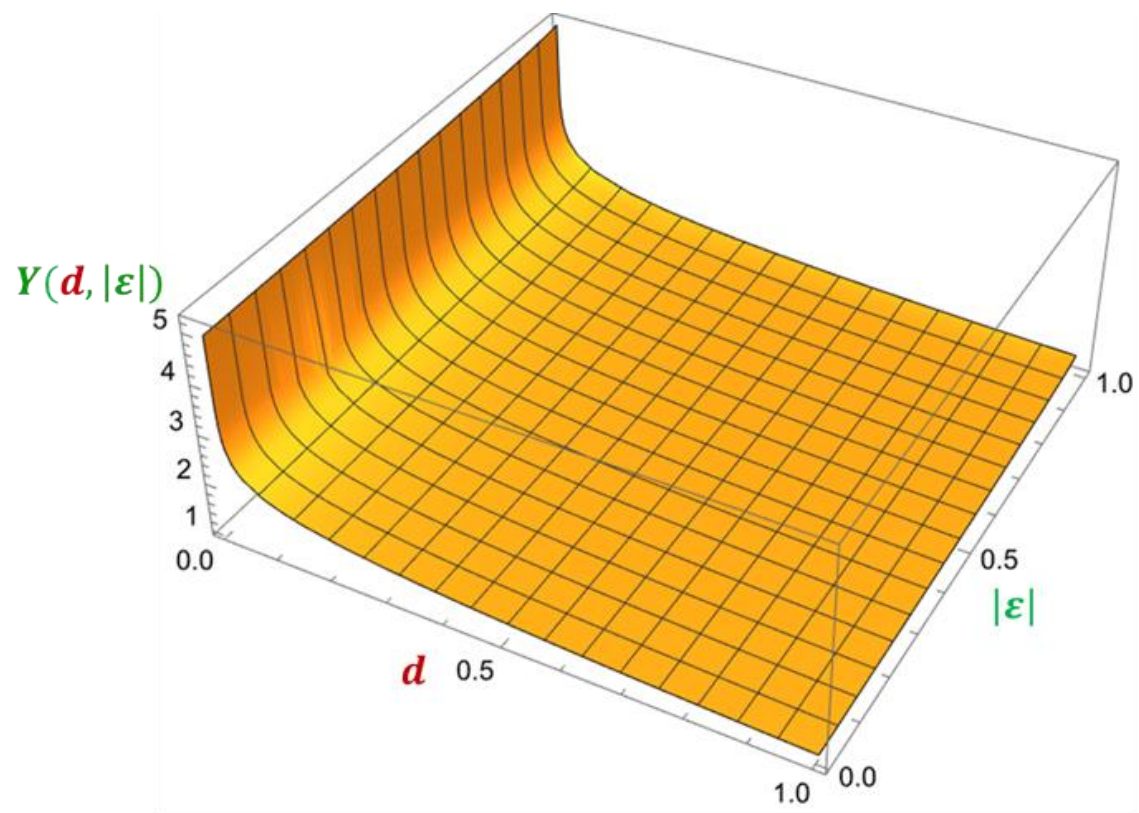

a

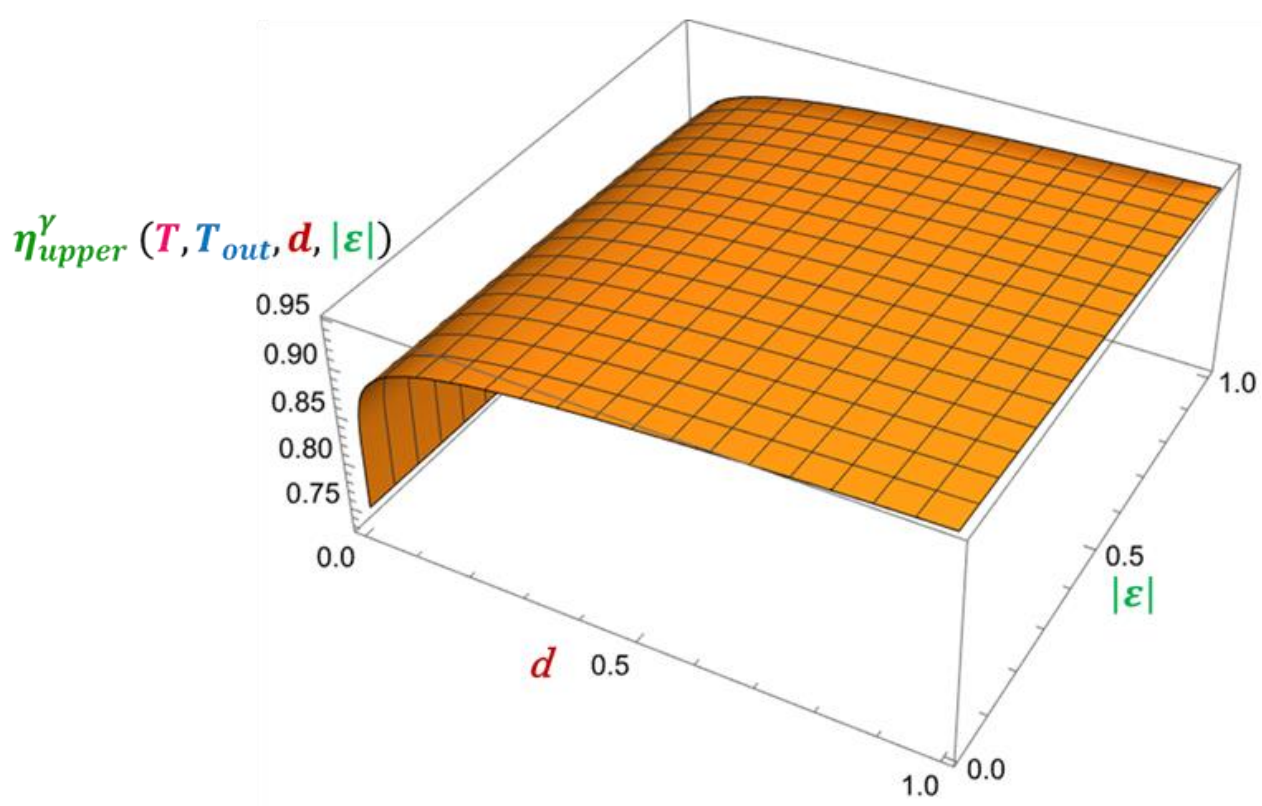

b

**Fig.4. Dependence of the factor $\boldsymbol{Y(|\varepsilon|, d)}$ and the ideal efficiency of a *light-powered system* $\boldsymbol{\eta^{\gamma}_{upper}(T = 5800\ \mathrm{K}, T_{out} = 300\ \mathrm{K}, d, |\varepsilon|)}$ on the dilution indicator $\boldsymbol{d}$ and light absorption ratio $\boldsymbol{|\varepsilon|}$. (a)** $Y(d, |\varepsilon|)$ and **(b)** $\eta^{\gamma}_{upper}(T = 5800\mathrm{K}, T_{out} = 300\mathrm{K}, d, |\varepsilon|)$.

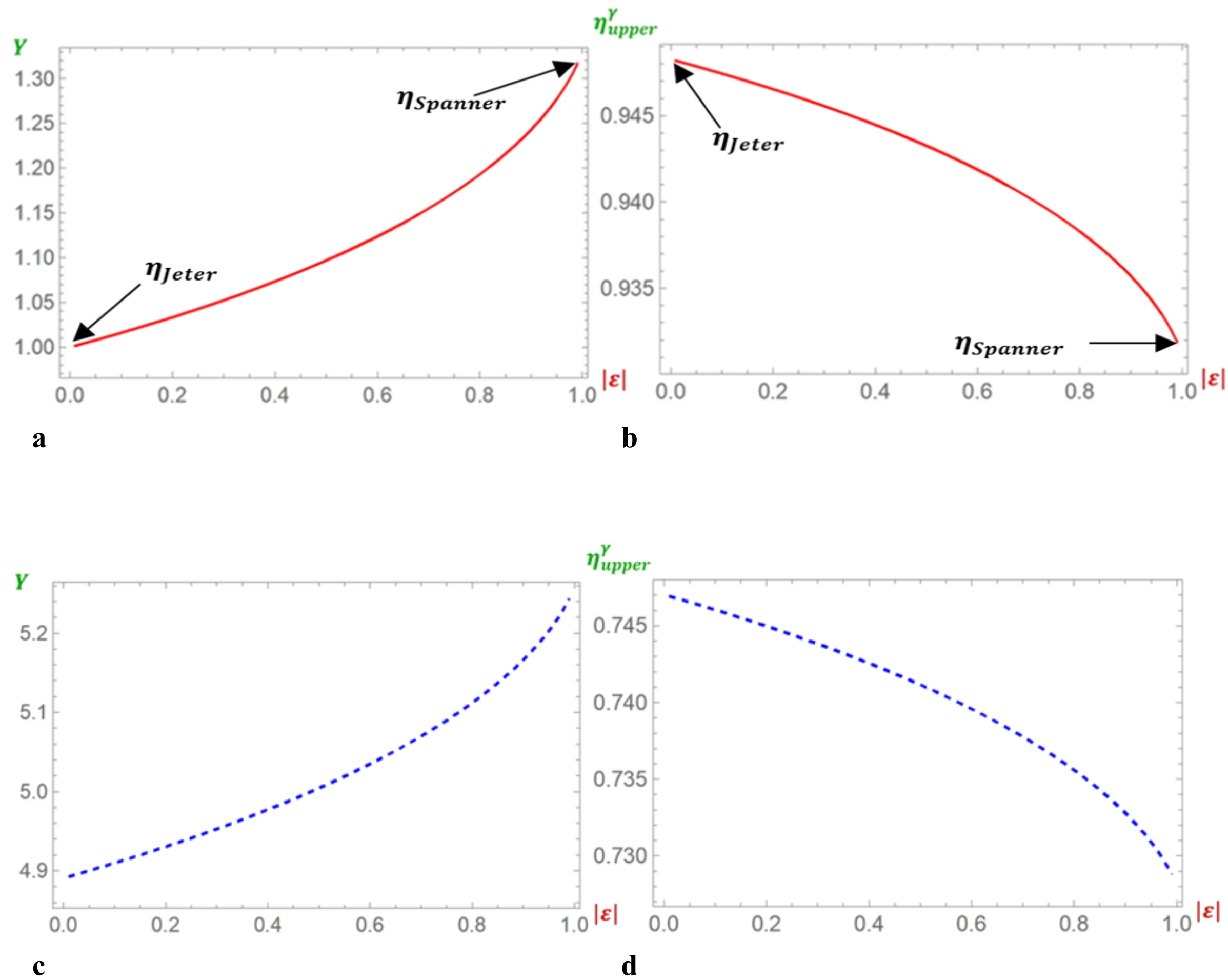

**Fig. 5. (a) Dependence of $\boldsymbol{Y(d = 1, |\varepsilon|)}$ on the light absorption ratio** $|\varepsilon|$: $Y(d = 1, |\varepsilon| \to 0) = 1$ corresponding to the Jeter efficiency (Carnot efficiency), $Y(d = 1, |\varepsilon| = 1)$ =4/3 corresponding to the Spanner efficiency. **(b) Dependence of $\boldsymbol{\eta^{\gamma}_{upper}(T = 5800\ \mathrm{K}, T_{out} = 300\ \mathrm{K}, d = 1, |\varepsilon|)}$ on $\boldsymbol{|\varepsilon|}$**: $\eta^{\gamma}_{upper}(T = 5800\ \mathrm{K}, T_{out} = 300\ \mathrm{K}, d = 1, |\varepsilon| \to 0) = 0.95$ and $\eta^{\gamma}_{upper}(T = 5800\ \mathrm{K}, T_{out} = 300\ \mathrm{K}, d = 1, |\varepsilon| \to 1) = 0.93$ corresponding to the Jeter and Spanner efficiencies at solar temperature $T = 5800\ \mathrm{K}$, respectively. **(c) Dependence of $\boldsymbol{Y(d = (R/D)^2, |\varepsilon|)}$ on $\boldsymbol{|\varepsilon|}$** in the case of diluted sunlight at the ground surface on the light absorption ratio $|\varepsilon|$: $Y(d = (R/D)^2, |\varepsilon| = 1) = 5.3$, $Y(d = (R/D)^2, |\varepsilon| \to 0) = 4.9$ **(d) Dependence of $\boldsymbol{\eta^{\gamma}_{upper}(T = 5800\ \mathrm{K}, T_{out} = 300\mathrm{K}, d = (R/D)^2, |\varepsilon|)}$ on $\boldsymbol{|\varepsilon|}$** in the case of diluted sunlight at the ground surface on the light absorption ratio $|\varepsilon|$: $\eta^{\gamma}_{upper}(T = 5800\mathrm{K}, T_{out} = 300\mathrm{K}, d = (R/D)^2, |\varepsilon| \to 0) = 0.75$ corresponding to the quasi-Carnot efficiency, $\eta^{\gamma}_{upper}(T = 5800\ \mathrm{K}, T_{out} = 300\ \mathrm{K}, d = (R/D)^2, |\varepsilon| = 1) = 0.73$.

From Eq. (5.10), the Claudius formula modified by dilution and absorption ratio becomes

$$\Delta S = Y(d, |\varepsilon|) \frac{\Delta U}{T}, \tag{5.11}$$

where $\Delta U$ is the change in internal energy of the radiation. Though discussions have been held to explain the reasons behind the differences in various ideal efficiencies of *light-powered systems*, such as the Jeter and Spanner efficiencies, a correct settlement has not yet been reached. The results in Fig.5 a and 5bshow that the essential difference between the Jeter and Spanner efficiencies lies in their light absorption ratios of the systems, represented by the variable $|\varepsilon|$. In summary, (1) the Jeter efficiency is the ideal efficiency under the first-order evaluation condition, $|\varepsilon| \to 0$, while the Spanner efficiency is the ideal efficiency beyond the first-order evaluation condition, $|\varepsilon| = 1$, and (2) Fig. 5a and 5b are valuable as the interpolations for $|\varepsilon|$ to any value within the range of$0 \leq |\varepsilon| \leq 1$. The next section discusses various types of ideal efficiencies, including the Petela and Landsberg efficiencies, and it explains how these efficiencies ultimately result in the most general formula, Eq. (6.12), which was constructed in this study.

## 6. Classification of the ideal efficiencies based on the energy–entropy flux conditions according to whether the first-order evaluation is applicable or not

Carnot efficiency of a heat-powered system (heat engine) holds only when the quasi-equilibrium condition between the heat bath and system, i.e., the first-order evaluability condition given by the temperature condition $\Delta T/T \ll 1$, is satisfied. Similarly, the quasi-Carnot efficiency of a *light-powered system* also holds only when the photon number condition $\Delta N/N \ll 1$ in radiation is met. Considering Eq. (5.6), the difference between the Jeter efficiency (quasi-Carnot efficiency) and Spanner efficiency can be attributed to the aforementioned difference in absorption ratio s $\varepsilon_{in} = |\varepsilon|$, especially for photosynthetic type *light-powered systems*, such as photosynthetic systems, rather than to the difference of manipulation processes using, for example, enclosed photon gas in an idealized cylinder–piston system explained in the previous studies**.**

In previous studies, three main types of ideal blackbody radiation-work conversion efficiencies were proposed: the Jeter efficiency $\eta_{Jeter}$, the Spanner efficiency $\eta_{Spanner}$, and the Landsberg and the Petela efficiency $\eta_{Landsberg-Petela}$ (Eq. (6.1)) [15]. However, the differences between them have not been settled, and efforts to understand them in a unified way are still ongoing [e.g., 21].

$$\eta_{Landsberg-Petela} = 1 - (4/3)T_{out}/T_{in} + (1/3)(T_{out}/T_{in})^4. \tag{6.1}$$

Landsberg and Tonge conducted a flux analysis [9] similar to the energy and entropy flow analysis presented in this paper and derived $\eta_{Landsberg}$ (Eq. (6.1)), often referred to as the Landsberg limit [40-45] Landsberg and Tonge conducted a flux analysis [9] similar to the energy and entropy flow analysis conducted in this paper and derived η_Landsberg (Eq. (6.1)), often referred to as the Landsberg limit [40-45]. However, their analytical approach and chosen conditions were almost unsystematic and unrealistic, respectively. This study systematically examines the classification conditions for these efficiencies and correctly classify the ideal energy efficiency of the light-powered system based on the applicability of the first-order evaluation, representing a quasi-equilibrium condition between the system and radiation (the energy source) or the environment (the sink).

Fig. 6(a) corresponds to the diagram in Ref. [9]. Based on the conditions shown in this figure, Landsberg and Tonge [9] derived the third term in the Petela–Landsberg efficiency (Eq. (6.1)), presuming that a system discards its entropy via both the blackbody radiation within the system and waste heat. The total radiation condition assumed by Landsberg and Tonge [9] can be met if the system environment (Fig. 6(a)) is a vacuum sink such as outer space. However, it remains unfulfilled in a medium such as the atmosphere. In such instances, blackbody radiation of roughly the same temperature flowing into the system from the surrounding medium, and the net radiation outflow is ultimately determined by its subtraction. This condition can be described by $(T_S - T_{out})/T_{out} = \Delta T/T_{out} \ll 1$, where $T_S$ and $T_{out}$ are the temperature of the system and its surrounding environment, respectively. Consequently, the net energy and entropy fluxes from the system to environment (medium),$E^{\gamma}_{out}(T_S)$ and $S^{\gamma}_{out}(T_S)$, are given by $E^{\gamma}_{out}(T_S) = E^{\gamma}_{(out)}(T_{out} + \Delta T) - E^{\gamma}_{(in)}(T_{out})$ and $S^{\gamma}_{out}(T_S) = S^{\gamma}_{(out)}(T_{out} + \Delta T) - S^{\gamma}_{(in)}(T_{out})$, respectively. The bracketed subscripts (out) and (in) represent the elementary outflux from and influx into the system before subtraction, respectively. Fig. 6(b) illustrates the correct flux analysis according to these equations. The energy and entropy fluxes of the photon gas due to blackbody radiation at temperature $T$ are obtained as $E^{\gamma}(T) = \sigma T^4$ and $S^{\gamma}(T) = 4/3\sigma T^3$, where $\sigma$ is the Stefan-Boltzmann constant given by $\sigma = (1/4)\rho_{BB} = (2\pi^5 k_B^4)/(15c^2h^3) = 5.67 \times 10^{-8}\mathrm{Wm^{-2}K^{-4}}$ ($\rho_{BB}$ is the blackbody radiation density). Therefore, the net radiation energy and entropy outflow (here more precisely, outflux) can be respectively expressed as:

$$E^{\gamma}_{out}(T_S) = \sigma(T_{out} + \Delta T)^4 - \sigma T^4_{out} = \sigma T^4_{out}(1 + \Delta T/T_{out})^4 - \sigma T^4_{out} \cong 4\sigma T^3_{out}\Delta T, \quad (6.2)$$

$$S^{\gamma}_{out}(T_S) = (4/3)\sigma T^3_{out}(1 + \Delta T/T_{out})^3 - (4/3)\sigma T^3_{out} \cong 4\sigma T^2_{out}\Delta T, \quad (6.3)$$

From Eqs. (6.2) and (6.3), we obtain

$$\frac{S^{\gamma}_{out}(T_S)}{E^{\gamma}_{out}(T_S)} = \frac{1}{T_{out}}, \quad (6.4)$$

which implies that $Y^{\gamma}_{out}(\Delta T/T_{out} \to 0) = 1$ is given by the first-order evaluation of $\Delta T/T_{out}$.

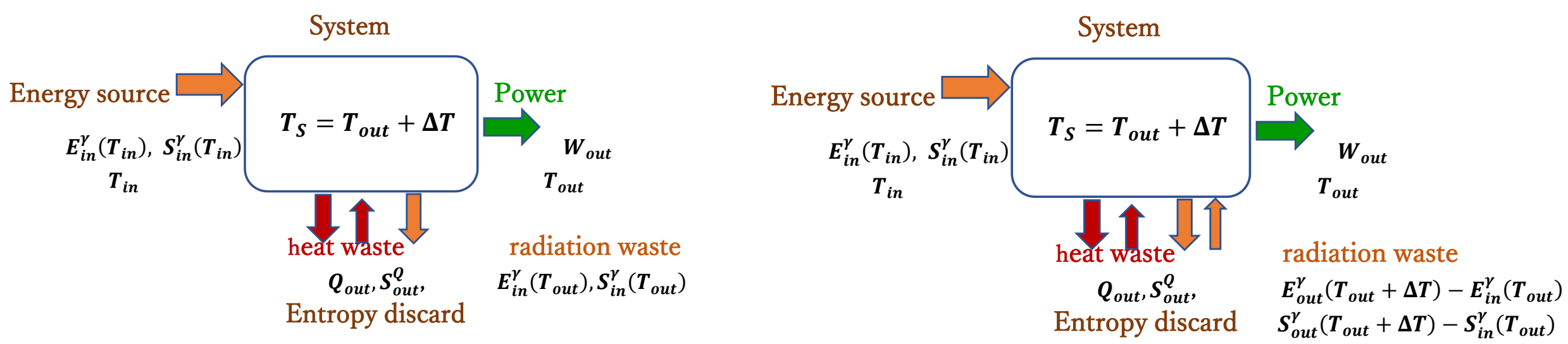

**Fig.6. Schematic of energy and entropy fluxes into and out of the system:** a) the analysis of Landsberg and Tonge [9], b) the analysis conducted in this study. In this figure, the temperature of the inflow and outflow radiation of the system, and the environment are represented by $T_{in}$ and $T_{out} + \Delta T,\ T_{out}$, respectively.

Eq. (5.9) also applies to outflow radiation from a system. The photon number reduction rates due to inflowing and outflowing radiation are denoted as $\varepsilon_{in}^{\gamma}$ and $\varepsilon_{out}^{\gamma}$, respectively. Thus, the equation for outflow radiation is obtained as

$$S_{out}^{\gamma} = Y\left(d_{out} = 1, \left|\varepsilon_{out}^{\gamma}\right|\right)\frac{E_{out}^{\gamma}}{T_{out}}. \tag{6.5}$$

Through calculations, we can derive

$$Y\left(d_{out} = 1, \left|\varepsilon_{out}^{\gamma}\right| \to 0\right) = 1. \tag{6.6}$$

The equation obtained by applying Eq.(6.6), which is derived from the first-order evaluation based on $\varepsilon_{out}^{\gamma}$=$\Delta N_{out}^{\gamma}/N_{out}^{\gamma}$, to equation (6.5) yields the same results as those using Eq. (6.4), which is derived from the first-order evaluation based on $\Delta T/T_{out}$. The reason for the same result is explained in Appendix A. Based on these findings, general flow conditions are classified, and the ideal efficiency equations for each case are presented below.
When performing a flow analysis, it is necessary to consider the heat and radiation flows separately.

### 6.1. Heat flow

In the case of *light-powered systems*, only heat flow is assumed to be waste heat used to discard entropy, and a first-order evaluation based on $\Delta T/T$ is assumed to be applicable, similar to the case of a heat-powered system (heat engine). Therefore, the following equation holds:

$$S_{out}^{Q} = \frac{\boldsymbol{E_{out}^{Q}}}{T_{out}}. \tag{6.7}$$

From here on, the entropy discarded via heat (waste heat) $Q_{out}$ will be written as $E_{out}^{Q}$.

### 6.2. Radiation flow

From Eq. (5.10), the following equation holds in its most general form

$$S_{in,out}^{\gamma} = Y\left(d_{in,out}, \left|\varepsilon_{in,out}^{\gamma}\right|\right)\frac{E_{in,out}^{\gamma}}{T_{in,out}}, \tag{6.8}$$

where $\varepsilon_{in,out}^{\gamma}$ denotes the decreasing ratio of photon numbers $\Delta N_{in,out}^{\gamma}/N_{in,out}^{\gamma}$ in the inflow and outflow radiation.

On this basis, the most general formula for the ideal efficiency of *light-powered systems* is formulated and specifically applied to the ideal efficiencies, $\eta_{Petela-Landsberg}$, $\eta_{Spanner}$, and $\eta_{Jeter}$ derived from previous studies to clarify their correct derivation conditions. Section 8 discusses and analyzes a simplified mathematical model that is suitable as a flowing radiation model for photosynthetic-type *light-powered systems*. To avoid complications, the variables $d$ and ε for the factor $Y(d, \varepsilon)$ are omitted, where appropriate and simply expressed as $Y$.

### 6.3. General unified formulation for the upper efficiency of ***light-powered system*** and derivations of several upper efficiencies

From the first law of thermodynamics (the law of conservation of energy), the efficiency of *light-powered systems* is defined as

$$\eta = \frac{E_{in}^{\gamma} - (E_{out}^{Q} + E_{out}^{\gamma})}{E_{in}^{\gamma}}, \tag{6.9}$$

where $E_{in}^{\gamma}$ represents the absorbed radiation energy in a system; $E_{out}^{Q}$ and $E_{out}^{\gamma}$ represent the released heat energy (waste heat) and emitted radiation from a system as the entropy discard, respectively. The respective weights (ratios) of entropy discard via heat and radiation $p_Q$ *and* $p_{\gamma}$, respectively, are defined as:

$$p_Q = \frac{S_{out}^{Q}}{S_{out}}, \; p_{\gamma} = \frac{S_{out}^{\gamma}}{S_{out}}, \tag{6.10}$$

From $S_{out} = S_{out}^{Q} + S_{out}^{\gamma}$, the equation $p_Q + p_{\gamma} = 1$ holds. Here, $S_{out}^{Q}$ and $S_{out}^{\gamma}$ represent the entropy discarded via heat and radiation, respectively. Using $Y$-factors defined by Eq. (6.8), $S_{out}^{Q}$, $S_{out}^{\gamma}$, and $S_{in}^{\gamma}$ are given by $S_{out}^{Q} = Y_{out}^{Q} \frac{E_{out}^{Q}}{T_{out}}$, $S_{out}^{\gamma} = Y_{out}^{\gamma} \frac{E_{out}^{\gamma}}{T_{out}}$ and $S_{in}^{\gamma} = Y_{in}^{\gamma} \frac{E_{in}^{\gamma}}{T_{in}}$, respectively.

Applying these formulae to Eq. (6.9), and using the definition of $p_Q$ and $p_{\gamma}$ given by Eq. (6.10), we obtain

$$\eta = \frac{E_{in}^{\gamma} - (E_{out}^{Q} + E_{out}^{\gamma})}{E_{in}^{\gamma}}$$

$$= 1 - \frac{(E_{out}^{Q} + E_{out}^{\gamma})}{E_{in}^{\gamma}}$$

$$= 1 - \frac{T_{out}\left(\frac{1}{Y_{out}^{Q}} S_{out}^{Q} + \frac{1}{Y_{out}^{\gamma}} S_{out}^{\gamma}\right)}{E_{in}^{\gamma}}$$

$$= 1 - \frac{T_{out}\left\{\frac{1}{Y_{out}^{Q}}\left(S_{out}^{Q} + S_{out}^{\gamma}\right) + \left(\frac{1}{Y_{out}^{\gamma}} - \frac{1}{Y_{out}^{Q}}\right) S_{out}^{\gamma}\right\}}{E_{in}^{\gamma}}$$

$$= 1 - \frac{T_{out}\left\{\frac{1}{Y_{out}^{Q}} S_{out} + \left(\frac{1}{Y_{out}^{\gamma}} - \frac{1}{Y_{out}^{Q}}\right) S_{out}^{\gamma}\right\}}{E_{in}^{\gamma}}$$

$$= 1 - \frac{T_{out}\left\{\frac{1}{Y_{out}^{Q}} (p_Q + p_{\gamma}) S_{out} + \left(\frac{1}{Y_{out}^{\gamma}} - \frac{1}{Y_{out}^{Q}}\right) p_{\gamma} S_{out}\right\}}{E_{in}^{\gamma}}$$

$$= 1 - \frac{T_{out} S_{out}\left\{\frac{p_Q}{Y_{out}^{Q}} + \frac{p_{\gamma}}{Y_{out}^{\gamma}}\right\}}{E_{in}^{\gamma}}$$

$$\leq 1 - \frac{T_{out} S_{in}^{\gamma} \left\{ \frac{p_Q}{Y_{out}^Q} + \frac{p_\gamma}{Y_{out}^\gamma} \right\}}{E_{in}^{\gamma}}. \quad \textbf{(6.11)}$$

The last inequality is due to $S_{out} \geq S_{in}^{\gamma}$ given by the second law of thermodynamics.

Using $S_{in}^{\gamma} = Y_{in}^{\gamma} \frac{E_{in}^{\gamma}}{T_{in}}$ and (6.11), the following general formulation is derived:

$$\eta_{upper} = 1 - \left( \frac{p_Q}{Y_{out}^Q} + \frac{p_\gamma}{Y_{out}^\gamma} \right) Y_{in}^{\gamma} \frac{T_{out}}{T_{in}}. \quad (6.12)$$

In Eq. (6.12), where the ideal efficiency $\eta_{upper}$ is given, $p_Q$ and $p_\gamma$ (Eq. (6.10)) become

$$p_Q = \frac{S_{out}^Q}{S_{in}^{\gamma}}, \ p_\gamma = \frac{S_{out}^{\gamma}}{S_{in}^{\gamma}}, \quad (6.13)$$

because $S_{out} = S_{in}^{\gamma}$ holds owing to the ideal condition. From its derivation process, we can see that equation (6.12) is not applicable when $p_Q = 0$, i.e., when thermal entropy discard is 0.

The general formula in Eq. (6.12) applies to both flowing radiation models and cylinder–piston models defined in Section7 for *light-powered systems*. For the thermal discarded entropy (waste heat), similar to previous studies, the first-order evaluability condition (quasi-equilibrium between the system and environment) is assumed to be satisfied ($\varepsilon_{out}^Q \to 0$), i.e., $Y_{out}^Q = 1$. The following part of this section outlines the method for calculating the ideal efficiency of *light-powered systems* using Eqs. (6.12) and (6.13), and demonstrates its application to various ideal photosynthetic efficiencies reported in previous studies. The ideal efficiency of any powered system, whether heat- or light-powered, is determined by the $Y$ factors $Y_{in}^{\gamma}$, $Y_{out}^{\gamma}$, $Y_{out}^{Q}$, which depend on $d$ and $|\varepsilon|$, as well as the $p$ factors $p_\gamma$, $p_Q$ ($p_\gamma + p_Q = 1$), based on Eq. (6.12).

However, the condition $p_\gamma = \frac{S_{out}^{\gamma}(T_{out})}{S_{in}^{\gamma}(T_{in})}$ applied to Eq. (6.12) is not unconditionally guaranteed to be physically reasonable and requires separate examination. The two issues of $\eta_{Petela-Landsberg}$ later shown exemplify this point.

### 6.3.1. Procedure 1

Based on the specific conditions of each analysis of *light-powered systems*, $E_{in}^{\gamma}, E_{out}^{\gamma}$、 $S_{in}^{\gamma}, S_{out}^{\gamma}$,

$p_\gamma = \frac{S_{out}^{\gamma}(T_{out})}{S_{in}^{\gamma}(T_{in})}, p_Q = 1 - p_\gamma = 1 - \frac{S_{out}^{\gamma}(T_{out})}{S_{in}^{\gamma}(T_{in})}$ and $Y_{in}^{\gamma}, Y_{out}^{\gamma}$ are obtained individually. Here, $Y_{out}^Q = 1$ is always assumed.

### 6. 3. 2. Procedure 2

$\eta_{upper}$ is obtained by substituting the obtained $p_\gamma, p_Q, Y_{in}^{\gamma}, Y_{out}^{\gamma}$ in Eq. (6.12). Following the above procedure, the ideal efficiencies of several *light-powered systems* in previous studies are obtained as examples.

### 6.4. Application Examples: The analysis of the ideal efficiencies in the several previous studies

#### 6.4.1. Derivation of Spanner efficiency

From set conditions given by Spanner [12], the following conditional expressions are derived:

$$p_\gamma = 0 \text{ and } \quad p_Q = 1, \tag{6.14}$$

and

$$Y_{out}^{Q} = 1, Y_{in}^{\gamma} = \frac{4}{3} \ \ (\varepsilon_{in}^{\gamma} = 1). \tag{6.15}$$

Eqs. (6.14) and (6.15) are substituted into Eq. (6.12) to obtain the following equation:

$$\eta_{upper} = 1 - \left(\frac{p_Q}{Y_{out}^{Q}} + \frac{p_\gamma}{Y_{out}^{\gamma}}\right) Y_{in}^{\gamma} \frac{T_{out}}{T_{in}}$$

$$= 1 - \frac{4}{3}\frac{T_{out}}{T_{in}} = \eta_{Spanner}. \tag{6.16}$$

A persistent doubt has remained about $\eta_{Spanner} < 0$ when$T_{in} < (4/3)T_{out}$[11,46]**.** Spanner tried to clarify this doubt in his original paper [12]. The doubt actually stems mainly from a generalized prejudice that for a heat engine, if the temperature of a heat source is even slightly above the ambient temperature, then work can always be extracted in principle. This notion assumes a first-order evaluable condition $\varepsilon \to 0$ (quasi-equilibrium between heat source and system). However, Spanner's efficiency, which does not assume this condition but $\varepsilon_\gamma = 1$, is consistent with the first and second laws of thermodynamics and is therefore not physically unreasonable, as discussed in detail in subsection (II) in Appendix B of this paper.

#### 6.4.2. Derivation of Jeter efficiency

The set conditions result in the following conditional expressions:

$$p_\gamma = 0 \text{ and } p_Q = 1, \tag{6.17}$$

and

$$Y_{out}^{Q} = 1, Y_{in}^{\gamma} = 1 \ (\varepsilon_{in}^{\gamma} \to 0). \tag{6.18}$$

Eqs. (6.17) and (6.18) are substituted into Eq. (6.12) to obtain the following equation:

$$\eta_{upper} = 1 - \left(\frac{1}{1} + \frac{0}{Y_{out}^{\gamma}}\right) 1 \frac{T_{out}}{T_{in}}$$

$$= 1 - \frac{T_{out}}{T_{in}} = \eta_{Jeter} \tag{6.19}$$

In their original paper [19], Jeter gave the wrong reason for Eq. (6.19) to hold. This is discussed in detail and presented the correct reason in subsection (III) in Appendix B of this paper.

As mentioned above, the original conditions for deriving the Jeter efficiency were that the entropy of $Y_{in}^{\gamma} = 1$ (the first-order evaluable condition for the inflow radiation) was discarded only via thermal processes, and that radiative discard was not assumed. However, even if there are both thermal and radiative entropy discarding, the Jetter efficiency is derived when the first-order evaluable conditions (quasi-reversible conditions) for both inflow and outflow radiation are satisfied. In fact, by substituting quasi-reversible conditions for the inflow and outflow radiative flows ($Y_{out}^{\gamma} = 1$ and $Y_{in}^{\gamma} = 1$, respectively) into the general equation (6.12) established in this study, we obtain the Jeter efficiency (Carnot efficiency).

### 6.4.3. Derivation of Landsberg efficiency and modified Landsberg efficiency

Landsberg's efficiency is referred to in many applied scientific papers [23-28,40-45], some of which refer to it as Landsberg's limit [40-45]. From the set conditions given by Landsberg and Tonge [9], the following conditional expressions can be obtained using our approach:

$$p_{\gamma} = \frac{S_{out}^{\gamma}(T_{out})}{S_{in}^{\gamma}(T_{in})} = \frac{(4/3)\sigma T_{out}^{3}}{(4/3)\sigma T_{in}^{3}} = \left(\frac{T_{out}}{T_{in}}\right)^{3}. \tag{6.20}$$

Thus,

$$p_{Q} = 1 - \left(\frac{T_{out}}{T_{in}}\right)^{3}. \tag{6.21}$$

Further,

$$Y_{out}^{Q} = 1,\ Y_{out}^{\gamma} = \frac{4}{3}\ \left(\varepsilon_{out}^{\gamma} = 1\right),\ Y_{in}^{\gamma} = \frac{4}{3}\ \left(\varepsilon_{in}^{\gamma} = 1\right). \tag{6.22}$$

Substituting Eq. (6.20)–(6.22) into Eq. (6.12), we obtain:

$$\eta_{upper} = 1 - \left(\frac{1-\left(\frac{T_{out}}{T_{in}}\right)^{3}}{1} + \frac{\left(\frac{T_{out}}{T_{in}}\right)^{3}}{\frac{4}{3}}\right) 4/3 \frac{T_{out}}{T_{in}}$$

$$= 1 - \frac{4}{3}\frac{T_{out}}{T_{in}} + \frac{1}{3}\left(\frac{T_{out}}{T_{in}}\right)^{4} = \eta_{Landsberg}. \tag{\textbf{6.23}}$$

However, there are two issues with $\eta_{Landsberg}$. The first issue is that, unless the system is in a vacuum, the correct flow conditions are represented by Figure 6(b) rather than Figure 6(a), which is the condition set by Landsberg and Tonge [9]. Thus, $Y_{out}^{\gamma} = 1\ \left(\varepsilon_{out}^{\gamma} \to 0\right)$, and the third term of $\eta_{Landsberg}$ obtained in Eq. (6.23) disappears, resulting in the Spanner efficiency. Keeping this in mind, subsequent research papers assumed that Landsberg's sink had an absolute temperature value of nearly zero so that there would be no back-radiation from the sink, as shown in Fig. 7a. However, unless the volume of the absolute-zero sink is sufficiently large, absorption of thermal radiation will raise its temperature above absolute zero. Therefore, this condition to be realistic—as illustrated in Fig. 7b—it would, for example, require a future lunar capsule city equipped with a light engine positioned high above the capsule to radiate entropy into the vastness of space.

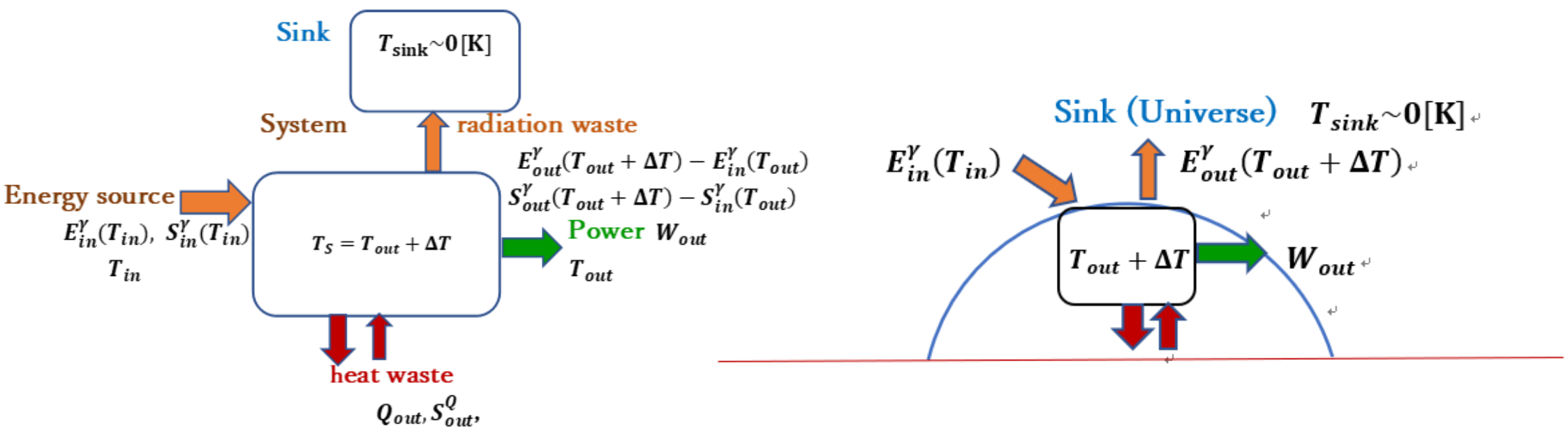

**a** **b**

**Fig.7. Schematic of energy and entropy fluxes into and out of the system with a sink of zero temperature:** a) Landsberg's sink [9] should have an absolute temperature value of nearly zero so that there is no back-radiation from the sink. (b) A future lunar capsule city should be equipped with a light-powered system that is positioned high above the capsule and radiates entropy into the vastness of space.

The second problem is that as soon as the temperature of the energy source $T_{in}$ drops below the ambient temperature $T_{out}$, both the Landsberg efficiency (derived from the above conditions) and the Petela efficiency (derived from the definition below) become positive again, as shown in Fig. 11a. This might violate the first law of thermodynamics (the law of conservation of energy). Regarding this point, the following comment was made in a previous paper [47]: "*whereas for* $T < T_0$ *(*$T_{in} < T_{out}$ *in this study) some of the environment internal energy is converted into work: it would then be more significant to define the efficiency relative to the energy lost by the environment*"

Taking this into consideration, this study developed a quantitative analysis based on equation (6.12) as follows:

From Eq. (6.21), the definition of $p_Q$ and $Y^{Q}_{out} = 1$, we can see

$$E^{Q}_{out}(T_{out}) = \left\{1 - \left(\frac{T_{out}}{T_{in}}\right)^3\right\}\frac{4}{3}\sigma T_{in}^3 T_{out} = \frac{4}{3}\left\{1 - \left(\frac{T_{out}}{T_{in}}\right)^3\right\}\sigma T_{in}^4 \frac{T_{out}}{T_{in}} \qquad (6.24)$$

holds.

When $T_{in} < T_{out}$, the following inequality holds.

$$p_{\gamma} = \left(\frac{T_{out}}{T_{in}}\right)^3 > 1\,. \qquad (6.25)$$

$$p_Q = 1 - \left(\frac{T_{out}}{T_{in}}\right)^3 < 0\,. \tag{6.26}$$

Seeing that $E_{out}^{Q}(T_{out}) < 0$ from Eq. (6.24) and Eq. (6.26), we can redefine the positive quantity $\tilde{E}_{in}^{Q}(T_{out})$ as follows.

$$\tilde{E}_{in}^{Q}(T_{out}) = -E_{out}^{Q}(T_{out}) = \frac{4}{3}\left\{\left(\frac{T_{out}}{T_{in}}\right)^3 - 1\right\}\sigma T_{in}^4 \frac{T_{out}}{T_{in}}. \tag{6.27}$$

Therefore, we can see that the energy source includes not only radiation energy $E_{in}^{\gamma}(T_{in})$, but also environmental heat energy $\tilde{E}_{in}^{Q}(T_{out})$ as shown in Fig.8a, and the energy efficiency is naturally defined as $\eta = \frac{W_{out}}{E_{in}^{\gamma}(T_{in})+\tilde{E}_{in}^{Q}(T_{out})}$ based on the standard definition of energy efficiency, which is defined by $\eta =$ (extracted work) / (absorbed energy).

$$\begin{aligned} W_{out} &= E_{in}^{\gamma}(T_{in}) + \tilde{E}_{in}^{Q}(T_{out}) - E_{out}^{\gamma}(T_{out}) \\ &= E_{in}^{\gamma}(T_{in}) - \left\{E_{out}^{Q}(T_{out}) + E_{out}^{\gamma}(T_{out}) \right\} \\ &= \sigma T_{in}^4 \left\{1 - \frac{4}{3}\frac{T_{out}}{T_{in}} + \frac{1}{3}\left(\frac{T_{out}}{T_{in}}\right)^4\right\} > 0 \end{aligned} \tag{6.28}$$

From Eq. (6.27), Eq. (6.28) and $E_{in}^{\gamma}(T_{in}) = \sigma T_{in}^4$,

$$\eta_{upper} = \frac{W_{out}}{E_{in}^{\gamma}(T_{in})+\tilde{E}_{in}^{Q}(T_{out})} = \frac{1-\frac{4}{3}\frac{T_{out}}{T_{in}}+\frac{1}{3}\left(\frac{T_{out}}{T_{in}}\right)^4}{1-\frac{4}{3}\frac{T_{out}}{T_{in}}+\frac{4}{3}\left(\frac{T_{out}}{T_{in}}\right)^4} \tag{6.29}$$

holds when $\boldsymbol{T_{in} < T_{out}}$.

Based on the above considerations, we can redefine Landsberg efficiency as follows:

$$\tilde{\eta}_{Landsberg} \begin{cases} = 1 - \frac{4}{3}\frac{T_{out}}{T_{in}} + \frac{1}{3}\left(\frac{T_{out}}{T_{in}}\right)^4 & \left(0 \le \frac{T_{out}}{T_{in}} \le 1\right) \\ = \frac{1-\frac{4}{3}\frac{T_{out}}{T_{in}}+\frac{1}{3}\left(\frac{T_{out}}{T_{in}}\right)^4}{1-\frac{4}{3}\frac{T_{out}}{T_{in}}+\frac{4}{3}\left(\frac{T_{out}}{T_{in}}\right)^4} & \left(1 \le \frac{T_{out}}{T_{in}}\right) \end{cases} \tag{6.30}$$

From Eq. (6.29), we can easily see that $\lim_{\boldsymbol{T_{in}}\to 0}\left(\tilde{\eta}_{Landsberg}\left(\frac{\boldsymbol{T_{out}}}{\boldsymbol{T_{in}}}\right)\right) = \frac{1}{4}$ under the condition of a constant environmental temperature $T_{out}$, and we can also see that $\frac{d}{dx}\tilde{\eta}_{Landsberg}(x) = \frac{\boldsymbol{4x^3(x-1)}}{\left(\boldsymbol{1-\frac{4}{3}x+\frac{4}{3}x^4}\right)^2}$ $(\boldsymbol{x = T_{out}/T_{in} > 1})$, and the behavior of $\tilde{\eta}_{Landsberg}(x)$ is as shown in Figure 8b. For $T_{out}/T_{in} > 1$, Fig.8b of $\tilde{\eta}_{Landsberg}$ is entirely different from Fig.11a of $\eta_{Landsberg}$.

The formula $\lim_{T_{in}\to 0}\left(\tilde{\eta}_{Landsberg}\left(\frac{T_{out}}{T_{in}}\right)\right) = \frac{1}{4}$ means that when the energy source $E_{in}$ has no radiation source $E_{in}^{\gamma}(T_{in})$ and only an environmental heat source $\tilde{E}_{in}^{Q}(T_{out})$, the ideal efficiency is 1/4. From this, it is found that the Landsberg efficiency condition for radiative entropy discard yields also an "environmental heat engine".

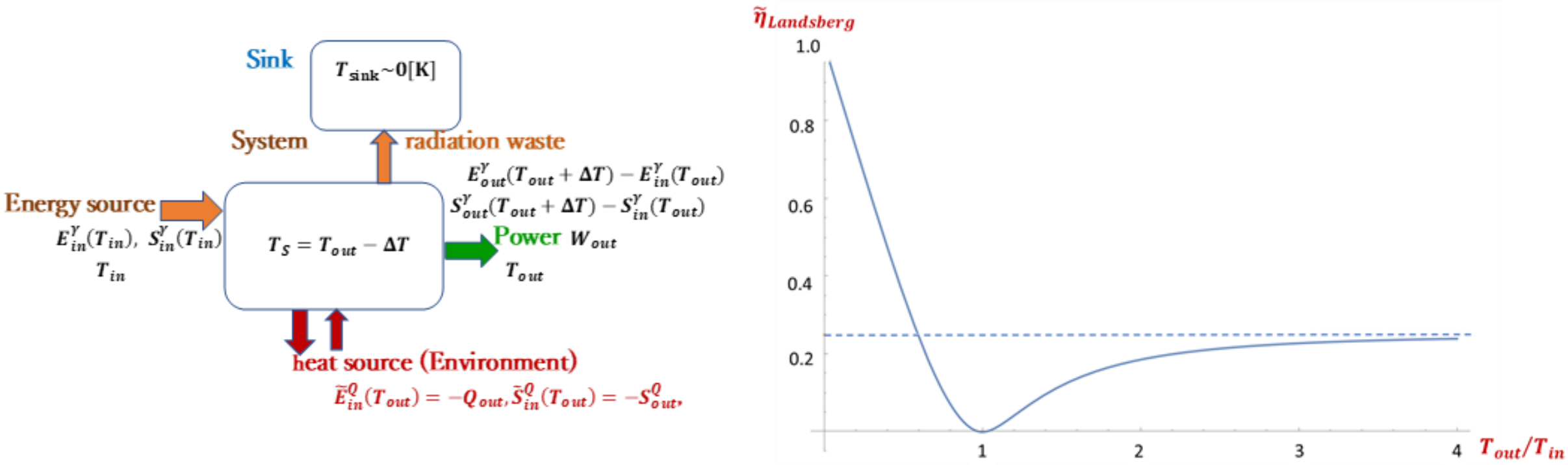

**a** **b**

**Fig.8. Schematic and graph for the case $T_{in} < T_{out}$: a)** In the case of $T_{in} < T_{out}$, $E_{out}^{Q}(T_{out}) < 0$ is not the heat for the entropy discard, but rather the heat for the energy source, which is expressed as $\tilde{E}_{in}^{Q}(T_{out}) = -E_{out}^{Q}(T_{out})$. **b) The graph with the horizontal axis as the $T_{out}/T_{in}$ and the vertical axis as the $\tilde{\eta}_{Landsberg}$: The behavior of** $\tilde{\eta}_{Landsberg}$, is modified from the original $\eta_{Landsberg}$ in the range of $T_{out}/T_{in} > 1$. The graph of $\tilde{\eta}_{Landsberg}$ approaches 1/4 as $T_{out}/T_{in}$ goes to infinity.

Although the quantitative analysis above tentatively resolved the two problems regarding $\eta_{Landsberg}$, we found that the conditions under which it is realized are very specific. Furthermore, it is very important for us to understand that the mechanism underlying its radiation to discard entropy is not suitable for photosynthetic, light-driven systems. In deriving the $\eta_{Landsberg}$, the radiative discard of entropy is assumed to be blackbody radiation of a photon gas within the system. However, in actual photosynthetic *light-powered systems*, radiation (induced and spontaneous) occurs when molecules (e.g., pigment molecules) transition from an excited state to a ground state. This radiation is responsible for the radiative discard of entropy. The reason for the above problem with the $\eta_{Landsberg}$ is that the radiative entropy discarding mechanism was not properly modeled, but was regarded as simple blackbody radiation from the system.

Section 8 derived a new ideal efficiency using a simple mathematical model based on Einstein's theory of radiation and absorption in a two-level system (composed of an excited state and a ground state), which assumes quantum transitions, to resolve those problems. The newly obtained ideal efficiency $\tilde{\eta}_{upper}(T_{in}, T_{out})$ was found to differ significantly from the Landsberg efficiency given by Eq. (6.23), and to behave in a very similar manner to the Carnot efficiency as shown in Fig. 11.b and Fig.12, despite their substantial difference in mathematical form (see Eq. (6.23) and Eq. (8.36)).

### 6.4.4. Derivation of Petela efficiency based on the standard definition of energy efficiency

The same efficiency as $\eta_{Landsberg}$ had been derived by Petela using a different method [15] before Landsberg. In the original paper, Petela derived the theoretical maximum efficiency from an exergy perspective using a $p$-$V$ graph in a piston-cylinder (closed radiation) model. This definition differs from the standard energy efficiency, which is $\eta =$ (extracted work) / (absorbed energy), and has been calculated as $\eta =$(extracted work) / (internal energy). Applying the standard definition to the Petela $p$-$V$ graph yields a different formula $\tilde{\eta}_{Petela} = \frac{1}{1-(T_0/T_1)^4}\{1-(4/3)T_0/T_1+(1/3)(T_0/T_1)^4\}$ (Eq.(C6)) than that of Eq. (6.23). As a result, the contradiction from the second issue mentioned regarding Eq. (6.23) is resolved, as seen in Fig. 14. Further details are provided in Appendix C.

The conditions of $Y$-factors and $p$-parameters for the various ideal efficiencies mentioned above are summarized in Table.

| $\boldsymbol{p_\gamma}$ | $\boldsymbol{p_Q}$ | $\boldsymbol{Y_{in}^{\gamma}}$ | $\boldsymbol{Y_{out}^{\gamma}}$ | $\boldsymbol{Y_{out}^{Q}}$ | $\boldsymbol{\eta_{upper}^{\gamma}}$ |
|---|---|---|---|---|---|
| 0 | 1 | $\frac{4}{3}(\varepsilon_{in}^{\gamma}=1)$ | | $1(\varepsilon_{out}^{Q}\to 0)$ | $\boldsymbol{\eta_{Spanner} = 1-(4/3)T_{out}/T_{in}}$ |
| 0 | 1 | $1\ (\varepsilon_{in}^{\gamma}\to 0)$ | | $1(\varepsilon_{out}^{Q}\to 0)$ | $\boldsymbol{\eta_{Jeter} = 1-T_{out}/T_{in}}$ |
| $p_\gamma$ | $1-p_\gamma$ | $1\ (\varepsilon_{in}^{\gamma}\to 0)$ | $1\ (\varepsilon_{out}^{\gamma}\to 0)$ | $1(\varepsilon_{out}^{Q}\to 0)$ | $\boldsymbol{\eta_{Jeter} = 1-T_{out}/T_{in}}$ |
| $\left(\frac{T_{out}}{T_{in}}\right)^3$ | $1-\left(\frac{T_{out}}{T_{in}}\right)^3$ | $\frac{4}{3}(\varepsilon_{in}^{\gamma}=1)$ | $\frac{4}{3}(\varepsilon_{out}^{\gamma}=1)$ | $1(\varepsilon_{out}^{Q}\to 0)$ | $\boldsymbol{\eta_{Landsberg-Petela} = 1-(4/3)T_{out}/T_{in} + (1/3)(T_{out}/T_{in})^4}$ |
| $\frac{4x^3}{(x+1)(x^2+1)}$ | $\frac{(1-x)(3x^2+2x+1)}{(x+1)(x^2+1)}$ | $\frac{4}{3}(\varepsilon_{in}^{\gamma}=1)$ | $\frac{4}{3}(\varepsilon_{in}^{\gamma}=1)$ | $1(\varepsilon_{out}^{Q}\to 0$ | $\boldsymbol{\tilde{\eta}_{Petela} = \frac{\eta_{Landsberg-Petela}}{1-(T_{out}/T_{in})^4}}$ |

**Table. Analysis of the ideal efficiency in all three cases, resulting from three $Y$ factors, $\boldsymbol{Y_{in}^{\gamma}, Y_{out}^{\gamma}, Y_{out}^{Q}}$ ($\boldsymbol{\varepsilon_{out}^{Q}\to 0,\ \varepsilon_{out}^{\gamma}=1}$ and $\boldsymbol{\varepsilon_{in}^{\gamma}\to 0}$ or $\boldsymbol{\varepsilon_{in}^{\gamma}=1}$), and two weights $\boldsymbol{p_\gamma, p_Q\ (p_\gamma+p_Q=1)}$:** $x = T_{out}/T_{in}$ is used in bottom row of this table. In the table it is shown that even if there are both thermal and radiative entropy discarding, the Jetter efficiency is derived when the first-order evaluable conditions (a quasi-reversible conditions) for both inflow and outflow radiation are satisfied.

## 7. Classification and comparative analysis of microscopic photosynthetic-type *light-powered systems* into piston-cylinder and flowing radiation models

Most previous analyses of the ideal efficiency o*f light-powered systems* have focused on a cylinder–piston system with a photon gas, similar to studies regarding a thermal engine using molecular gases. On the other hand, the Landsberg efficiency is based on a flux analysis similar to the flow analysis performed in this study. However, assumptions from these previous studies, including the Landsberg analysis, have not been scrutinized for applicability to photosynthetic *light-powered systems*. This section reviews and analyzes these models and demonstrates their unsuitability for microscopic photosynthetic *light-powered systems*.

Most studies on *light-powered systems* relied on the piston-cylinder model (closed photon gas model) until Landsberg's study [9]. The debate over the amount of work that can be extracted from thermal radiation arises partly from historically posing this question in two distinct contexts [20]:

A. The radiation (photon gas) enclosed in the idealized cylinder–piston system (Fig. 9a)

B. The radiation (photon flow) passing through the idealized powered system (Fig. 9b)

This study proposes that photosynthesis is an open, not closed, light-powered system, with radiation serving as a dynamic flow of energy rather than a static phenomenon. This is demonstrated in this paper through a comparative analysis from two perspectives, A and B, using essential conceptual diagrams (Figs. 9a and 9b).

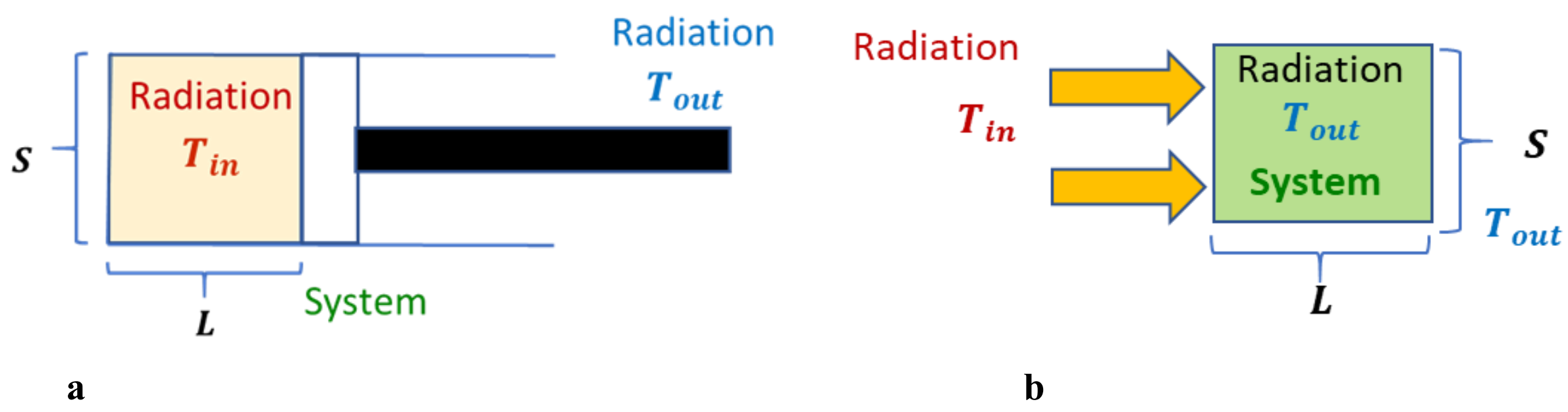

**Fig. 9. (a) Piston-cylinder model (closed photon gas model) and (b) Flowing radiation model (open photon gas model):** a) Blackbody radiation is enclosed at temperature $T_{in}$ in a piston-cylinder system, and the ideal efficiency $\eta_{upper}$ is analyzed using on the $p$-$V$ graph. For example, Petela's research is also based on this model [15]. b) Blackbody radiation at temperature $T_{in}$ is flowing in a system, and the ideal efficiency $\eta_{upper}$ is analyzed using the formula given by Eq. (6.12) in this work.

Standard statistical mechanics, including the second law of thermodynamics, at the macroscopic scale may not be applicable to photosynthetic systems at the microscopic (spatial and temporal) scale. In the light of this assertion, it is prudent to examine the two types of photosynthetic systems (A and B) mentioned above, using the following two necessary conditions:

( i ) The condition for $G$ (the number of quantum states of a photon) should be

$$G \geq 1. \tag{7.1}$$

( ii ) The condition for Δλ (the uncertainty of wavelength of light) to identify the wavelength should be

$$\Delta\lambda \leq \lambda. \tag{7.2}$$

The number of quantum states permissible for a photon, $G$, is given by

$$G = \frac{2\Delta p^3 \Delta q^3}{h^3} = \frac{2\Omega p^2 \Delta p \Delta q^3}{h^3} = \frac{8\pi p^2 \Delta p \Delta q^3}{h^3}, \tag{7.3}$$

where the factor 2 is the number of spin degrees-of-freedom of a photon, $\Omega$ is the solid angle as seen from the light-powered system, and here it is assumed to be $4\pi$ here, not for direct sunlight but for atmospheric scattered sunlight.

Substituting $\Delta q^3 = V, p = h/\lambda,\ \Delta p = (h/\lambda^2)\Delta\lambda$ into Eq. (7.3), we obtain

$$G = 8\pi \frac{V}{\lambda^4} \Delta\lambda. \tag{7.4}$$

By solving Eq. (7.4) for $\Delta\lambda$, we obtain

$$\Delta\lambda = \frac{\lambda^4 G}{8\pi V}. \tag{7.5}$$

Using Eq. (7.5), the condition ( i ) $G \geq 1$ is as follows,

$$\Delta\lambda \geq \frac{\lambda^4}{8\pi V}. \tag{7.6}$$

From Eqs. (7.2) and (7.6) we obtain $\lambda \geq \Delta\lambda \geq \lambda^4/(8\pi V)$, and finally the following inequality is derived:

$$V \geq \frac{\lambda^3}{8\pi}. \tag{7.7}$$

The inequality in Eq. (7.7) intuitively implies that the spatial scale of the system is larger than the wavelength of the irradiated light. It can be interpreted through the position–momentum uncertainty relation as follows: From the uncertainty relation between the momentum $p = h/\lambda$ and length $\ell$ for a photon gas,

$$\frac{\Delta\lambda}{\lambda^2} \cdot \Delta\ell \geq \frac{1}{(4\pi)} \sim 1 \tag{7.8}$$

is obtained. Using the inequality (7.8) and the wavelength and length identification conditions $\Delta\lambda \leq \lambda$ and $\Delta\ell \leq \ell$, we obtain

$$1 \geq \frac{\Delta\lambda}{\lambda} \geq \frac{\lambda}{\Delta\ell} = \frac{\lambda}{\ell} \cdot \frac{\ell}{\Delta\ell} \geq \frac{\lambda}{\ell}, \tag{7.9}$$

and finally, we obtain the following inequality:

$$\ell \geq \lambda. \tag{7.10}$$

According to the inequality (7.10), the spatial length of a system must exceed the wavelength of the incident light for wavelength identification. We examine whether the inequality condition (7.7) is satisfied in each case when the two above-mentioned analytical models A and B are applied to the light-harvesting system (LHC) of natural photosynthesis. This will be discussed in the following subsections:

### 7.1. Piston-cylinder model (closed photon gas model)

As shown in Fig. 9a, let $L$ be the length of the light-harvesting system (including the reaction center) in the direction of the incident light, and let $S$ be the cross-sectional area (perpendicular to the light) receiving the light. The closed system model (Fig. 9a) fails to meet the inequality (7.7) derived from the necessary conditions (i) and (ii). This discrepancy is evident for the estimated value $L$ = 100 nm and the actual estimated cross-sectional area of the light-harvesting complex (LHC), $S = (10 \text{ nm})^2 = 1 \times 10^{-16} \text{ m}^2$, which yields

$$V = SL = 1 \times 10^{-23} \text{ m}^3. \tag{7.11}$$

When the mean value of photosynthetically active radiation (PAR), $\lambda = 550 \text{ nm} = 5.5 \times 10^{-7}$ m, is substituted into the right-hand side of (7.7), we obtain

$$\frac{\lambda^3}{8\pi} = \frac{(5.5\times10^{-7}\text{m})^3}{8\pi} = \frac{(5.5)^3}{8\pi} \times 10^{-21}\text{m}^3 = 6.62 \times 10^{-21}\text{m}^3. \tag{7.12}$$

Eqs. (7.11) and (7.12) do not satisfy the inequality (7.7). Thus, the box model A is not physically plausible on the scale of the actual photosynthetic LHC absorbing PAR owing to the significantly smaller LHC length than the PAR wavelength. If the length $L$ is 100 μm, which is three orders of magnitude larger than the estimated $L$ = 100 nm of the actual LHC, then the inequality (7.7) can be satisfied.

### 7.2. Flowing radiation model (open photon gas model)

To satisfy inequality (7.7) derived from the necessary conditions (i) and (ii), a proper definition of $\ell$ contained in the radiation volume $V = S\ell$ is required. In case A, where the photon gas is confined in a closed system, inequality (7.7) is not satisfied because $\ell = L$ is the size of the photosynthetic light-harvesting system. In case B, the photon gas flows through an open system, with the length $\ell$ representing the photon gas flowing through the system during the time interval $\Delta t$ required to receive eight photons for one glucose molecule production. This understanding applies to the analysis of direct solar radiation at the surface of the Earth (Fig. 9b). In this case, the length $\ell_{direct}$ is expressed as $\ell_{direct} = c\Delta t$. The time required to produce one molecule of glucose, determined from actual data, is ~1 ms. Thus, $\ell_{direct}$ using $\Delta t$ = 1 ms and $S = 1 \times 10^{-16} \text{ m}^2$ can be estimated as

$$V_{direct} = S\ell_{direct} = Sc\Delta t = 3 \times 10^{-11}\text{m}^3, \tag{7.13}$$

The validity of the statistical mechanical analysis of the radiation (photon gas) incident on microscopic systems such as the LHC on the surface of the Earth should be evaluated based on the number of photon quantum states $G$ and the number of microscopic states $W$ for direct solar radiation, which has a lower $G$ and then more severe conditions, rather than for solar radiation scattered in the atmosphere. Under the condition of direct solar radiation on the ground, the solid angle $\Omega_{direct}$ is not 4π, but $\Omega_{direct} \cong 2\pi(R/D)^2$ , and consequently, $G$ given by Eq. (7.4) also became $G_{direct} \cong 2 \times 2\pi(R/$

$D)^2\, V_{direct} \Delta\lambda_{direct}/\lambda^4$. Finally, Eq. (7.6) changed to

$$\Delta\lambda_{direct} \geq \frac{\lambda^4}{4\pi (R/D)^2 V_{direct}}, \tag{7.14}$$

and finally, the condition (7.7) is modified to

$$V_{direct} \geq \frac{\lambda^3}{4\pi (R/D)^2}. \tag{7.15}$$

By substituting $\lambda = 550$ nm, the radius of the Sun as $R = 7.0 \times 10^8$ m, and the distance between the Sun and Earth $as\ D = 1.5 \times 10^{11}$ m in the right-hand side of Eq. (7.15), the following lower bound condition is obtained for $V_{direct}$:

$$V_{direct} \geq 6.1 \times 10^{-16}\ \mathrm{m}^3. \tag{7.16}$$

Evidently, Eq. (7.13) satisfies the inequality (7.16) modified from inequality (7.7).

The number of quantum states of a photon, $G(\lambda, T_D(\lambda))$, at the effective temperature $T_D(\lambda)$ is

$$G(\lambda, T_D(\lambda)\,) = \frac{N}{f(\lambda, T_D(\lambda)\,)}, \tag{7.17}$$

where

$$f\big(\lambda, T_D(\lambda)\big) = \frac{1}{e^{hc/(\lambda k_B T_D(\lambda))} - 1}. \tag{7.18}$$

The effective temperature $T_D(\lambda)$ in Eq. (7.18) was formulated in Part 1 of this study [29,30] as

$$T_D^{\gamma}(\lambda) = \frac{hc/\lambda k_B}{\ln\left\{\frac{1}{d}\left\{\exp\left(\frac{hc}{\lambda k_B T_{sun}}\right) - 1\right\} + 1\right\}}, \tag{7.19}$$

where the dilution indicator was expressed as $d = (R/D)^2$ using the radius $R$ of the Sun and the Sun–Earth distance $D$. The number of the photons $N$ can be given by

$$N = \frac{\widetilde{N}}{a}, \tag{7.20}$$

where $\widetilde{N}$ is the number of photons required to produce one molecule of glucose ($\widetilde{N}$ = 8 has been adopted from previous research in the field of photosynthesis), and $a$ is the light absorption ratio. Here, 0.8 is used as the average of this value in PAR. By substituting $\widetilde{N} = 8$ and $a = 0.8$ to Eq. (7.20), we obtain

$$N = 10. \tag{7.21}$$

Using Eq. (7.18) and (7.21), we obtain

$$G(\lambda, T_D(\lambda)\,) = 10\big(e^{hc/(\lambda k_B T_D(\lambda))} - 1\big). \tag{7.22}$$

By substituting $\lambda = 5.5 \times 10^{-7}$m, $T_D(\lambda = 550\ \mathrm{nm})) = 1720$ [K] calculated by Eq. (7.19) and the specific values of Planck constant $h$, light velocity $c$, Boltzmann constant $k_B$, we obtain

$$G(\lambda = 550 \text{ nm}, T_D(\lambda = 550 \text{ nm})) = 4.0 \times 10^7. \tag{7.23}$$

$\Omega_{direct} \neq 4\pi$, but $\Omega_{direct} \cong 2\pi(R/D)^2$, and consequently, $G$ given by Eq. (7.4) also becomes

$$G_{direct}(\lambda = 550 \text{ nm}, T_D(\lambda = 550 \text{ nm})) = 4.0 \times 10^7 \times \frac{\Omega_{direct}}{4\pi}$$

$$= 4.0 \times 10^7 \times \frac{2\pi(R/D)^2}{4\pi}$$

$$= 440. \tag{7.24}$$

From $W = {}_{G-1+N}C_N$, i.e., the number of cases in which $N$ indistinguishable Bose particles can be arranged in $G$ distinguishable quantum states,

$$W_{direct}(\lambda = 550 \text{ nm}, T_D(\lambda = 550 \text{ nm}))$$

$$= {}_{G_{direct}(\lambda = 550\text{nm}, T_D(\lambda = 550\text{nm}))-1+N}C_N$$

$$= {}_{449}C_{10} = 8.3 \times 10^{19}. \tag{7.25}$$

This value is sufficiently large to be the subject of statistical mechanics.

The above analysis indicates that by considering the length of radiation ($L$) during the time interval for photons in the photosynthetic system to produce one glucose molecule, both the issues of (1) ***the inapplicability of statistical mechanics*** and (2) ***the unidentifiability of wavelengths in PAR*** can be resolved in the flowing radiation model (open photon gas model).

Landsberg and Tonge [9] derived the Petela efficiency based on the Flowing radiation model (Fig. 9b). However, their analysis has two issues: 1) as discussed in Section 6, the radiative discarded entropy in their conditions is emitted by the blackbody radiation within the system. Unless the environment of a system is vacuum, the quasi-equilibrium condition must be satisfied, resulting in $Y_{out}^{\gamma} = 1$, which contradicts their assumption $Y_{out}^{\gamma} = 4/3$. 2) as shown in Fig.11a, the derived ideal efficiency $\eta_{Petela-Landsberg}$ returns positive as soon as the temperature of the energy source $T_{in}$ becomes lower than the ambient temperature $T_{out}$. This suggests that work can be extracted from radiation flowing from low to high temperatures, which contradicts the first law of thermodynamics. These contradictions indicate that $\eta_{Petela-Landsberg}$ is not a physically reasonable or correct ideal efficiency. Therefore, a simplified model analysis, considering the mechanisms of photosynthetic-type *light-powered systems,* was conducted in this study. It is presented in the next section. Einstein's absorption and emission model for a two-level system (ground and excited states) [48] is applied to photosynthetic type *light-powered systems* for the mathematical model analysis, resolving the two issues in $\eta_{Petela-Landsberg}$.

## 8. A new ideal efficiency using a simple mathematical model based on Einstein's radiation absorption and emission theory

### 8.1. Formulation introducing Einstein coefficients (A, B)

In this study, a simplified mathematical model was constructed, and the essential behavior of the ideal efficiency due to radiative entropy discard was analytically extracted.

In this mathematical model, the radiative discarded entropy is not emitted by the blackbody radiation within the system, as Landsberg assumed for deriving $\eta_{Landsberg}$, but by the newly emitted photons during the relaxation of excited pigment electrons to the ground state, as illustrated in Fig. 10.

In the mathematical model used for analysis, Einstein's theory of absorption and emission of monochromatic light for a two-level system of ground and excited states [48,49] is extended to absorption and emission of radiation of all wavelengths and simplified for the theoretical analysis. Natural photosynthesis involves complex processes in transferring light absorbed by the light-harvesting pigment complex to the reaction center, but this mechanism is simplified here. Two-level systems are assumed for each frequency $\nu$ of the pigment, and a common value is used for the number $n_2$ of pigment molecules in the ground state. The theoretical analysis is performed using this mathematical model.

Henceforth, the frequency representation of light is used in this paper instead of the wavelength. All previous relations can be mechanically replaced by $\nu = c/\lambda$. We first define the energy $E_{in}^{\gamma}(\nu, T_{in})$, $E_{out}^{\gamma}(\nu)$ and their associated entropy $S_{in}^{\gamma}(\nu, T_{in})$, $S_{out}^{\gamma}(\nu)$ absorbed and emitted by the system at the corresponding temperature under ideal conditions. For this, we use the number of photons absorbed and emitted per unit frequency width, $N_{out}^{\gamma}(\nu, T_{out})$ and $N_{out}^{\gamma}(\nu, T_{out})$, respectively.

#### 8.1.1. Radiation energy (absorption, emission)

・Absorption energy：

$$E_{in}^{\gamma}(\nu, T_{in})d\nu = h\nu N_{in}^{\gamma}(\nu, T_{in})d\nu, \tag{8.1}$$

・Emission energy：

$$E_{out}^{\gamma}(\nu, T_{out})d\nu = h\nu N_{out}^{\gamma}(\nu, T_{out})d\nu. \tag{8.2}$$

#### 8.1.2. Radiation entropy (absorption, emission)

After rewriting $\Delta N^{\gamma}$ in Eq. (4.3) to $N_{in}^{\gamma}$,

・Absorption entropy:

$$S_{in}^{\gamma}(\nu,T_{in})d\nu = k_B \ln\left(1+\frac{1}{f(\nu,T_{in})}\right)\{1+\gamma(f(\nu,T_{in}),\varepsilon(\nu))\}N_{in}^{\gamma}(\nu,T_{in})d\nu, \tag{8.3}$$

・Emission entropy：

$$S_{out}^{\gamma}(\nu,T_{out})d\nu = k_B \ln\left(1+\frac{1}{f(\nu,T_{out})}\right)\{1+\gamma(f(\nu,T_{out}),\varepsilon(\nu))\}N_{out}^{\gamma}(\nu,T_{out})d\nu. \tag{8.4}$$

By applying the distribution function $f(\nu,T)=\frac{1}{e^{h\nu/k_BT}-1}$ of the blackbody radiation at temperature $T$, we obtain

$$S_{in}^{\gamma}(\nu,T_{in})d\nu = \frac{h\nu}{T_{in}}\{1+\gamma(f(\nu,T_{in}),\varepsilon(\nu))\}N_{in}^{\gamma}(\nu,T_{in})d\nu, \tag{8.5}$$

$$S_{out}^{\gamma}(\nu,T_{out})d\nu = \frac{h\nu}{T_{out}}\{1+\gamma(f(\nu,T_{out}),\varepsilon(\nu))\}N_{out}^{\gamma}(\nu,T_{out})d\nu. \tag{8.6}$$

We formulate $N_{in}^{\gamma}(\nu,T_{in}), N_{out}^{\gamma}(\nu,T_{out})$ by Eqs. (8.7)-(8.10) using the coefficients $A$ and $B$ defined in Einstein's theory of absorption and emission of monochromatic light, and then performed the model analysis based on Eq. (6.12). In this paper, $\tilde{A}$ and $\tilde{B}$ represent the $A$ and $B$ coefficients per unit frequency width, respectively. For simplicity, this mathematical model assumes that the coefficients $B_{21}, B_{12}$ and then $\tilde{B}_{21}$ and $\tilde{B}_{12}$are constants independent of the frequency corresponding to the difference between the excited and ground states. This allows us to analytically extract the essential behavior of the ideal efficiency due to the radiative entropy discard**.** Based on this assumption, the following equations are set up.

$$N_{in}^{\gamma}(\nu,T_{in})d\nu = n_2\tilde{B}_{21}\rho(\nu,T_{in})d\nu dt, \tag{8.7}$$

$$N_{out}^{\gamma}(\nu,T_{out})d\nu = N_{out}^{\gamma(B)}(\nu,T_{out})d\nu + N_{out}^{\gamma(A)}(\nu)d\nu, \tag{8.8}$$

where $N_{out}^{\gamma(B)}(\nu,T_{out})d\nu$ and $N_{out}^{\gamma(A)}(\nu)d\nu$ correspond to spontaneous and stimulated emissions, respectively, and are given by

$$N_{out}^{\gamma(B)}(\nu,T_{out})d\nu = n_1(\nu,T_{in})\tilde{B}_{12}\rho(\nu,T_{out})d\nu dt, \tag{8.9}$$

$$N_{out}^{\gamma(A)}(\nu)d\nu = n_1(\nu,T_{in})\tilde{A}_{12}d\nu dt. \tag{8.10}$$

where $n_1(\nu,T)$ and $n_2$ are the numbers of pigments in the excited and ground states, respectively; $n_2$ is set as a constant, which is independent of both temperature and frequency, in this simplified model; and $n_1(\nu,T_{in})$ is given by Eq. (8.15). The original Einstein's coefficients $B_{21}, B_{12}$ have dimensions (units) of $[\mathrm{J^{-1}m^3s^{-2}}]$. However, here, $d\nu$ with a dimension of $[\mathrm{s^{-1}}]$ is explicitly included in the formulation as a preliminary (preparatory) step for the subsequent frequency integral; thus, these are denoted as $\tilde{B}_{21}, \tilde{B}_{12}$with the dimensions of $[\mathrm{J^{-1}m^3s^{-1}}]$. Similarly, $A_{12}$ with the original dimensions of $[\mathrm{s^{-1}}]$ is denoted as dimensionless $\tilde{A}_{12}$. $\rho(\nu,T)$ is the energy density per unit frequency of blackbody radiation at temperature $T$, given by the following formula, with the dimension (in units) of $[\mathrm{Jm}^{-3}\mathrm{s}]$:

$$\rho(\nu,T)d\nu = \frac{1}{V}h\nu g(\nu)d\nu\frac{1}{e^{h\nu/k_BT}-1} = \frac{8\pi h\nu^3}{c^3}\frac{1}{e^{h\nu/k_BT}-1}d\nu. \tag{8.11}$$

$$(g(\nu) = \frac{8\pi\nu^2 V}{c^3}d\nu)$$

Thus, Eq. (8.7) is written as

$$N_{in}^{\gamma}(\nu,T_{in})d\nu = n_2\tilde{B}_{21}\frac{8\pi h\nu^3}{c^3}\frac{1}{e^{h\nu/k_BT_{in}-1}}d\nu dt, \tag{8.12}$$

and the following relationship between the coefficients $A_{12}$ and $B_{12}$ is also obtained from Einstein's analysis under equilibrium conditions. $A_{12}$ and $B_{12}$ are the probability coefficients (Einstein coefficients) for relaxation from the excited state 1 to the ground state 2.

$$A_{12}(\nu) = \frac{8\pi h\nu^3}{c^3}B_{12}. \tag{8.13}$$

From Eqs. (8.8)–(8.10), we obtain

$$N_{out}^{\gamma}(\nu,T_{out})d\nu = n_1(\nu,T_{in})_1\tilde{B}_{12}\frac{8\pi h\nu^3}{c^3}\frac{1}{e^{h\nu/k_BT_{out}-1}}d\nu dt + n_1(\nu,T_{in})\frac{8\pi h\nu^3}{c^3}\tilde{B}_{12}d\nu dt$$

$$= n_1(\nu,T_{in})\tilde{B}_{12}dt\frac{8\pi h\nu^3}{c^3}\frac{1}{e^{h\nu/k_BT_{out}-1}}(1 + e^{h\nu/k_BT_{out}} - 1)d\nu$$

$$= n_1(\nu,T_{in}\tilde{B}_{12}dt\frac{8\pi h\nu^3}{c^3}\frac{e^{h\nu/k_BT_{out}}}{e^{h\nu/k_BT_{out}-1}}d\nu. \tag{8.14}$$

The constitutive ratio of the number ($n_1(\nu,T_{in})$) of excited state pigment molecules to that of ground state molecules is traditionally given by the Boltzmann-type factor and more accurately by the absorption-radiation entropy analysis derived in Part 1 of this study [29,30], which is also applicable to the thermal non-equilibrium state. The temperature is expressed through the dilution indicator $d$ and the absorption ratio $|\varepsilon|$ as an indicator beyond the first-order evaluation based on photon number. The conditions $d = 1$ and $|\varepsilon| = 1$ result in the following formula, which is consistent with the conventional formula.

$$n_1(\nu,T_{in}) = e^{-h\nu/k_BT_{in}}n_2. \tag{8.15}$$

The following equation is obtained for the final form of $N_{out}^{\gamma}(\nu)d\nu$.

$$N_{out}^{\gamma}(\nu,T_{out},T_{in})d\nu = e^{-h\nu/k_BT_{in}}n_2\tilde{B}_{12}dt\frac{8\pi h\nu^3}{c^3}\frac{e^{h\nu/k_BT_{out}}}{e^{h\nu/k_BT_{out}-1}}d\nu$$

$$= n_2\tilde{B}_{12}dt\frac{8\pi h\nu^3}{c^3}\frac{1}{e^{h\nu/k_BT_{out}-1}}e^{\frac{h\nu}{k_BT_{out}}(1-\frac{T_{out}}{T_{in}})}d\nu. \tag{8.16}$$

Using $N_{in}^{\gamma}(\nu, T_{in})d\nu$ (Eq. (8.12)) and $N_{out}^{\gamma}(\nu, T_{out}, T_{in})d\nu$ (Eq. (8.16)), the evaluation at the total frequency ν (by frequency integration) of the absorbed and emitted radiation energies ($E_{in}^{\gamma}(\nu)d\nu$ and $E_{out}^{\gamma}(\nu)d\nu$, respectively) and the associated absorption and discard entropies ($S_{in}^{\gamma}(\nu)d\nu$ and $S_{outo}^{\gamma}(\nu)d\nu$, respectively) are formulated as follows:

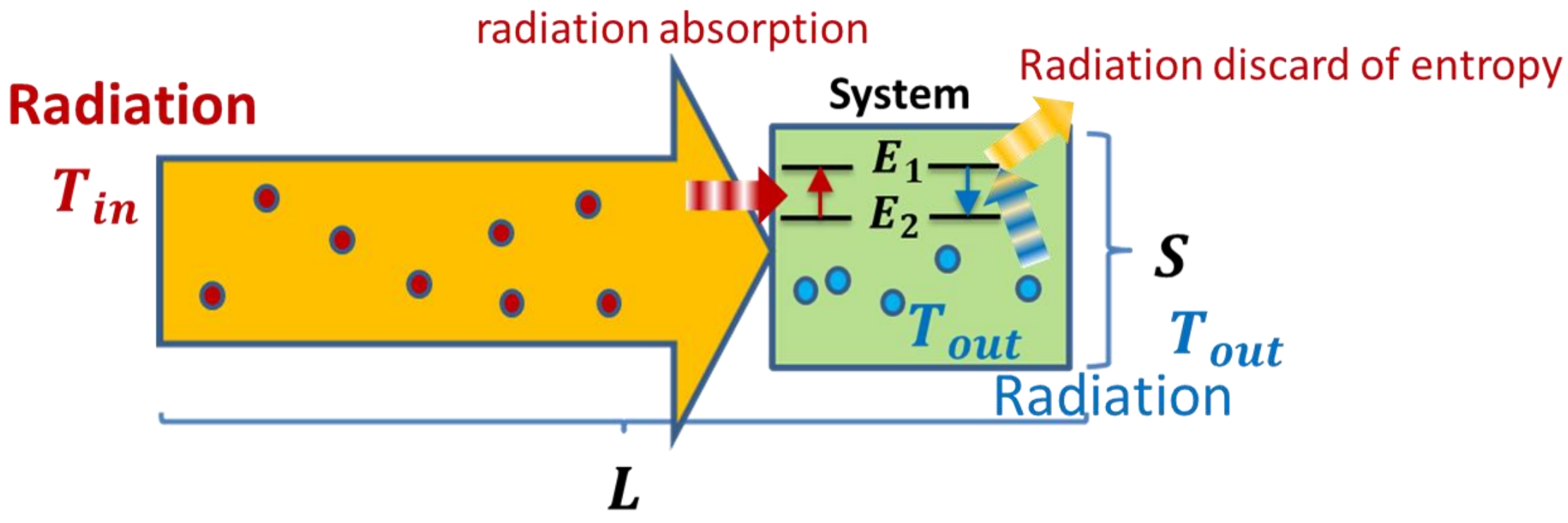

**Fig. 10. Schematic of the model based on Einstein's theory of radiation absorption and emission:** In this model, the radiative discarded entropy is not emitted by the blackbody radiation within the system, as Landsberg assumed for deriving $\boldsymbol{\eta_{Landsberg}}$[40], but by the newly emitted photons during the relaxation of excited pigment electrons to the ground state in the system.

### 8.2. Evaluation in integral form over all frequencies ν

#### 8.2.1. Radiation energies (absorption, emission)

・Absorption energy $E_{in}^{\gamma}(T_{in})$ :

$$E_{in}^{\gamma}(T_{in}) = \int_0^{\infty} h\nu\, N_{in}^{\gamma}(\nu, T_{in})d\nu$$

$$= \int_0^{\infty} h\nu n_2 \tilde{B}_{21} \frac{8\pi h\nu^3}{c^3} \frac{1}{e^{h\nu/k_B T_{in}-1}} dt d\nu \quad \text{(from Eq. (8.12))}$$

$$= \tilde{B}_{21} n_2 dt \frac{8\pi h^2}{c^3} \int_0^{\infty} \frac{\nu^4}{e^{a_{in}\nu}-1} d\nu \qquad (a_{in} \equiv h/k_B T_{in})$$

$$= \tilde{B}_{21} n_2 dt \frac{8\pi h^2}{c^3} \frac{1}{{a_{in}}^5} \int_0^{\infty} \frac{\mu^4}{e^{\mu}-1} d\mu \quad \text{(Replaced by } \mu = a_{in}\nu)$$

$$= n_2 \tilde{B}_{21} dt \left(\frac{8\pi k_B^5}{c^3 h^3}\right) T_{in}^5 \Gamma(5)\, \zeta(5). \qquad \textbf{(8.17)}$$

・Emission energy $E_{out}^{\gamma}(T_{out})$ :

$$E_{out}^{\gamma}(T_{out}) = \int_0^{\infty} h\nu N_{out}^{\gamma}(\nu, T_{out}, T_{in})\, d\nu$$

$$= \int_0^{\infty} h\nu\, n_2 \tilde{B}_{12} dt \frac{8\pi h\nu^3}{c^3} \frac{1}{e^{h\nu/k_B T_{out}-1}} e^{\frac{h\nu}{k_B T_{out}}(1-\frac{T_{out}}{T_{in}})} d\nu. \quad \text{(From Eq. (8.14))}$$

$$= \tilde{B}_{21} n_2 dt \frac{8\pi h^2}{c^3} \int_0^{\infty} \frac{\nu^4}{e^{a_{out}\nu}-1} e^{a_{out}\nu\left(1-\frac{T_{out}}{T_{in}}\right)} d\nu \qquad (a_{out} \equiv h/k_B T_{out})$$

$$= \tilde{B}_{21} n_2 dt \frac{8\pi h^2}{c^3} \frac{1}{{a_{out}}^5} \int_0^\infty \frac{\mu^4}{e^\mu - 1} e^{\mu(1-\frac{T_{out}}{T_{in}})} d\mu \quad \text{(Replaced by } \mu = a_{out}\nu\text{)}$$

$$= n_2 \tilde{B}_{12} dt \left(\frac{8\pi k_B^5}{c^3 h^3}\right) T_{out}^5 \Gamma(5)\, \zeta(5, T_{out}/T_{in}). \tag{8.18}$$

where the function $\zeta(n,x)$ is called the Hurwitz zeta function, and it is explained by Eqs. (8.26) and (8.27) in the next subsection 8.2.2.

### 8.2.2. Radiation entropies (absorption, emission discard)

・Absorption entropy $S_{in}^{\gamma}(T_{in})$：

For simplicity, we assume here that $\varepsilon_\nu^{in} = -1$ at all frequencies, as in all previous studies using blackbody radiation [e.g. 9,12-15,19].

$$S^\gamma(T_{in}) = \int_0^\infty S_{in}^\gamma(\nu, T_{in}) N_{in}^\gamma(\nu, T_{in})\, d\nu$$

$$= \int_0^\infty \{1 + \gamma(f_\nu(T_{in}), \varepsilon_\nu^{in} = -1)\} \frac{h\nu}{T_{in}} N_{in}^\gamma(\nu, T_{in}) d\nu. \tag{8.19}$$

Some calculations yielded the following results:

$$1 + \gamma(f_\nu(T), \varepsilon_\nu = 1) = e^{a\nu} - \frac{e^{a\nu}-1}{a\nu} \ln(e^{a\nu} - 1) \quad (a \equiv h/k_B T). \tag{8.20}$$

Substituting equations (8.14) and (8.20) into equation (8.19) yields

$$S_{in}^\gamma(T_{in}) = \int_0^\infty S_{in}^\gamma(\nu, T_{in}) N_{in}^\gamma(\nu, T_{in})\, d\nu$$

$$= \int_0^\infty \left\{e^{a_{in}\nu} - \frac{e^{a_{in}\nu}-1}{a_{in}\nu} \ln(e^{a_{in}\nu} - 1)\right\} \frac{h\nu}{T_{in}} n_2 \tilde{B}_{21} \frac{8\pi h\nu^3}{c^3} \frac{1}{e^{a_{in}\nu}-1} dt d\nu \quad (a_{in} \equiv h/k_B T_{in})$$

$$= \tilde{B}_{21} n_2 dt \frac{8\pi h^2}{c^3} \frac{1}{{a_{in}}^5} \frac{1}{T_{in}} \int_0^\infty \left\{e^\mu - \frac{e^\mu - 1}{\mu} \ln(e^\mu - 1)\right\} \frac{\mu^4}{e^\mu - 1} d\mu \quad \text{(Replacing by } \mu = a_{in}\nu\text{)}$$

$$= n_2 \tilde{B}_{21} dt \left(\frac{8\pi k_B^5}{c^3 h^3}\right) T_{in}^4 \frac{5}{4} \Gamma(5)\, \zeta(5)\ . \tag{8.21}$$

where the following well-known functions are used:

$$\Gamma(5) = 4! = 24. \tag{8.22}$$

$$\zeta(5) = \frac{1}{\Gamma(5)} \int_0^\infty d\mu \frac{\mu^4}{e^\mu - 1}. \tag{8.23}$$

The expansion in Eq. (8.21) used the following variant of the equation which is generally obtained by integration by parts:

$$\int_0^\infty \mu^{n-1} \ln\left(\frac{e^\mu}{e^\mu - 1}\right) d\mu = \left[\frac{\mu^n}{n} \{\mu - \ln(e^\mu - 1)\}\right]_0^\infty - \int_0^\infty \frac{\mu^n}{n} \frac{d}{d\mu} \left\{\ln\left(\frac{e^\mu}{e^\mu - 1}\right)\right\} d\mu$$

$$= \frac{1}{n} \Gamma(n+1)\, \zeta(n+1). \tag{8.24}$$

・Emission entropy $S_{out}^\gamma(T_{out})$：

In this model, radiative discarded entropy is emitted during the relaxation of the excited pigment electron to the ground state, not through blackbody radiation within the system. Because all photons generated in this process are emitted, the non-first-order evaluation index $|\varepsilon| = -\varepsilon_{out}^{\gamma} = 1$ can be naturally assumed.

$$S_{out}^{\gamma}(T_{out}) = \int_0^{\infty} S_{out}^{\gamma}(\nu, T_{out}) N_{out}^{\gamma}(\nu, T_{out}, T_{in}) d\nu$$

$$= \int_0^{\infty} \{1 + \gamma(f_\nu(T_{out}), |\varepsilon| = 1)\} \frac{h\nu}{T_{out}} N_{out}^{\gamma}(\nu, T_{out}, T_{in}) d\nu$$

$$= \int_0^{\infty} \left\{ e^{a_{out}\nu} - \frac{e^{a_{out}\nu} - 1}{a_{out}\nu} \ln(e^{a_{out}\nu} - 1) \right\} \frac{h\nu}{T_{out}} n_2 \tilde{B}_{12} dt \frac{8\pi h \nu^3}{c^3} \frac{e^{a_{out}\nu(1 - T_{out}/T_{in})}}{e^{a_{out}\nu} - 1} d\nu$$

$$(a_{out} \equiv h/k_B T_{out})$$

$$= \frac{1}{T_{out}} n_2 \tilde{B}_{12} dt \frac{8\pi h^2}{c^3} \frac{1}{a_{out}^5} \int_0^{\infty} \left\{ e^{\mu} - \frac{e^{\mu} - 1}{\mu} \ln(e^{\mu} - 1) \right\} \frac{\mu^4}{e^{\mu} - 1} e^{\mu(1 - T_{out}/T_{in})} d\mu$$

(Replacing by $\mu = a_{out}\nu$)

$$= n_2 \tilde{B}_{12} dt \frac{8\pi h^2}{c^3} \frac{1}{T_{out}} \left( \frac{k_B T_{out}}{h} \right)^5 \int_0^{\infty} \left\{ \frac{\mu^4}{e^{\mu} - 1} + \mu^3 \ln\left( \frac{e^{\mu}}{e^{\mu} - 1} \right) \right\} e^{\mu(1 - T_{out}/T_{in})} d\mu$$

$$= n_2 \tilde{B}_{12} dt \left( \frac{8\pi k_B^5}{c^3 h^3} \right) T_{out}^4 \left\{ \int_0^{\infty} \frac{\mu^4}{e^{\mu} - 1} e^{\mu(1 - T_{out}/T_{in})} d\mu + \int_0^{\infty} \mu^3 \ln\left( \frac{e^{\mu}}{e^{\mu} - 1} \right) e^{\mu(1 - T_{out}/T_{in})} d\mu \right\}$$

$$= n_2 \tilde{B}_{12} dt \left( \frac{8\pi k_B^5}{c^3 h^3} \right) T_{out}^4 \Gamma(5)\, \zeta(5, T_{out}/T_{in}) \left\{ 1 + \frac{F(4, T_{out}/T_{in})}{\Gamma(5)\, \zeta(5, T_{out}/T_{in})} \right\}, \tag{8.25}$$

where $\zeta(n, x)$ and $F(n, x)$ are expressed below.

The function $\zeta(n, x)$, called the Hurwitz zeta function, is shown in the integral form as

$$\zeta(n, x) = \int_0^{\infty} \mu^{n-1} \left( \frac{e^{\mu}}{e^{\mu} - 1} \right) e^{\mu(1 - x)} d\mu, \tag{8.26}$$

and in the series representation as

$$\zeta(n, x) = \sum_{k=0}^{\infty} \frac{1}{(k + x)^n}. \tag{8.27}$$

The function $F(n, x)$ is shown in the integral form as

$$F(n, x) \equiv \int_0^{\infty} \mu^{n-1} \ln\left( \frac{e^{\mu}}{e^{\mu} - 1} \right) e^{\mu(1 - x)} d\mu \tag{8.28}$$

and the following series representation was obtained in this study:

$$F(n, x) = \Gamma(n) \sum_{k=1}^{\infty} \frac{1}{k(k - 1 + x)^n}. \tag{8.29}$$

### 8.2.3 Derivation of a new ideal efficiency

If the excited state (state 1) of the light-powered system is not degenerate, then the transition constants $\tilde{B}_{21}$ (from the ground state (state 2) to the excited state (state 1)) and $\tilde{B}_{12}$ (from excited state (state 1) to ground state (state 2)) are equal, i.e.,

$$\tilde{B}_{12} = \tilde{B}_{12}. \tag{8.30}$$

Thus, $Y_{in}^{\gamma}(T_{in})$ , $Y_{out}^{\gamma}(T_{out}/T_{in})$ and $p_{\gamma}$ , which are required to obtain the ideal efficiency $\eta_{upper}(T_{in}, T_{out})$ given by Eq. (6.12), are derived as follows:

$$Y_{in}^{\gamma} = T_{in}\frac{S_{in}^{\gamma}(T_{in})}{E_{in}^{\gamma}(T_{in})} = T_{in}\frac{n_2\tilde{B}_{21}dt\left(\frac{8\pi k_B^5}{c^3h^3}\right)T_{in}^4\frac{5}{4}\Gamma(5)\,\zeta(5)}{n_2\tilde{B}_{21}dt\left(\frac{8\pi k_B^5}{c^3h^3}\right)T_{in}^5\Gamma(5)\,\zeta(5)} = \frac{5}{4}\ , \tag{8.31}$$

$$\begin{aligned} Y_{out}^{\gamma}(T_{out}/T_{in}) &= T_{out}\frac{S_{out}^{\gamma}(T_{out})}{E_{out}^{\gamma}(T_{out})} \\ &= T_{out}\frac{n_2\tilde{B}_{12}dt\left(\frac{8\pi k_B^5}{c^3h^3}\right)T_{out}^4\Gamma(5)\,\zeta(5,T_{out}/T_{in})\,\left\{1+\frac{F(4,T_{out}/T_{in})}{\Gamma(5)\,\zeta(5,T_{out}/T_{in})}\right\}}{n_2\tilde{B}_{12}dt\left(\frac{8\pi k_B^5}{c^3h^3}\right)T_{out}^5\Gamma(5)\,\zeta(5,T_{out}/T_{in})} \\ &= 1+\frac{F(4,T_{out}/T_{in})}{\Gamma(5)\,\zeta(5,T_{out}/T_{in})}, \end{aligned} \tag{8.32}$$

$$p_{\gamma} = \frac{S_{out}^{\gamma}(T_{out})}{S_{in}^{\gamma}(T_{in})} = \frac{Y_{out}^{\gamma}\frac{E_{out}^{\gamma}}{T_{out}}}{Y_{in}^{\gamma}\frac{E_{in}^{\gamma}}{T_{in}}} = \frac{Y_{out}^{\gamma}}{Y_{in}^{\gamma}}\left(\frac{T_{out}}{T_{in}}\right)^{-1}\frac{E_{out}^{\gamma}}{E_{in}^{\gamma}}\ . \tag{8.33}$$

From Eqs. (8.17) and (8.18), we obtain:

$$\frac{E_{out}^{\gamma}}{E_{in}^{\gamma}} = \frac{n_2\tilde{B}_{12}dt\left(\frac{8\pi k_B^5}{c^3h^3}\right)T_{out}^5\Gamma(5)\,\zeta(5,T_{out}/T_{in})}{n_2\tilde{B}_{21}dt\left(\frac{8\pi k_B^5}{c^3h^3}\right)T_{in}^5\Gamma(5)\,\zeta(5)} = \left(\frac{T_{out}}{T_{in}}\right)^5\frac{\zeta(5,T_{out}/T_{in})}{\zeta(5)}. \tag{8.34}$$

Eqs. (8.31), (8.32), and (8.34) are substituted into Eq. (8.33) to obtain $p_{\gamma}$ as follows:

$$p_{\gamma} = \frac{4}{5}Y_{out}^{\gamma}\left(\frac{T_{out}}{T_{in}}\right)^4\frac{\zeta(5,T_{out}/T_{in})}{\zeta(5)}\quad . \tag{8.35}$$

Substituting the above related equations into the ideal efficiency formula ($\eta_{upper}(T_{in}, T_{out})$) (Eq. (6.12)), we obtain:

$$\begin{aligned} \tilde{\eta}_{upper}(T_{in}, T_{out}) &= 1-\left(\frac{p_Q}{Y_{out}^Q}+\frac{p_{\gamma}}{Y_{out}^{\gamma}}\right)Y_{in}^{\gamma}\frac{T_{out}}{T_{in}} \\ &= 1-\left\{1+\frac{p_{\gamma}}{Y_{out}^{\gamma}}\left(1-Y_{out}^{\gamma}\right)\right\}\frac{5}{4}\frac{T_{out}}{T_{in}} \\ &= 1-\frac{5}{4}\frac{T_{out}}{T_{in}}+\frac{F(4,T_{out}/T_{in})}{\Gamma(5)\zeta(5)}\left(\frac{T_{out}}{T_{in}}\right)^5. \end{aligned} \tag{8.36}$$

**Fig. 11** shows $\tilde{\eta}_{upper}$ and $\eta_{Petela-Landsberg}$. The graph of $\tilde{\eta}_{upper}$ (Fig. 11b) is indistinguishable from that of Carnot efficiency $\eta_{Carnot} = 1 - T_{out}/T_{in}$. However, the application of Taylor expansion around $x = T_{out}/T_{in} = 0$, for example, exhibits a slightly different relationship:

$$\tilde{\eta}_{upper} = 1 - \left(\frac{5}{4} - \frac{1}{4\zeta(5)}\right)x + \frac{\left(-180+15\pi^2+\pi^4+45\psi^{(2)}(2)\right)}{360\zeta(5)}x^5 + O(x^6), \quad (8.37)$$

where $\psi(x)$ is the Polygamma function. The value of the first-order coefficient of $x$ in Eq. (8.37) is (5/4- 1/4ζ(5)) = 1.00890, which is only 0.009 different from 1. The behavior of the difference between the ideal efficiency obtained by the mathematical formula $\tilde{\eta}_{upper}$ in this study and Carnot efficiency $\eta_{Carnot}$ is shown as the $T_{out}/T_{in}$ dependence in Fig.12.

As shown in Fig. 11a, $\eta_{Petela-Landsberg}$ reaches zero when the temperature of the energy source $T_{in}$ becomes equal to the ambient temperature $T_{out}$, and then, returns to positive, as soon as $T_{in}$ becomes lower than $T_{out}$, which contradicts the first law of thermodynamics (conservation of energy). However, $\tilde{\eta}_{upper}(T_{in}, T_{out})$ shown in Fig. 11b, derived from this model, turns negative after reaching zero at $T_{in} = T_{out}$. This made work extraction impossible for $T_{in} < T_{out}$, which is physically reasonable, and then resolved the second contradiction of the Landsberg efficiency $\eta_{Landsberg}$. On the other hand, the first contradiction has been resolved by the mechanism of the simplified model above, that the radiation waste entropy is not emitted by the blackbody radiation within the system, but by the newly emitted photons during the relaxation of the excited pigment electrons to the ground state. From Fig, 11c, we can see $\tilde{\eta}_{upper} > \eta_{Petela-Landsberg}$ at any temperature $T_{in}$ and $T_{out}$ in the realistic condition $T_{in} > T_{out}$. This can be naturally understood from the fact that the difference between $\eta_{Carnot}$, which is very close to $\tilde{\eta}_{upper}$, and $\eta_{Petela-Landsberg}$ is obtained by $\eta_{Carnot} - \eta_{Petela-Landsberg} = \left(1 - \frac{T_{out}}{T_{in}}\right) - \left\{1 - \frac{4}{3}\left(\frac{T_{out}}{T_{in}}\right) + \frac{1}{3}\left(\frac{T_{out}}{T_{in}}\right)^4\right\} = \frac{1}{3}\left(\frac{T_{out}}{T_{in}}\right)\left\{1 - \left(\frac{T_{out}}{T_{in}}\right)^3\right\} > 0 \ (T_{in} > T_{out} > 0)$ This tells us something very important: The ideal efficiency of a light-powered system with radiative entropy discard, such as photosynthesis and solar cells, might be a new efficiency $\tilde{\eta}_{upper}$ nearly equal to the Carnot efficiency, which is higher than the Landsberg efficiency currently cited in many papers in the field. This means that, as with heat engines, the ideal efficiency for technological development of light-powered system engines will be the Carnot efficiency.

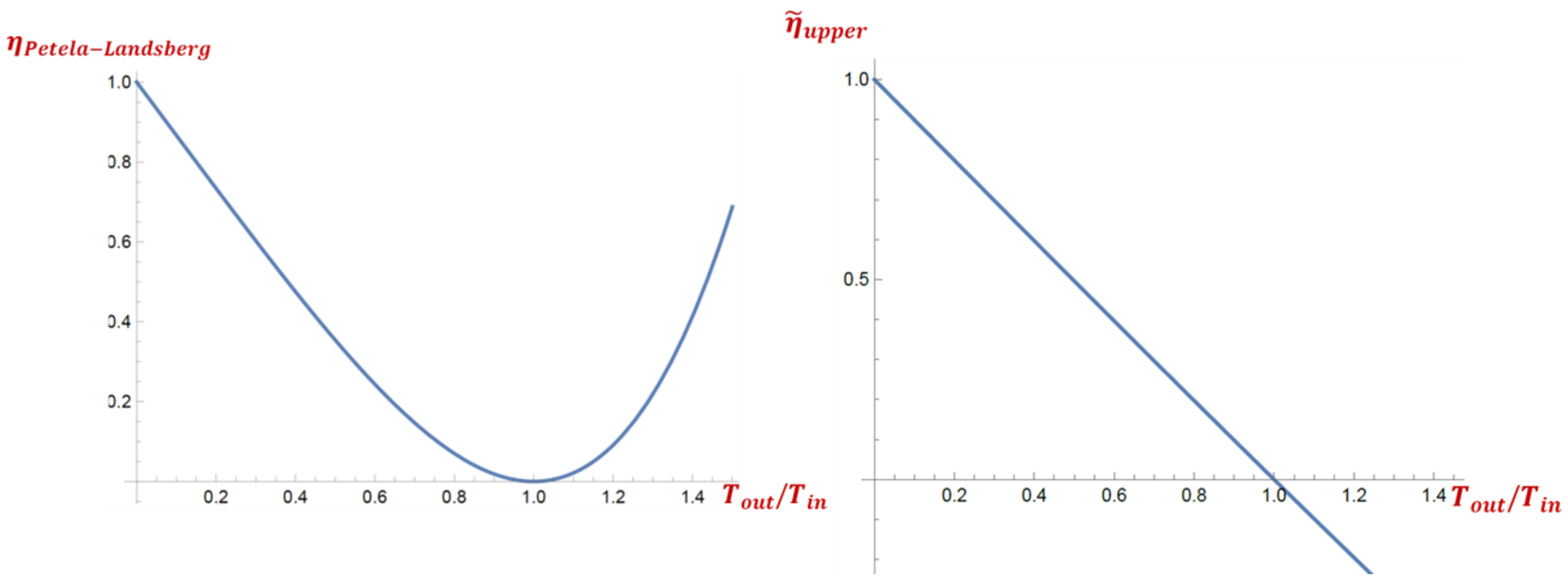

**a** **b**

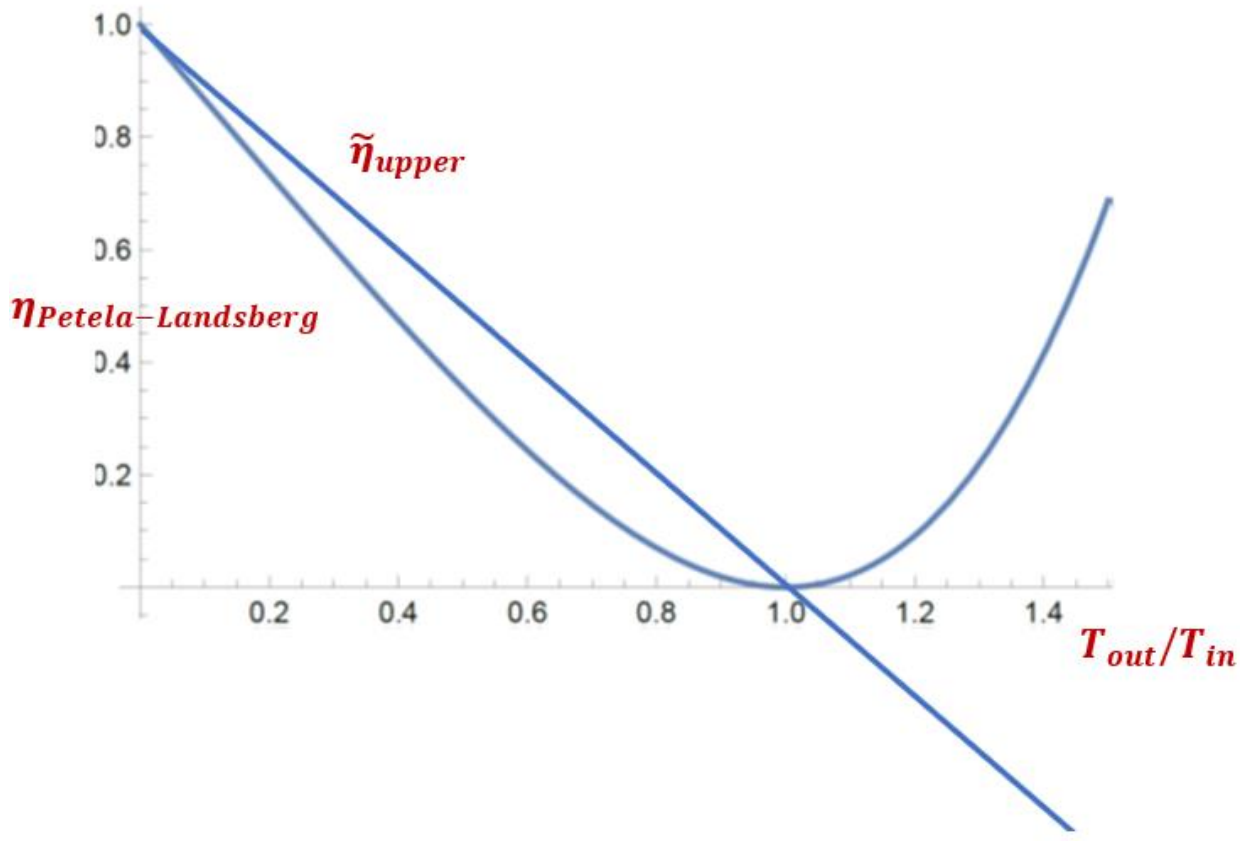

c

**Fig. 11. Two graphs with the horizontal axis as the $T_{out}/T_{in}$ and the vertical axis as the ideal efficiency:** (a) $\eta_{Petela-Landsberg}$ derived in the previous works [9,15] and (b) The ideal efficiency $\tilde{\eta}_{upper}$ derived in this study. (c) Comparison $\tilde{\eta}_{upper}$ and $\eta_{Petela-Landsberg}$: We can see $\tilde{\eta}_{upper} > \eta_{Petela-Landsberg}$ at any temperature $T_{in}$ and $T_{out}$ in the realistic condition $T_{in} > T_{out}$.

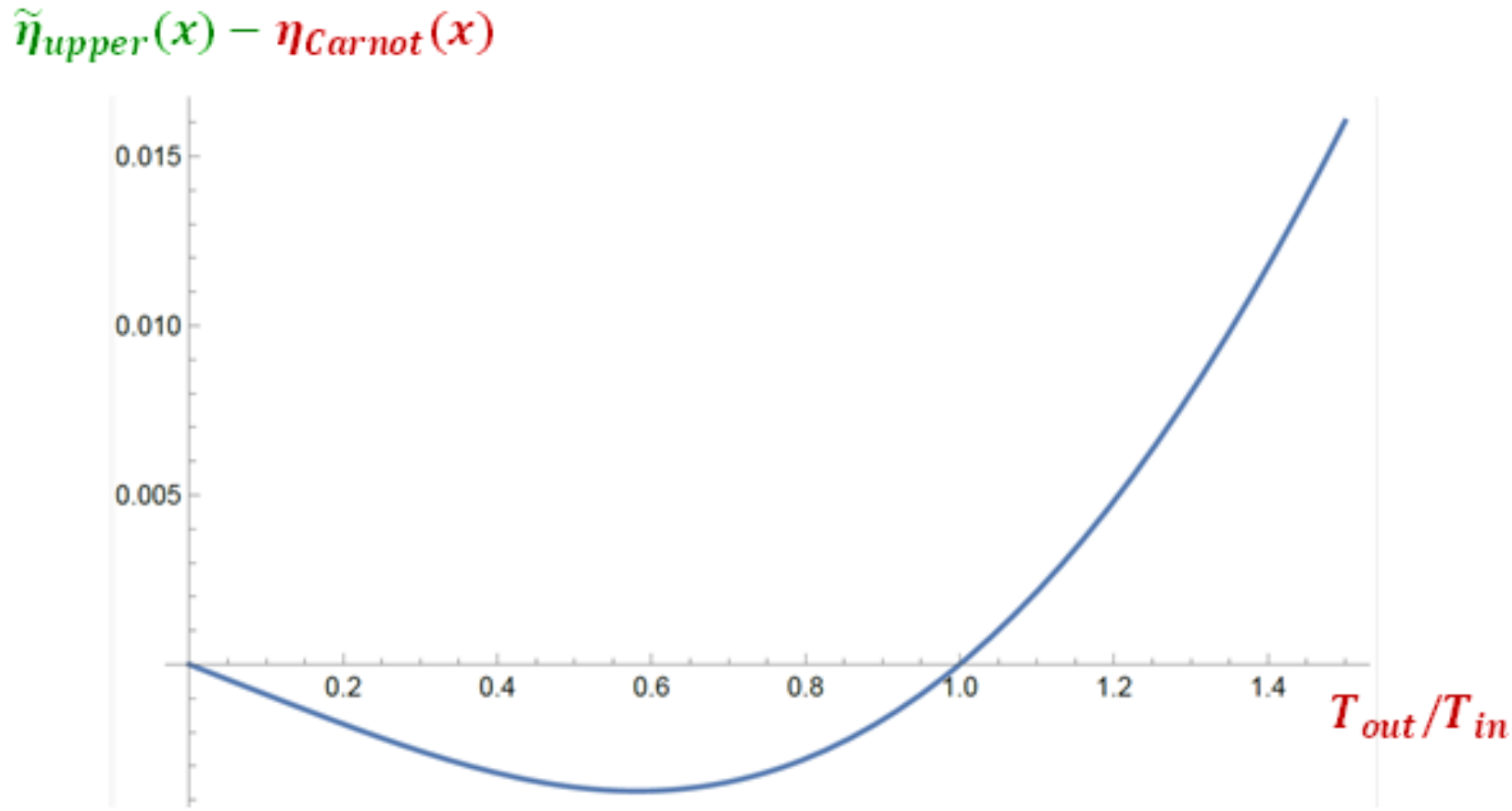

**Fig. 12. The graph of the behavior of the difference between $\tilde{\eta}_{upper}$ and $\eta_{Carnot}$**
The horizontal axis is the $T_{out}/T_{in}$ and the vertical axis is $\tilde{\eta}_{upper} - \eta_{Carnot}$ (the difference between the ideal efficiency obtained by the mathematical formula $\tilde{\eta}_{upper}$ in this study and Carnot efficiency $\eta_{Carnot}$).

## 9. Conclusions and future work

This study formulated the ideal efficiency and Boltzmann-type factor for photosynthetic systems under varied irradiation conditions using energy–entropy flow analysis. In this study, the generalized formula for the ideal efficiency of *light-powered systems* is formulated in the following steps and the theoretical maximum efficiencies derived in previous studies were organized.

First, the following general formula for the ideal efficiency was formulated, assuming that entropy is discarded only by waste heat. (Eq. (5.6))

$$\eta^{\gamma}_{upper}(T, T_{out}, d, |\varepsilon|) = 1 - \frac{T_{out}}{T} Y(d, |\varepsilon|), \tag{9.1}$$

where $d$ is the dilution indicator**,** which is 1 without dilution effect, and the absorption ratio $|\varepsilon|$.
Second, by extending it to cases where entropy is discarded through radiation along with waste heat, this study has derived the most general formula for the ideal efficiency of a *light-powered system* as follows. (Eq. (6.12))

$$\eta_{upper} = 1 - \left(\frac{p_Q}{Y^Q_{out}} + \frac{p_\gamma}{Y^\gamma_{out}}\right) Y^\gamma_{in} \frac{T_{out}}{T_{in}}, \tag{9.2}$$

where $Y^Q_{out}$and $Y^\gamma_{out}$ are the $Y$ factors for the entropy discarded via heat (**thermal discarded entropy)** and radiation (**radiative discarded entropy)** from the system, respectively, and $p_Q$ and $p_\gamma$ are their respective weights (ratios); $Y^\gamma_{in}$ corresponds to the entropy flowing via the blackbody radiation into the system. This formula correctly classified several ideal efficiencies, including Landsberg, Jeter, Spanner, and Petela. It also derived the modified Landsberg efficiency $\tilde{\eta}_{Landsberg}$ (given by Eq. (6.30) and shown in Fig.8b) and the modified Petela efficiency $\tilde{\eta}_{Petela}$ (given by Eq. (C.6) and shown in Fig.14), as well as the Jeter efficiency under different conditions (shown in Table). It has given us with a unified understanding of these efficiencies.
Third, this study also classified previous research on *light-powered systems* into (a) the piston-cylinder model (closed photon-gas model) and (b) the flowing radiation model (open photon-gas model), and showed the suitability of the latter for microscopic *light-powered systems* by deriving the following inequality as a condition for identifying the wavelength in the microscopic scale. (Eq. (7.7))

$$V \geq \frac{\lambda^3}{8\pi}. \tag{9.3}$$

where $V$ and $\lambda$ are the volume of a system and the wavelength of the monochromatic light irradiated, respectively.
Fourth, this study clarified that the ideal efficiency derived from the flowing radiation model as proposed by Landsberg and Tonge [9] has two issues, and solved them by quantitative analysis on a simplified mathematical model based on Einstein's emission and absorption theory. In this model, the radiative discarded entropy along with the waste heat, is not emitted by the blackbody radiation within the system, as assumed by Landsberg for the derivation of $\eta_{Petela-Landsberg}$[9], but by the newly emitted photons during the relaxation of excited pigment electrons to the ground state in the system. This is a more realistic condition. Consequently, the new formula of the ideal efficiency $\tilde{\eta}_{upper}(T_{in}, T_{out})$ for the flowing radiation model suitable for microscopic *light-powered systems* was derived as follows. (Eq. (8.36))

$$\tilde{\eta}_{upper}(T_{in}, T_{out}) = 1 - \frac{5}{4}\frac{T_{out}}{T_{in}} + \frac{F(4, T_{out}/T_{in})}{\Gamma(5)\zeta(5)}\left(\frac{T_{out}}{T_{in}}\right)^5. \tag{9.4}$$

The ideal efficiency $\tilde{\eta}_{upper}(T_{in}, T_{out})$ obtained is found to be very similar to Carnot efficiency.

From Fig, 11c, we can see $\tilde{\eta}_{upper} > \eta_{Petela-Landsberg}$ at any temperature $T_{in}$ and $T_{out}$ in the realistic condition $T_{in} > T_{out}$. This tells us something very important: The ideal efficiency of a light-powered system with radiative entropy discard, such as photosynthesis and solar cells, might be a new efficiency $\tilde{\eta}_{upper}$ nearly equal to the Carnot efficiency, which is higher than the Landsberg efficiency currently cited in many papers in the field. This means that, as with heat engines, the ideal efficiency for technological development of light-powered system engines will be the Carnot efficiency.

This fundamental study has provided a basic and correct understanding of the theoretical efficiency of *light-powered systems* such as next-generation solar power and artificial photosynthesis, which are crucial for a future decarbonized society.

Based on this study, the following further considerations and analyses are to be undertaken with the collaborators.

(1) Based on Einstein's theory of radiation and absorption in a two-level system, we will develop a more realistic and general mathematical model analysis. We will then conduct analyses that can be applied to actual natural and artificial light-powered systems.

(2) This study has confirmed that standard statistical mechanics on the macroscopic scale can be applied to photosynthetic systems on the microscopic scale, at least as far as radiation is concerned, through the proper use of the flowing radiation model (open photon gas model). Verification on the matter side (LHC:Light Harvesting Complex) is planned for the future in collaboration with specialists in photosynthesis research. In particular, we will perform the practical entropy analyses that take into account the various relaxation processes of the excited states of the LHC in actual photosynthesis. We will investigate the specific mechanisms of entropy generation (e,g.,[50]), such as quantum coherence and excited energy transfer (based also on the work of the present author [51]), and elucidate the efficient transfer of excited energy to the reaction center.

**Glossary**

LHC: Light Harvesting Complex

*Light-powered* system: In this study, this is defined not only as a system that outputs electrical energy, such as solar power generation, photovoltaics and solar cells, etc., but also as a system from which any kind of physical work is extracted using light energy, including natural and artificial photosynthesis

**Data availability**

No database was used and no new data were generated for this article.

## ACKNOWLEDGMENTS

I would like to express my gratitude to Prof. Ayumi Tanaka and Prof. Iwao Yamazaki for their valuable discussions on photosynthesis systems and optical science. Additionally, I extend my thanks to Prof. Yousuke Ohyama and Prof. Fumio Hiroshima for their insightful mathematical discussions, and Dr. Nobuki Maeda, Prof. Takashi Uchida, and Prof, Shigenori Tanaka for their valuable contributions to the physical discussions.

**Appendix A**

The condition $Y_{out}^{\gamma}(\Delta T/T_{out} \rightarrow 0) = 1$ obtained from the first-order evaluation of $\Delta T/T_{out}$ in Eq. (6.4) is the same as the condition $Y_{out}^{\gamma}\left(\Delta N_{out}^{\gamma}/N_{out}^{\gamma} \rightarrow 0\right) = 1$ obtained from the first-order evaluation of $\varepsilon_{out}^{\gamma} = \Delta N_{out}^{\gamma}/N_{out}^{\gamma}$, as shown below.

The number of photons of blackbody radiation at temperature $T$ [K] is given by

$$N_{\gamma}(T) = \int_0^{\infty} d\nu g(\nu) \frac{1}{e^{h\nu/k_B T - 1}}, \quad \text{(A.1)}$$

where the number of quantum states of a photon $g(\nu)$ per unit frequency is given by

$$g(\nu)d\nu = \frac{8\pi\nu^2 V}{c^3} d\nu. \quad \text{(A.2)}$$

Thus, from Eqs. (A.1) and (A.2), $N_{\gamma}(T)$ can be expanded as

$$N_{\gamma}(T) = \frac{8\pi V}{c^3} \int_0^{\infty} d\nu \frac{\nu^2}{e^{h\nu/k_B T - 1}}$$

$$= \frac{8\pi V}{c^3} \int_0^{\infty} d\nu \frac{\nu^2}{e^{a\nu}} \qquad (a \equiv h/k_B T)$$

$$= \frac{8\pi V}{c^3} \frac{1}{a^3} \int_0^{\infty} d\mu \frac{\mu^2}{e^{\mu} - 1} \qquad \text{(Replaced by } \mu = a\nu)$$

$$= \frac{8\pi V}{c^3} \left(\frac{k_B T}{h}\right)^3 \int_0^{\infty} d\mu \frac{\mu^2}{e^{\mu} - 1}$$

$$= \frac{8\pi V k_B^3}{c^3 h^3} T^3 \Gamma(3)\zeta(3), \quad \text{(A.3)}$$

and

$$\Delta N_{\gamma}(T) = \left\{\frac{\partial}{\partial T} N_{\gamma}(T)\right\} \Delta T$$

$$= \left\{\frac{\partial}{\partial T}\left(\frac{8\pi V k_B^3}{c^3 h^3} T^3 \Gamma(3)\zeta(3)\right)\right\} \Delta T$$

$$= \frac{8\pi V h}{c^3 k_B} 3\Gamma(3)\zeta(3) T^2 \Delta T. \quad \text{(A.4)}$$

Eqs. (A.3) and (A.4) give

$$\frac{\Delta N_{\gamma}(T)}{N_{\gamma}(T)} = 3\frac{\Delta T}{T}. \quad \text{(A.5)}$$

Eq. (A.5) shows the equivalence between $\Delta T/T \rightarrow 0$ and $\varepsilon_{\gamma}(\nu, T) = \Delta N_{\gamma}(T)/N_{\gamma}(T) \rightarrow 0$.
Therefore, the following equation holds:

$$Y(\Delta T/T \rightarrow 0) = Y\left(\varepsilon_{\gamma}(\nu, T) \rightarrow 0\right). \quad \text{(A.6)}$$

In this study the frequency dependence of the light absorption ratio $\left|\varepsilon_{\gamma}(\nu, T)\right|$ is ignored for simplicity and $|\varepsilon_\gamma(\nu, T)| = |\varepsilon_\gamma(T)|$ is used as an approximation for the analysis.

**Appendix B**

Although Eq. (6.12) is derived from the second law of thermodynamics, the condition $p_\gamma = \frac{S^\gamma_{out}(T_{out})}{S^\gamma_{in}(T_{in})}$ applied to it is not guaranteed to be physically reasonable a priori and requires separate evaluation. Three additional considerations are given below as examples.

( I ) For the general formula represented by Eq. (6.12)

$$\eta_{upper} = 1 - \left(\frac{p_Q}{Y^Q_{out}} + \frac{p_\gamma}{Y^\gamma_{out}}\right) Y^\gamma_{in} \frac{T_{out}}{T_{in}}, \tag{B.1}$$

If $Y^\gamma_{in}(\varepsilon^\gamma_{in} = 1) = 4/3$ and $Y^\gamma_{out}(\varepsilon^\gamma_{out} = 1) = 4/3$ are imposed, then the entropy discard condition $p_\gamma = 1, p_Q = 0$ changes $\eta_{Petela-Landsberg} = 1 - \frac{4}{3}\frac{T_{out}}{T_{in}} + \frac{1}{3}\left(\frac{T_{out}}{T_{in}}\right)^4$ to $\eta_{Jeter} = 1 - T_{out}/T_{in}$, but leads to a physically meaningless result, because from Eq. (6.13), $p_\gamma = 1$ gives $p_\gamma = \frac{S^\gamma_{out}}{S^\gamma_{in}} = 1$, i.e., $S^\gamma_{out} = S^\gamma_{in}$, which leads to $T_{out} = T_{in}$ and then to $\eta_{Jeter} = 0$. .

( II ) There is still a critical debate about the fact that even though $T_{in} \geq T_{out}$, the condition $T_{in} < (4/3)T_{out}$ gives $\eta_{Spanner} < 0$ [11], and Spanner himself referred to this as an anomalous result, and tried to explain the reason for it in his original paper [12]. This sense of contradiction stems from the prejudice that has been generalized from the fact that in the case of a heat engine, work can always be extracted if the temperature of the heat source is even slightly higher than the ambient temperature. This prejudice is based on the fact that in the heat engine we usually assume the first-order evaluable condition $\varepsilon \to 0$, i.e., a quasi-equilibrium between the heat source and the system. The behavior of Spanner efficiency, which does not assume this condition $\varepsilon_\gamma = 0$, but the condition $\varepsilon_\gamma = 1$, is consistent with the first and second laws of thermodynamics and is never physically unreasonable. This is discussed in more detail below.

The energy efficiency becomes $\eta = \{E^\gamma_{in} - (E^Q_{out} + E^\gamma_{out})\}/E^\gamma_{in}$, and the following two conditions are necessary to extract the work from the first and second laws of thermodynamics:

$$E^Q_{out} + E^\gamma_{out} \leq E^\gamma_{in} \quad \text{(from the first law of thermodynamics)}, \tag{B.2}$$

$$S^Q_{out} + S^\gamma_{out} \geq S^\gamma_{in} \quad \text{(from the second law of thermodynamics)}. \tag{B.3}$$

In the case of the Spanner efficiency, the following conditions are given:

$$p_\gamma = 0 \text{ and } \quad p_Q = 1, \tag{B.4}$$

and

$$Y^Q_{out} = 1, Y^\gamma_{in} = \frac{4}{3} \ (\varepsilon^\gamma_{in} = 1). \tag{B.5}$$

From Eqs. (B2) and (B.4), the following can be obtained:

$$E_{out}^{Q} \leq E_{in}^{\gamma} \quad \text{(from the first law of thermodynamics)}, \tag{B.6}$$

From Eqs. (B.3)–(B.5), the following can be derived:

$$E_{out}^{Q}/T_{out} \geq (4/3)E_{in}^{\gamma}/T_{in} \quad \text{(from the second law of thermodynamics)}. \tag{B.7}$$

Finally, (B.6) and (B.7) indicate necessity of $(4/3)E_{in}^{\gamma}T_{out}/T_{in} \leq E_{out}^{Q} \leq E_{in}^{\gamma}$. Thus, the following condition becomes necessary:

$$T_{in} \geq (4/3)T_{out}. \tag{B.8}$$

As mentioned above, the condition (B.8) obtained from both the first and second laws of thermodynamics gives the condition for $\eta_{Spanner} \geq 0$, and therefore, it is never physically unreasonable to obtain $\eta_{Spanner} < 0$ when $T_{in} < (4/3)T_{out}$.

On the other hand, in the case of Carnot efficiency, because $Y_{out}^{Q} = 1, Y_{in}^{Q} = 1$ are assumed, conditions (B6) and (B7) become $E_{out}^{Q} \leq E_{in}^{Q}$ and $E_{out}^{Q}/T_{out} \geq E_{in}^{Q}/T_{in}$, respectively. Then, the condition (B.8) becomes

$$T_{in} \geq T_{out}. \tag{B.9}$$

Condition (B.9) has given us the incorrect prejudice that work can always be extracted as long as there is even a small temperature difference between any energy source and any powered system. This is the source of our sense of contradiction for the aforementioned behavior of the Spanner efficiency. If the condition $Y_{out}^{Q} = 1, Y_{in}^{\gamma} = 1$ ($\varepsilon_{in}^{\gamma} \to 0$) is also imposed for a light-powered system, then the condition for work extraction is (B.9), which is the same as that for Carnot efficiency, i.e., the Jeter efficiency $\eta_{Jeter}$.

(Ⅲ) The formula $\eta_{Jeter} = 1 - T_{out}/T_{Sun}$ is the same as that for Carnot efficiency. However, the mechanism underlying the obtained $\eta_{Jeter}$, explained by Jeter in their original paper [19], is incorrect as follows:

According to Jeter [19], the entropy of solar radiation in the upper atmosphere, before it enters the atmosphere and undergoes atmospheric scattering, is equal to the entropy at the surface of the Sun. Therefore, the ideal efficiency of a light-powered system using sunlight that is not scattered by the atmosphere on the Earth, such as sunlight outside the atmosphere or direct sunlight on the ground, is equal to Carnot efficiency due to the solar surface temperature, i.e., $\eta_{Jeter} = 1 - T_{out}/T_{Sun}$.

The claim that the radiative temperature of sunlight outside the atmosphere is $T_{Sun}$= 5800 K is consistent with the result of the present author's study in Appendix C of [29,30]. However, the conclusion that this gives Carnot efficiency is not correct. In addition to this condition, the correct condition, under which the ideal efficiency $\eta_{upper}$ of the light-powered system is $\eta_{Jeter}$ is to use $Y_{out}^{Q} = Y_{out}^{\gamma} = Y_{in}^{\gamma} = 1$ in Eq. (6.12) constructed in this study. If we use the incident flux $E_{in}^{\gamma} = \sigma T_{Sun}^{4}$ assumed in the literature [19], i.e., $Y_{in}^{\gamma} = 4/3$, then the ideal efficiency using direct solar radiation without scattering on the Earth becomes the Spanner efficiency, $\eta_{Spanner} = 1 - (4/3)T_{out}/T_{Sun}$.

## Appendix C. Derivation of the ideal efficiency from the standard definition of energy efficiency based on the $p$-$V$ graph from which Petela derived $\eta_{Petela}$

In previous studies [15,16], the maximum efficiency was derived using the piston-cylinder radiation model (closed photon gas model) with $p$-$V$ graphs (Fig. 13), based on Petela's work. The process is assumed to be in radiative equilibrium at any instant, with a defined temperature, indicating a quasi-static process. The radiation in the piston-cylinder system undergoes a reversible, adiabatic expansion from state 1 (where the temperature of the energy source, such as the Sun, is $T_1$) to state 0 (where the temperature matches the ambient temperature $T_0$) (Fig. 13). The maximum energy efficiency for this whole cycle is given by

$$\eta_{max} = \frac{W_{max}}{U_1}, \tag{C.1}$$

where $U_1$ is the initial internal energy of the photon gas. Eq. (C.1) differs from the standard definition of energy efficiency, which is $\eta = (\text{extracted work})/(\text{absorbed energy})$, and has been calculated using the formula $\eta = (\text{extracted work})/(\text{internal energy})$. In their original work, Petela probably derived the ideal efficiency $\eta_{Petela} = 1 - (4/3)T_0/T_1 + (1/3)(T_0/T_1)^4$ in the piston-cylinder model (closed radiation model) as the theoretical maximum efficiency from the point of view of exergy, using the $p$-$V$ graph, in accordance with the above-mentioned definition.

The ideal efficiency $\eta_{upper}$ calculated using the standard definition, based on the $p$-$V$ graph used by Petela, differs from $\eta_{Petela}$, and is found to be physically reasonable. (Details are given below).

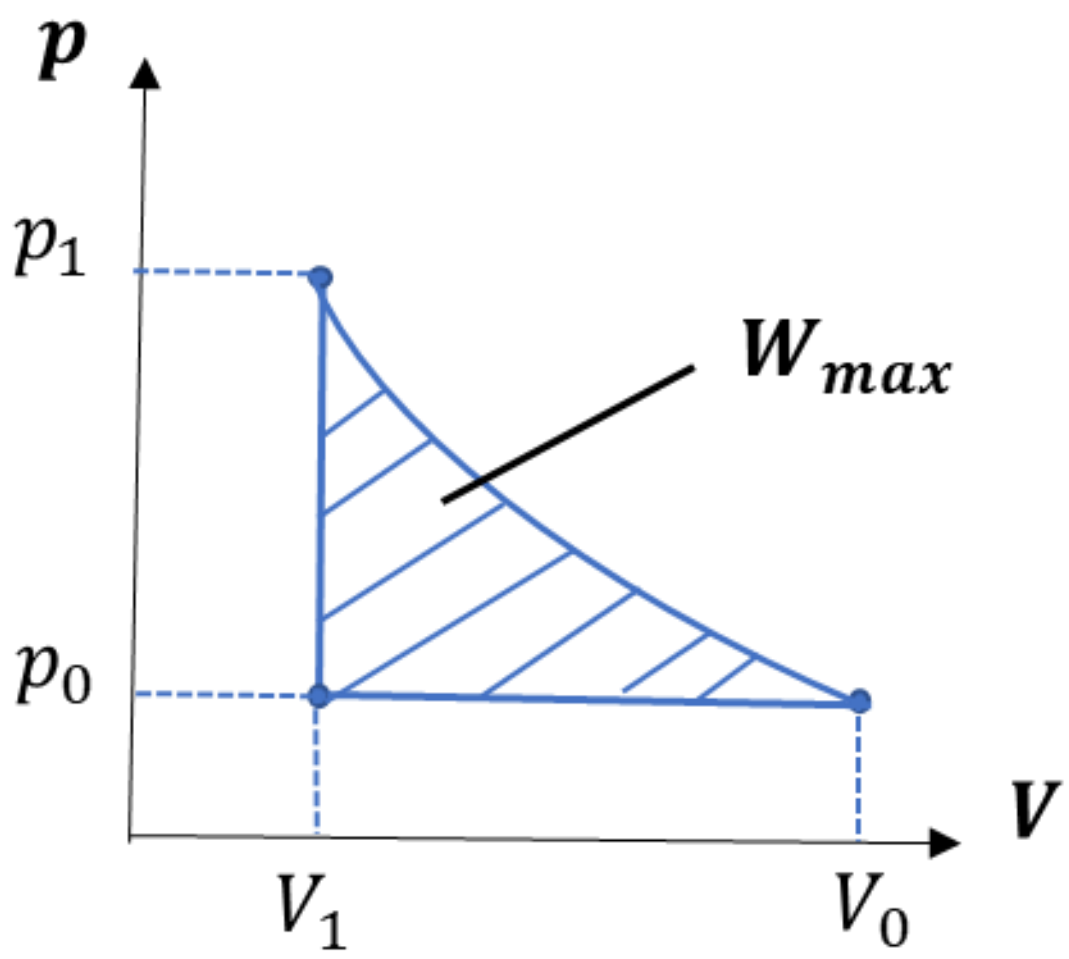

**Fig. 13. $p$-$V$ graph of a photon gas in a cylinder with a piston [15,16]**

The work represented by the shadow is done by expanding photon gas from state 1 to state 0. The photon gas is allowed to expand reversibly and adiabatically from state 1, where T1 is the equivalent blackbody temperature of the Sun (5800 K) and T0 is the ambient temperature.

The shaded area shows the work $W_{max}$ extracted to the outside by the ideal piston-cylinder system (closed photon gas) and is given by

$$W_{max} = \int_{V_1}^{V_0} pdV - p_0(V_0 - V_1), \quad \text{(C.2)}$$

where the first term is the work done by the photon gas on the outside, represented as an integration in the adiabatic expansion process, the second term is the work done by the outside on the photon gas. The entropy and internal energy of radiation at temperature $T$ are given by

$$S^{\gamma}(T,V) = \frac{4}{3}AVT^3 \quad \text{(C.3)}$$

and

$$U^{\gamma}(T,V) = AVT^4, \quad \text{(C.4)}$$

respectively. Here, $A = \pi^2 k_B^4/(15c^3\hbar^3)$ is a constant. Therefore, $VT^3$ is constant in the adiabatic process, i.e., the entropy preserving process. Using Eqs. (C.2) and (C.3) and $p^{\gamma} = \left(\frac{1}{3}\right)U^{\gamma}/V$ for the pressure of a photon gas, we obtain the formula $W_{max} = U_1 V_1[1 - (4/3)T_0/T_1 + (1/3)(T_0/T_1)^4]$ and, consequently, $\eta_{max} = \frac{W_{max}}{U_1} = 1 - (4/3)T_0/T_1 + (1/3)(T_0/T_1)^4$ [15,16].

However, this is not the correct upper efficiency, although it can be referred to as exergy. Formulating this as the ratio of the maximum work extracted to the absorbed energy $W_{max}/(U_1 - U_0)$ which is the standard definition of ideal efficiency, yields the following equation:

$$\eta_{upper} = \frac{W_{max}}{U_1 - U_0}$$

$$= \frac{u_1 V_1}{(u_1 - u_0)V_1}\{1 - (4/3)T_0/T_1 + (1/3)(T_0/T_1)^4\} \quad . \quad \text{(C.5)}$$

The final correct upper bound is obtained as

$$\eta_{upper} = \frac{1}{1-(T_0/T_1)^4}\{1 - (4/3)T_0/T_1 + (1/3)(T_0/T_1)^4\} = \tilde{\eta}_{Petela}. \quad \text{(C.6)}$$

If the ratio $T_0/T_1$ between the radiation temperature of the energy source (relatively high temperature; $T_1$) and the temperature of a system (equal to the ambient (relatively low) temperature $T_0$) is expressed as $x$, then $\eta_{upper}(x)$ is expressed as a function of $x$ as follows:

$$\tilde{\eta}_{Petela}(x) = \frac{1}{1-x^4}\{1 - (4/3)x + (1/3)x^4\}$$

$$= \frac{1}{3}\frac{1}{(x+1)(x^2+1)}(1-x)(x^2 + 2x + 3) \quad . \quad \text{(C.7)}$$

According to Eq. (C.7), $\tilde{\eta}_{Petela}(x = 1) = 0$ and $\tilde{\eta}_{Petela}(x \to 0) = 1$, lead to the following physically reasonable equations:

$$\tilde{\eta}_{Petela}(T_1 = T_0) = 0 \quad , \tag{C.8}$$

$$\tilde{\eta}_{Petela}(T_1 \to \infty) = 1 \quad . \tag{C.9}$$

The differential of $\tilde{\eta}_{Petela}$ with respect to $x$ is

$$\frac{d}{dx}\tilde{\eta}_{Petela} = -\frac{4}{3}\frac{1}{\{(x+1)(x^2+1)\}^2}(3x^2 + 2x + 1). \tag{C.10}$$

This shows that $\tilde{\eta}_{Petela}(x)$ is monotonically decreasing in the range $0 \leq x$. The numerical analysis graph of $\tilde{\eta}_{Petela}(x)$ is shown in Fig.14, which indicates that the derived ideal efficiency $\tilde{\eta}_{Petela}(T_0/T_1)$ becomes negative after $T_1 = T_0$. Thus, work cannot be extracted from the radiation flow from low to high temperatures, indicating that the point of contradiction in the original $\eta_{Petela}(T_0/T_1)$ has been resolved.

Next, this $\tilde{\eta}_{Petela}(T_0/T_1)$ is verified by applying it to Eq. (6.12) as the general formula of the ideal efficiency constructed in this study. Based on the assumptions in the previous studies, the temperatures $T_1$ and $T_0$ are considered as $T_{in}$ and $T_{out}$, respectively, and the following conditions are set:

$$Y_{out}^{Q} = 1, Y_{out}^{\gamma} = Y_{in}^{\gamma} = 4/3. \tag{C.11}$$

Eq. (C.11) is applied to Eq. (6.12) to obtain

$$\eta_{upper} = 1 - \frac{4}{3}\frac{T_{out}}{T_{in}} + p_\gamma \frac{1}{3}\frac{T_{out}}{T_{in}}. \tag{C.12}$$

By matching Eqs. (C.7) and (C.12), the respective weights (ratios) of the entropy discard via heat and radiation ($p_Q$ and $p_\gamma$, respectively) are obtained as follows:

$$p_\gamma = \frac{4x^3}{(x+1)(x^2+1)}, \tag{C.13}$$

$$p_Q = \frac{(1-x)(3x^2+2x+1)}{(x+1)(x^2+1)}, \tag{C.14}$$

where $x = T_{out}/T_{in}$.

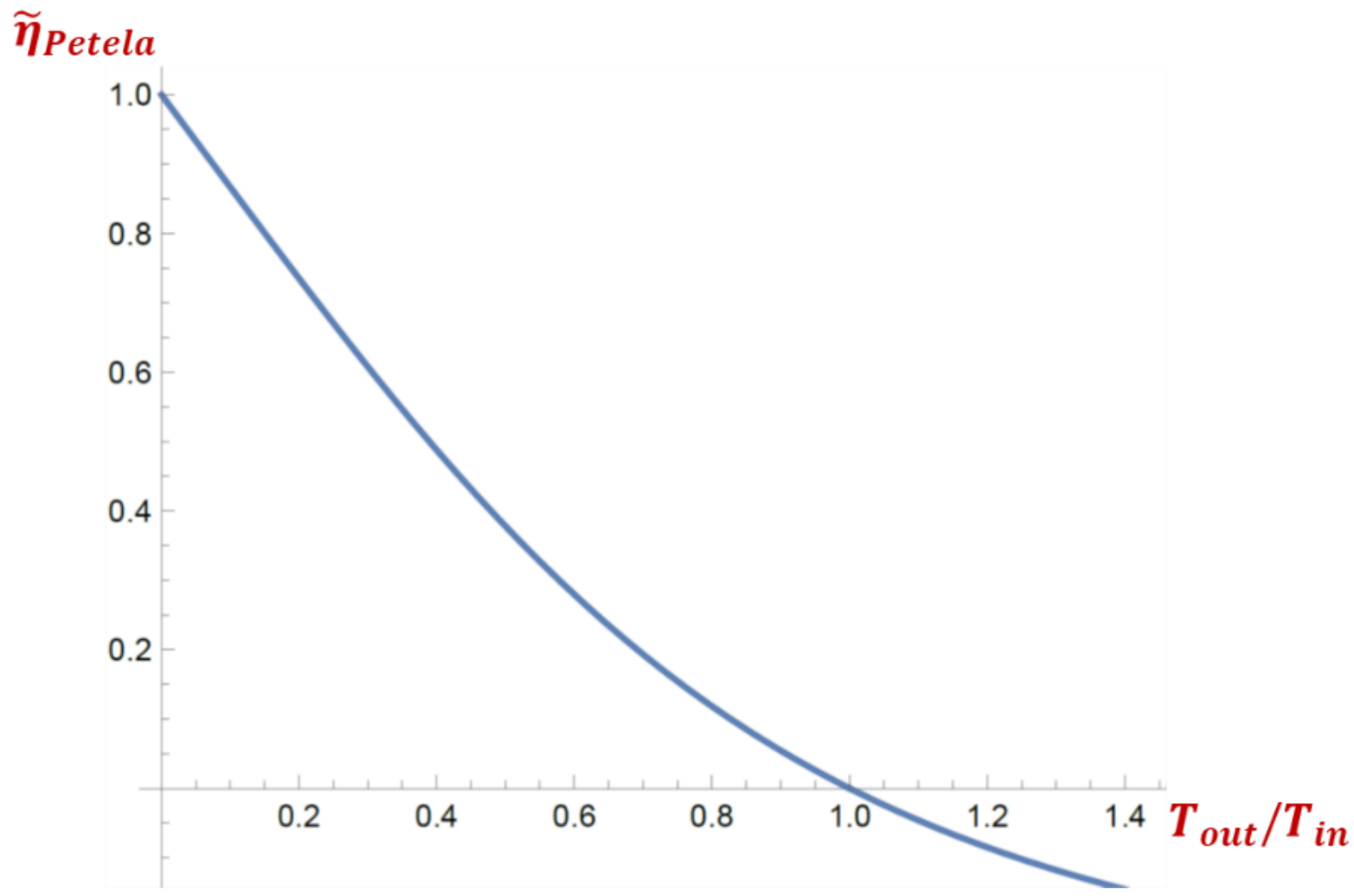

**Fig. 14. The upper efficiency $\tilde{\eta}_{Petela}(T_0/T_1)$ obtained by the standard definition $\eta =$ (extracted work) / (absorbed energy) based on the $p$-$V$ graph from which Petela derived $\eta_{Petela}$**
The graph is shown with the horizontal axis as $T_{out}/T_{in}$ and vertical axis as the ideal efficiency $\tilde{\eta}_{Petela}(T_0/T_1)$ decreases monotonically with $T_{out}/T_{in}$, and becomes negative after the temperature of the energy source $T_1$ becomes equal to the ambient temperature $T_0$.